\RequirePackage{ifpdf}
\documentclass[12pt]{JHEP3}
\pdfoutput=1

\usepackage{graphicx}
\usepackage{cite}
 \usepackage{ae}
 \usepackage{aecompl}
\usepackage{amsmath,epsfig}
\usepackage{amssymb,amsfonts}
\usepackage{latexsym}
\usepackage{epsfig}
\usepackage{subfigure}

\relax
\renewcommand{\theequation}{\arabic{section}.\arabic{equation}}

\def\be{\begin{equation}}
\def\ee{\end{equation}}

\newcommand{\la}{\lambda}

\newcommand{\bear}{\begin{eqnarray}}
\newcommand{\bea}{\begin{eqnarray}}
\newcommand{\eear}{\end{eqnarray}}
\newcommand{\eea}{\end{eqnarray}}
\newbox\pippobox

\def\II{\relax{\rm I\kern-.18em I}}

\def\l{\lambda}
\def\m{\mu}
\def\n{\nu}

\def\s{\sigma}
\def\pa{\partial}
\def\t{\theta}

\def\a{\alpha}
\def\b{\beta}

\def\tr{\ensuremath{\mathrm{Tr}}}

\def\t{\tau}

\def\d{\delta}

\newcommand{\mA}{\mathcal{A}}

\newcommand{\mE}{\mathcal{E}}

\newcommand{\mG}{\mathcal{G}}
\newcommand{\mV}{\mathcal{V}}

\newcommand{\mO}{\mathcal{O}}
\newcommand{\mP}{\mathcal{P}}

\newcommand{\nn}{\nonumber}

\newcommand{\eps}{\epsilon}

\newcommand{\order}[1]{${\cal O}\left(#1 \right)$}
\newcommand{\morder}[1]{{\cal O}\left(#1 \right)}

\renewcommand{\la}{\lambda}

\renewcommand{\l}{\lambda}
\renewcommand{\L}{\Lambda}
\newcommand{\LUV}{\Lambda_\mathrm{UV}}
\newcommand{\LIR}{\Lambda_\mathrm{IR}}
\renewcommand{\t}{\tau}
\newcommand{\Lt}{\Lambda_\tau}

\newcommand{\xBZ}{x_\mathrm{BZ}}
\newcommand{\vs}{u}
\newcommand{\gf}{w}
\newcommand{\h}{\kappa}
\newcommand{\Awf}{A}
\newcommand{\aslash}[1]{ \rlap{/}{#1} }

\title{Massive holographic QCD\\ in the Veneziano limit}

\author{Matti J\"arvinen\\ 
 ~\\
 \href{http://www.lpt.ens.fr/}{Laboratoire de Physique Th\'eorique},
\'Ecole Normale Sup\'erieure \& Institut de Physique Th\'eorique Philippe Meyer, 24 rue Lhomond, 75231 Paris Cedex 05,
France 
\\
{
\centering 
and 

}
\\ 
\href{http://hep.physics.uoc.gr/}
 {Crete Center for Theoretical Physics}, Department of Physics, University of Crete,
 71003 Heraklion, Greece}

\abstract{
QCD at finite, flavor independent 
quark mass is analyzed by using bottom-up holography in the Veneziano limit, where the backreaction of quarks to the gluon dynamics is fully included. 
The dependence on the quark mass of observables such as the bound state masses, the chiral condensate, the S-parameter, and the critical temperatures is studied. Many of the results are argued to be universal, i.e., independent of the details of the holographic model, and compared to explicit computations in the V-QCD models. 
The effect of adding four-fermion operators in QCD is also discussed.
}


\preprint{CCTP-2015-03\\ CCQCN-2015-61}

\begin{document}

\maketitle 


\section{Introduction and summary} 
 
\begin{figure}[!tb]
\begin{center}
\includegraphics[width=0.65\textwidth]{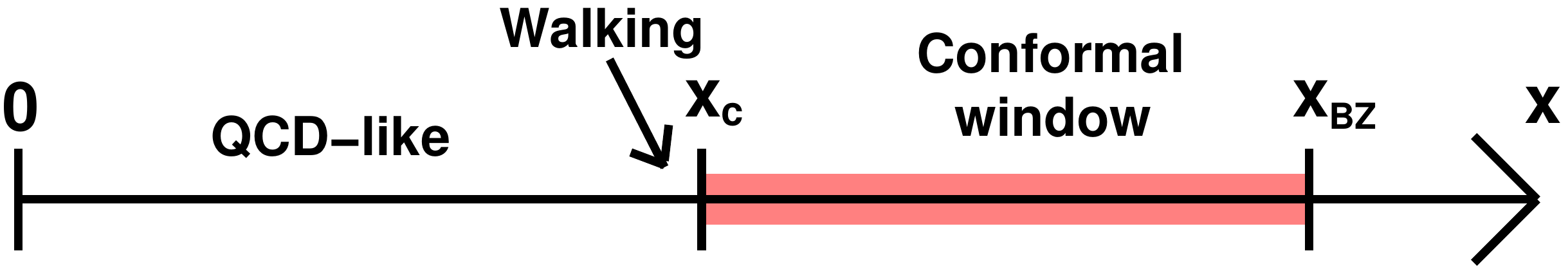}
\end{center}
\caption{The phases of QCD in the Veneziano limit (at zero temperature and quark mass) as a function of $x=N_f/N_c$.}
\label{fig:xphases}\end{figure}

QCD displays an interesting phase structure as the number of flavors $N_f$ and number of colors $N_c$ are varied. 
It is natural to discuss the phase diagram in 
the Veneziano limit~\cite{veneziano,vu1}:
\be \label{Vlimit}
 N_c \to \infty\ , \quad N_f \to \infty \ , \quad x \equiv \frac{N_f}{N_c}\ \ \mathrm{fixed} \ , \quad g^2 N_c \ \ \mathrm{fixed}\ ,
\ee
as a function of the variable $x$ which has become continuous in this limit. The ``standard'' expectation for the diagram is shown in Fig.~\ref{fig:xphases} (at zero temperature and quark mass). We restrict to the interval $0<x< 11/2 \equiv \xBZ$ where the theory is asymptotically free. Various regimes can be identified:
\begin{itemize}
 \item The QCD regime $0<x<x_c$ where the infrared (IR) dynamics is similar to ordinary QCD (having $N_c=3$ and a few light quarks), with confinement and chiral symmetry breaking.
 \item The walking regime with $0<x<x_c$ and $x_c-x \ll 1$ where the coupling constant of the theory varies very slowly, i.e., ``walks'', over a large range of energies.
 \item The conformal window $x_c<x<x_\mathrm{BZ}$ where the theory runs to an IR fixed point (IRFP), and chiral symmetry is intact.
\end{itemize}

The existence of the conformal window is solid in the Banks-Zaks (BZ) limit $x\to \xBZ$, because the value of the coupling is parametrically small and perturbation theory is trustable~\cite{bankszaks}. It is also credible that dynamics at small $x \lesssim 1$ is similar to ordinary QCD -- the $\sim 1/N_c^2$ corrections arising in the Veneziano limit are not expected to change the picture qualitatively. But the nature of the ``conformal transition'' at $x=x_c$, and the behavior of the theories near the transition, is an open question. 

The phase diagram of Fig.~\ref{fig:xphases} with a walking regime is obtained in the Dyson-Schwinger approach~\cite{ds}. The transition is then of the BKT type~\cite{bkt}, associated with the so-called Miransky scaling law~\cite{miransky}. The existence of the walking regime is important, since theories in this region may have properties, which are desirable for technicolor candidates~\cite{holdom,walk2} (see also the reviews~\cite{review}). The location of the transition has been estimated by using different approaches~\cite{cw}. However, it has also been suggested that the transition is discontinuous, such that the dynamics ``jumps'' and walking is absent~\cite{Sannino:2012wy}.
There is an ongoing effort to clarify these issues by using first-principles lattice simulations~\cite{lattice,Lombardo:2014mda}, but as it turns out, obtaining reliable results in the transition region is difficult.

The holographic V-QCD models~\cite{jk} also have the phase diagram of Fig.~\ref{fig:xphases}, including a BKT transition an walking. The ``V'' in V-QCD refers to the Veneziano limit of~\eqref{Vlimit}, where the models are defined. V-QCD is based on two building blocks, the improved holographic QCD (IHQCD)~\cite{ihqcd} for the gluon sector, and a method for adding flavor by inserting a pair of space filling $D4-\overline{D4}$ branes~\cite{Bigazzi:2005md,ckp}. This means that the gluon sector is described by a five dimensional Einstein-dilaton gravity, and the flavor sector is described by a tachyonic Dirac-Born-Infeld action. The sectors are fully backreacted in the Veneziano limit. The structure of V-QCD at finite temperature and chemical potential has been studied in~\cite{alho,Alho:2013hsa,Iatrakis:2014txa,pionloop}. Two-point correlators and bound state masses were analyzed in~\cite{letter,Arean:2013tja}.

There have also been numerous other holographic models addressing the phenomena expected in QCD at finite values of $x$. Walking gauge theories have been modeled by using as a starting point the traditional bottom-up models for example in~\cite{butc,dietrichholo,Sparam}. Walking within the top-down framework has been found and studied in~\cite{nunez}. The conformal transition~\cite{son} was studied in a top-down setup in~\cite{kutasov}, and by using a tachyon-Dirac-Born-Infeld action in~\cite{kutasovdbi}. IHQCD has been used to study walking and IR conformal theories by modifying the holographic RG flow ``by hand'', without inclusion of dynamical fermionic degrees of freedom \cite{Jarvinen:2009fe,Alanen:2011hh}. Walking dynamics and the conformal transition have also been studied in Dynamic AdS/QCD~\cite{Alvares:2012kr,Evans:2014nfa}, which has partially similar ingredients as V-QCD.

Including effects due to finite quark masses in holographic models for QCD is important for various reasons. First of all ordinary QCD, which describes the strong interactions observed in nature, has finite quark masses. Therefore, proper understanding of the dependence of the model on the quark masses should lead to better models for ordinary QCD.  

Knowledge on the dependence of quark masses is also useful when analyzing the data from lattice simulations, which are often carried out at unphysically large quark masses due to technical reasons, and therefore extrapolation in the quark mass is needed. For lattice studies which aim at uncovering the phase diagram of QCD as a function of $N_f$, extrapolations in (typically flavor independent) quark mass are particularly important in the probably most interesting region near the conformal transition where lattice simulations are demanding. For example, the so-called ``hyperscaling relations''~\cite{DelDebbio:2010ze} in the conformal window have turned out to be useful close to the conformal transition~\cite{Lombardo:2014pda}. Holography may further help to analyze the physics of the transition, because it can provide a unified description of the phases in the vicinity of $x=x_c$ where most important observables can be computed rather easily for all values of the quark mass.

Finally, it is also important to ensure that the behavior of the holographic model is realistic for all values of the quark mass. Studying limiting behavior (such as the limit of large quark mass) may lead to nontrivial constraints on the model. Fixing the behavior of the model to agree with QCD in limiting cases where field theory computations are tractable, is also expected to improve the model for all values of the quark mass and other parameters.

In this article we study in detail the dependence of holographic QCD on a flavor-independent quark mass $m_q$ in the Veneziano limit. We argue that many of the basic results, such as the dependence of the energy scales on $x$ and $m_q$, are essentially universal, i.e., independent of the details of the holographic model (given some natural assumptions which will be specified below). Additional scaling results for meson and glueball masses, the decay constants, the chiral condensate, the S-parameter, and the critical temperatures are derived for V-QCD analytically at large and small $m_q$ and in the different regimes for $x$. Explicit results are computed numerically in V-QCD and are shown to agree with the expectations from the analytic studies. 

The numerical analysis of~\cite{jk,alho,Arean:2013tja} was mostly done at $m_q=0$, but some observables were already computed for a few values of $m_q$ and for a limited range of $x$. Here we extend these results to cover all relevant regimes (with $0<x<\xBZ$) on the $(x,m_q)$-plane. Importantly, this extension does not require tuning the models or adding any new terms in the V-QCD action, as the value of the quark mass is determined through the boundary conditions of the model. Actually, we will use the choice for the V-QCD action with various potential terms in the action defined exactly as in~\cite{Arean:2013tja}.

We carry out a particularly detailed analysis of the dependence of the chiral condensate on $m_q$, again comparing analytic formulas with explicit numerical results in V-QCD. It is then shown how this analysis is used to study the deformation of QCD by a four-fermion operator $\sim (\bar qq)^2$. Understanding the effect of four-fermion operators is interesting for purely theoretical reasons, but also because such terms naturally arise in technicolor models due to the so-called extended technicolor interactions~\cite{review}.

This article is organized as follows. In the remaining part of introduction we summarize our main results, and discuss the status and future of the exploration of V-QCD. In Sec.~\ref{sec:vqcd}, we give a brief review of the V-QCD models. In Sec.~\ref{sec:scaling}, we discuss the universal scaling results for the energy scales of (holographic) QCD in the Veneziano limit, without explicitly referring to V-QCD. These results are confirmed numerically for V-QCD in Sec.~\ref{sec:massscales}, and extended to include the $m_q$-dependence of the bound state masses and decay constants. In Sec.~\ref{sec:condensate}, we analyze the mass dependence of the chiral condensate, and prove the Gell-Mann-Oakes-Renner (GOR) relation for the pion masses for the fully backreacted case. The results of Sec.~\ref{sec:condensate} are then used to study four-fermion deformations in Sec.~\ref{sec:4f}. The mass dependence of the S-parameter and the pion decay constant is analyzed in Sec.~\ref{sec:S}. Finally, a finite temperature is 
switched on in Sec.~\ref{sec:ft}, where the scaling laws for critical temperatures are derived and compared to numerical results. The Appendix provides technical details for the derivation of the results.

The article is long but many of its sections are largely independent, so a reader only interested in a specific topic may want to jump directly to the corresponding section. There are, however, exceptions: Sec.~\ref{sec:massscales} requires studying Sec.~\ref{sec:scaling} first, and Sec.~\ref{sec:4f} requires Sec.~\ref{sec:condensate}.

\subsection{Summary of results}

Let us then summarize the main results of this article. We identify three different regimes on the ($x$,$m_q$)-plane, shown schematically in Fig.~\ref{fig:scalingregions}, where the dependence of the model on the quark mass is qualitatively different. In regime~A, the quark mass is a small perturbation. Regime~B is the ``hyperscaling'' region where the couplings constant walks, and the amount of walking is controlled by the quark mass\footnote{In this article the term ``walking regime'' always refers to the region with $x_c-x \ll 1$ where walking is found at zero quark mass, even though the coupling constant also walks in regime~B.}. Regime~C refers to the limit of large quark mass (in units of the scale of the ultraviolet (UV) RG flow $\LUV$). The white regions in the plot indicate crossovers between the regimes~A, B, and~C -- there are no phase transitions at finite quark mass and zero temperature. 

In this article, we discuss in detail how the diagram arises quite in general in holographic\footnote{As such, the structure of Fig.~\ref{fig:scalingregions} is not surprising, and one can argue how it arises from QCD by using arguments directly based on field theory~\cite{DelDebbio:2010ze,Dietrich:2009ns}.} models. We show how scaling results for the energy and mass scales in the various regimes can be obtained by using relatively simple assumptions and straightforward analysis (explicit results are given in Eqs.~\eqref{regimeBlowx}, \eqref{regimeBhighx}, and~\eqref{largemqscalingt}).
The most important findings are:
\begin{itemize}
 \item The physics at small $m_q$, including regimes~A and~B of Fig.~\ref{fig:scalingregions}, is universal, i.e., qualitative features are independent of the details of the holographic model. 
 That is, the results are proven for all V-QCD models with such ``regular'' potentials that there is a BKT transition and therefore the model displays the structure of Fig.~\ref{fig:scalingregions}, 
 but they also hold in other models which involve a BKT transition triggered by the same mechanism as in V-QCD (details on this mechanism are given in Sec.~\ref{sec:scaling} and in Appendix~\ref{app:scaling}).
 \item At large $m_q$ (in regime~C) the results are model dependent, and can directly be compared to field theory results which are tractable in the limit of large $m_q$. We demonstrate that V-QCD  reproduces most important features such as the decoupling of the massive quarks. Remarkably, we find that a specific subclass of the V-QCD models, which was found to be closest to QCD by analyzing the asymptotic meson spectra in~\cite{Arean:2013tja}, also works best at large quark mass. These are the models with ``potentials I'' below.
\end{itemize}
We compute numerically the meson and glueball masses in V-QCD, and show that they agree with the generic scaling results in all regimes.

Notice that Fig.~\ref{fig:scalingregions} only shows changes in the dependence of the quark mass. There are also other features: In regime~A, the theories close to $x=x_c$ are walking and therefore much different from the theories at low values of $x$, reflecting the behavior of the $m_q=0$ backgrounds, which are perturbed by the quark mass. Within regime~B, the scaling laws change in a nonanalytic manner exactly at $x=x_c$ (see Eqs.~\eqref{regimeBlowx} and~\eqref{regimeBhighx}).

We also analyze in detail how the chiral condensate depends on the quark mass. The main results are the following:
\begin{itemize}
 \item In the QCD and walking regimes, there is an interesting spiral structure, which we call the Efimov spiral. We give an asymptotic formula~\cite{Iqbal:2011in} of the spiral at small $m_q$ in Eq.~\eqref{spiraleqs}, and compare this to data in Fig.~\ref{fig:spiral}. We demonstrate how the spiral is related to the subleading Efimov vacua\footnote{The terminology refers to the analogous Efimov effect in the formation of three-body bound states of identical bosons~\cite{Efimov:1970zz}.} of the theory and to the Miransky scaling law in the walking regime.
 \item The standard holographic proof of the GOR relation is extended to the case of V-QCD, which requires handling of the full backreaction and the logarithmic corrections which appear in near the UV boundary due to the RG flow. The relation is given in Eq.~\eqref{GORtext} and checked numerically in Fig.~\ref{fig:GOR}. 
\end{itemize}

We use the results for the chiral condensate to check how Witten's method~\cite{Witten:2001ua} for adding multi-trace deformations in holographic models by modifying the UV boundary conditions works for V-QCD. The resulting phase diagram for the case of a double-trace (four-fermion) deformation $\propto g_2 (\bar q q)^2$ at zero quark mass is shown in Fig.~\ref{fig:4fphases}  in the $(x,g_2)$-plane. For positive coupling $g_2$, we see no change with respect to $g_2=0$, whereas negative $g_2$ induces an instability.

The mass dependence of the S-parameter, the pion decay constant, and related quantities is studied in detail.
\begin{itemize}
 \item As any finite $m_q$ is turned on in the conformal window, the S-parameter discontinuously jumps from zero to a \order{N_f N_c} number. Except for this discontinuity, the dependence on $x$ and $m_q$ is weak.
 \item We demonstrate a novel power-law scaling of the subleading terms in the S-parameter at small quark mass in the conformal window and in the walking regimes (see Figs.~\ref{fig:massdeplog} and~\ref{fig:massdepwalk}). It is argued that the power can be expressed in terms of the dimension of $\mathrm{Tr}\, F^2$ at the IRFP as in Eqs.~\eqref{massdepCW} and~\eqref{massdepwalking}.
 \item We analyze the $x$-dependence of the S-parameter by writing it as a series over the contributions from the low-lying vector and axial vector meson poles (see Eq.~\eqref{Sseries}). It is shown that the increase of the S-parameter with $x$ in the QCD and walking regimes can be attributed to slower convergence of the series.
\end{itemize}

Finally we analyze the dependence of the critical temperatures of the deconfinement transition and various crossovers on $m_q$ and $x$. The scaling laws of \eqref{TcinregA} -- \eqref{TcinregC} and \eqref{Tqclaw} -- \eqref{Tmqlaw} are demonstrated numerically in Fig.~\ref{fig:FT}. The second order chiral transition, which is found at zero quark mass in the walking regime~\cite{alho}, transforms into a crossover as finite $m_q$ is turned on. We also discuss how the thermodynamics approaches that of the YM limit ($x \to 0$) as the quark mass is taken to infinity so that the quarks are decoupled.

\subsection{Outlook}

This article is part of an ongoing program~\cite{jk,alho,letter,Arean:2013tja,Alho:2013hsa,Iatrakis:2014txa,pionloop} for studying the properties of the V-QCD models, and in more general the structure of QCD in the Veneziano limit. There are several possible future directions to explore. As a continuation of this study, one could consider the case of flavor dependent quark masses, which would be interesting in order to construct more realistic models for ordinary QCD, where all quark masses are unequal. In holography, this means that the background solutions are nontrivial dependence on the flavor indices, and consequently are described in terms of a non-Abelian Dirac-Born-Infeld action. The precise definition of such a non-Abelian action is not known, so requiring the physics to be correct might lead to interesting constraints for it.

The study of the CP-odd terms of the V-QCD action is in progress at the moment~\cite{cpodd}. These terms govern the physics of the axial anomaly and the theta angle of QCD. Similarly to the analysis at zero values of theta, requiring the regularity of the solutions in the IR, and among other things the correct behavior of the asymptotic meson spectra at finite values of theta result in constraints for the potentials appearing in the CP-odd terms, in addition to the constraints obtained in the YM limit~\cite{ihqcd,data,cs}. After analyzing these constraints, we may make physically reasonable choices for these potentials, and in particular for their asymptotic behavior near the UV boundary and deep in the IR.

So far the potentials of the V-QCD have not been tuned to fit any nonperturbative QCD data, and therefore all predictions of the model are qualitative. But once the CP-odd sector of the action has been fixed, one can start fitting the parameters of the potentials both to experimental results for QCD (such as meson masses) and to lattice data (both for QCD at finite $N_f/N_c$~\cite{lattice} and for YM~\cite{panero,Lucini:2012gg}). The hope is that the overall fit, together with the other constraints for the potentials, fixes the predictions of the model to a good accuracy for all relevant values of the parameters (such as $x$, the quark mass, temperature, and chemical potential). Consequently the model would give a effective description of QCD with real predictive power, rather than being just a toy model.

\section{V-QCD} \label{sec:vqcd} 

In this section, we will briefly introduce a class of bottom-up models for QCD, which we call V-QCD~\cite{jk}. The V in the name refers to the fact that the models are defined in the Veneziano limit:
\be
 N_c \to \infty\quad \mathrm{and} \quad N_f \to \infty\ , \quad \mathrm{with}\quad x\equiv \frac{N_f}{N_c} \quad \mathrm{and} \quad g^2 N_c  \quad \mathrm{fixed} \ .
\ee

V-QCD is based on two ``building blocks''. The first block (the glue sector of V-QCD) is IHQCD~\cite{ihqcd} which is a bottom-up model for Yang-Mills (YM) theory inspired by five-dimensional noncritical string theory. The second block (the flavor sector of V-QCD) is a framework for adding flavor via tachyonic Dirac-Born-Infeld actions~\cite{Bigazzi:2005md,ckp}. This framework has been tested previously~\cite{ikp} in the probe (or 't Hooft) limit, i.e., without including the backreaction of the flavor branes to the background. However in V-QCD, and more generally in the Veneziano limit, the flavor and glue sectors are fully backreacted.

\subsection{The V-QCD action}

Let us then discuss briefly the dictionary of V-QCD. The most relevant fields are 
\begin{itemize}
 \item The dilaton $\phi$. The exponential $e^{-\phi}$ is dual to the operator $\tr F^2$. We will denote $\l=e^\phi$ below. As this notation indicates, its background value is identified as the 't Hooft coupling on the field theory side.
 \item The tachyon $T_{ij}$ which is a $N_f \times N_f$ matrix in flavor space. The combination $T+T^\dagger$ is dual to the operator $\bar q_i q_j$ whereas $T-T^\dagger$ is dual to $\bar q_i \gamma_5 q_j$. For the background solutions considered here we will take $T_{ij} = \t(r) \delta_{ij}$ so the flavor structure does not appear explicitly.
 \item The left- and right-handed gauge fields $A^{L/R}_\mu$ which are dual to $\bar q \gamma_\mu(1\pm\gamma_5) q$. They are also matrices in flavor space but we have hidden the flavor indices here. These fields evaluate to zero for the backgrounds considered in this article.
\end{itemize}
In addition, the scale factor $A$ of the metric
\be
ds^2=e^{2 \Awf(r)} (-dt^2/f(r) +d\mathbf{x}^2+f(r) dr^2)\,
\label{bame}
\ee
of the vacuum solution is identified as the logarithm of the energy scale on the field theory side. The blackening factor $f$ in the metric may be either identically equal to one or a nontrivial function of $r$.
Our convention will be that the UV boundary lies at $r=0$, and the bulk coordinate therefore runs from zero to infinity (when $f(r)\equiv 1$) or up to a horizon at a finite value of $r$ (when $f(r)$ has a nontrivial profile). The metric will be close to the AdS metric in the UV: $A \sim -\log(r/\ell)$, where $\ell$ is the (UV) AdS radius. In the UV, $r$ is therefore identified roughly as the inverse of the energy scale of the dual field theory.

The action for the V-QCD model consists of three terms:
\be
 S = S_g + S_f + S_a
\ee
where $S_g$, $S_f$, and $S_a$ are the actions for the glue, flavor and CP-odd sectors, respectively. Explicit expressions for the first two terms will be given below. As discussed in \cite{jk}, only these terms contribute in the vacuum structure of the theory if the phases of the quark mass matrix and the theta angle vanish. The last term $S_a$ is important for the realization of the theta angle and the axial anomaly of QCD~\cite{ckp}. This term has been written down explicitly in~\cite{Arean:2013tja}, and it will be zero for all configurations discussed in this article.

The glue action is that of IHQCD~\cite{ihqcd}. It includes five dimensional Einstein gravity and the dilaton $\l=e^{\phi}$:
\be
S_g= M^3 N_c^2 \int d^5x \ \sqrt{-\det g}\left(R-{4\over3}{
(\partial\lambda)^2\over\lambda^2}+V_g(\lambda)\right) \ .
\label{vg}\ee

The flavor action is the generalized Sen's action~\cite{ckp,sen} (see also~\cite{sstachyon}),
\be
S_f= - \frac{1}{2} M^3 N_c\  {\mathbb Tr} \int d^5x\,
\left(V_f(\l,T^\dagger T)\sqrt{-\det {\bf A}_L}+V_f(\l, TT^\dagger)\sqrt{-\det {\bf A}_R}\right)\ ,
\label{generalact}
\ee
where the quantities inside the square roots are defined as
\begin{align}
{\bf A}_{L\,MN} &=g_{MN} + \gf(\l,T) F^{(L)}_{MN}
+ {\h(\l,T) \over 2 } \left[(D_M T)^\dagger (D_N T)+
(D_N T)^\dagger (D_M T)\right] \ ,\nonumber\\
{\bf A}_{R\,MN} &=g_{MN} + \gf(\l,T) F^{(R)}_{MN}
+ {\h(\l,T) \over 2 } \left[(D_M T) (D_N T)^\dagger+
(D_N T) (D_M T)^\dagger\right] \ ,
\label{Senaction}
\end{align}
with the covariant derivative
\be
D_M T = \partial_M T + i  T A_M^L- i A_M^R T\ .
\ee
The trace $\mathbb Tr$ is over the flavor indices -- recall that the fields  $A_{L}$, $A_{R}$ as well as $T$ are $N_f \times N_f$ matrices in the flavor space. 

It is not known, in general, how the determinants over the Lorentz indices in~\eqref{generalact} should be defined when the arguments~\eqref{Senaction} contain non-Abelian matrices in flavor space. However, for our purposes such definition is not required: our background solution will be proportional to the unit matrix $\mathbb{I}_{N_f}$, as the quarks will be all massless or all have the same mass $m_q$.
In such a case, the fluctuations of the Lagrangian are unambiguous up to quadratic order.

The form of the tachyon
potential that we will use for the derivation of  the spectra is
\be
V_f(\l,TT^\dagger)=V_{f0}(\l) e^{- a(\l) T T^\dagger} \ .
\label{tachpot}
\ee
This is the string theory tachyon potential where the constants have been allowed to depend on the dilaton $\la$.
For the vacuum solutions (with flavor independent quark mass) we will take $T =\tau(r) \mathbb{I}_{N_f}$ where $\tau(r)$ is real, so that
\be
 V_f(\l,T)=V_{f0}(\l) e^{- a(\l)\tau^2} \ .
\label{vf}\ee
The coupling functions $\h(\l,T)$ and $\gf(\l,T)$ are allowed in general to depend on $T$,  through such combinations that the expressions~\eqref{Senaction} transform covariantly under flavor symmetry. In this article, we will take them to be independent of $T$, emulating the known string theory results. Under these assumptions, and for the vacuum solutions (so that the gauge fields also vanish) the flavor action simplifies to
\be
S_f= -  M^3 N_cN_f\int d^5x\,\sqrt{-\det g}\ V_{f0}(\l)\, e^{- a(\l)\tau^2}  \sqrt{1+g^{rr}\,\kappa(\l)\, (\t')^2} \ .
\ee

\subsection{Potentials and the holographic RG flow} 

In order to fully fix the action, the potentials $V_g(\l)$, $V_{f0}(\l)$, $a(\l)$, $\h(\l)$, and $\gf(\l)$ need to be specified. It turns out~\cite{jk} that $V_g(\l)$ must satisfy the same constraints as in IHQCD~\cite{ihqcd}. The other potentials will be subject to analogous constraints. We review the main idea here, and the details can be found in~\cite{jk,Arean:2013tja}.

First, identification of the field $\l$ as the 't Hooft coupling and the scale factor $A$ as the logarithm of the field theory energy scale defines the holographic renormalization group (RG) flow and the holographic beta function for the coupling as in IHQCD,
\be
 \beta_h(\la) = \frac{\la'(r)}{A'(r)}
\ee
with the understanding that the fields are evaluated on the $r$-dependent background solution. In IHQCD, the dilaton potential $V_g$ can be directly mapped to the holographic beta function~\cite{ihqcd} at any value of $r$, that is, at any energy scale. 
In V-QCD, there is an additional field, the tachyon, whose background value is linked to the running quark mass. Therefore one may define a holographic gamma function, which controls the holographic RG flow of the quark mass~\cite{jk}. The mapping between the beta and gamma functions and the potentials is, in general, more complicated than in IHQCD, but simplifies in the UV.

The behavior of the potentials in the UV (where $\l \to 0$) is then restricted by requiring that the holographic beta and gamma functions match with their QCD counterparts in the UV. For the UV structure to be consistent, all potentials are chosen to be analytic at $\l=0$, and the series coefficients can be related to those of the perturbative beta and gamma functions in QCD. It turns out that the dilaton potential $V_g$ is consequently mapped to the perturbative beta function of YM theory as in IHQCD. Due to the backreaction, the beta function of QCD (in the Veneziano limit) is mapped to the combination $V_g-x V_{f0}$. This mapping leaves one undetermined parameter, the UV normalization of $V_{f0}$, which we call $W_0$. The gamma function of QCD fixes the UV behavior of the ratio $a/\h$. Here we will match the expansions of the potentials around $\l=0$ up to two loops for the beta function and up to  one loop for the gamma function of QCD.

Using perturbative QCD to determine the UV behavior of the potentials may be surprising, since holography is in general not expected to work at small values of the coupling $\l$. The idea is, however, that by using this procedure correct boundary conditions for the more interesting IR dynamics are obtained. Notice also that the procedure can be seen as a rather mild generalization of what is usually done in bottom-up holography. For example, the bulk mass of the tachyon is typically required to satisfy the relation $-m^2\ell^2 = \Delta(4-\Delta)$ (at least in the UV), where $\ell$ is the UV AdS radius and $\Delta$ is either the dimension of the quark mass or the chiral condensate. Here this relation is effectively generalized to include loop effects, i.e., the perturbative anomalous dimension of the quark mass, which is roughly mapped to the first few coefficients in the expansion of the bulk mass of the tachyon at $\l=0$. 

In this article, we will carry out one more check which demonstrates that our UV boundary conditions make sense. Namely, we prove that the GOR relation holds even in the backreacted case and that the UV RG flow of the quark mass and the condensate, imposed by the matching to perturbative QCD, cancels in this relation, as it should (see Sec.~\ref{sec:gmor}).

The choice of the potentials in the IR is more relevant since it affects how the nonperturbative physics of QCD is modeled. Since we are working with bottom-up models, there is a lot of freedom in choosing the potentials, and it is important to choose them such that the IR physics resembles that of QCD. The mapping to the beta functions is not useful at large values of the coupling because of scheme dependence. The asymptotic behavior of the potentials at large $\l$ can however be constrained heavily 
by comparing to several different observables, most importantly the asymptotics of the spectra at large mass. 

The IR behavior of the dilaton potential $V_g$, i.e., its asymptotics as $\l \to \infty$, can be fixed by requiring (among other things) confinement and correct asymptotic behavior of the glueball spectrum at large excitation numbers. The remaining parameters have been fitted to YM data~\cite{data}. Similarly, asymptotics of the meson spectra sets strict constraints on the large $\l$ asymptotics of the other potentials in the V-QCD action~\cite{Arean:2013tja}. The remaining degrees of freedom will be fitted to experimental and lattice data in future studies.

In this article, we will be using the choices ``potentials~I'' and ``potentials~II'' which are exactly the ones given in~\cite{letter,Arean:2013tja}, and are defined explicitly in Appendix~\ref{app:numerics}. The potentials~I reproduce  more accurately qualitative features of QCD. This choice has the UV parameter $W_0$ set to a constant value $3/11$, and assumes $\gf=\h$. The potentials~II are included in order to study the model dependence of the results.  This choice has the UV parameter $W_0$ set to a value which guarantees ``automatically'' the Stefan-Boltzmann normalization of pressure at high temperatures~\cite{alho}, and assumes $\gf=1$. Most of the results derived in this article are qualitative (e.g., scaling laws of various observables) and therefore insensitive to the details of the potentials.

\subsection{Background solutions} \label{sec:bg}

Let us discuss some general features of the background solutions of the V-QCD models,  first restricting to the standard case, which has a phase diagram similar to what is usually expected to arise in QCD. Such a phase diagram is obtained if the potentials are chosen as discussed above.

\subsubsection{Zero temperature}

In this article we will mostly discuss solutions at zero temperature. In this case, the blackening factor $f$ in~\eqref{bame} is trivial, $f \equiv 1$.
To find the background, we consider $r$-dependent Ans\"atze for  $\l$, and $\Awf$. As pointed out above, we assume that the quark mass is flavor independent, and therefore  take $T=\t(r)\mathbb{I}_{N_f}$. We also set all other fields to zero, and look for solutions to the equations of motion (EoMs). The models are expected to have two classes of (zero temperature) vacuum solutions \cite{jk}:
\begin{enumerate}
 \item Backgrounds with nontrivial $\l(r)$, $\Awf(r)$ and with zero tachyon $\t(r)=0$. These solutions have zero quark mass and intact chiral symmetry.
 \item Backgrounds with nontrivial $\l(r)$, $\Awf(r)$ and $\t(r)$. These solution have broken chiral symmetry. As usual, the quark mass $m_q$ and the chiral condensate are identified as the coefficients of the normalizable and non-normalizable tachyon modes in the UV.
\end{enumerate}
In the first case, the EoMs can be integrated analytically into a single first order equation, which can easily be solved numerically. The regular solution ends on an IRFP, where the dilaton approaches a constant value, and the geometry is asymptotically AdS$_5$. In the second case, one needs to solve a set of coupled differential equations numerically. The regular solution ends in a ``good'' IR singularity~\cite{gubser}, where both the dilaton and the tachyon diverge. This kind of singularity supports extension to finite temperature and is repulsive: perturbations around the regular solution develop a nonanalyticity before reaching the singularity, which signals the fact that IR boundary conditions are uniquely fixed.

Let us first recall what happens at zero quark mass.  The ratio $x=N_f/N_c$ is constrained to the range $0 \le x <11/2 \equiv \xBZ$ where the upper bound was normalized to the Banks-Zaks (BZ) value in QCD, where the leading coefficient of the $\beta$-function turns positive. The standard\footnote{For some choices of potentials, also different structure can appear. In particular, there is the possibility that chiral symmetry breaking is absent in the regime of very small $x$~\cite{alho}.} phase diagram at zero quark mass has two phases separated by a phase transition at some $x=x_c$ within this range.
\begin{itemize}
 \item When $x_c\le x<\xBZ$, chiral symmetry is intact. The dominant vacuum solution is in the first class with the tachyon vanishing identically and an IRFP. Therefore the geometry is asymptotically AdS$_5$ in the IR.
 \item When $0<x<x_c$, chiral symmetry is broken. The dominant vacuum therefore is in the second class with nonzero tachyon even though the quark mass is zero. 
  The geometry ends at a good IR singularity in the IR. 
\end{itemize}

For potentials satisfying some reasonable requirements~\cite{jk,Arean:2013tja}, the phase transition at $x=x_c$ is due to an instability of the tachyon at the IRFP~\cite{son,Jensen:2010ga}. That is, the solution with vanishing tachyon is unstable if the bulk mass of the tachyon at the IRFP $ -m_*^2 \ell_*^2$, where $\ell_*$ is the IR AdS radius, violates the Breitenlohner-Freedman (BF) bound~\cite{Breitenlohner:1982jf}:
\be
 - m_*^2 \ell_*^2 = \Delta_*(4-\Delta_*) \le 4 \ ,
\ee
where $\Delta_*$ is the dimension of the quark mass at the fixed point.
Notice that when the bound is violated, $\Delta_*$ becomes complex.

As a consequence of violating the BF bound, the phase transition at $x=x_c$ (which is only present at zero quark mass) involves BKT \cite{bkt} or Miransky \cite{miransky} scaling, for values of $x$ right below the critical one. The order parameter for the transition, the chiral condensate $\sigma \sim \langle \bar q q \rangle$ vanishes exponentially,
\be \label{condscaling}
 \sigma \sim \exp\left(-\frac{2 K}{\sqrt{x_c-x}}\right)
\ee
as $x \to x_c$ from below. Here the constant $K$ is positive.

At finite quark mass, the BKT transition disappears: the background is always in the second class with finite tachyon, and the dominant vacua at all values of $x$ are smoothly connected. Therefore the geometry ends in a ``good'' singularity in the IR. In particular, the IR geometry changes in a discontinuous manner (from an IRFP to the good singularity) when a small quark mass is turned on in the conformal window ($x_c<x<\xBZ$). This discontinuity causes interesting behavior of observables which will be discussed in the following sections.

The models may also have unstable subdominant vacua when $0<x<x_c$. We will discuss such vacua in more detail in Sec.~\ref{sec:condensate}.

\subsubsection{Finite temperature}

At finite temperature, one can first identify two types of background geometries~\cite{alho}:
\begin{enumerate}
 \item The thermal gas solutions which have the same (and therefore temperature independent) $r$-dependence as the zero temperature solutions, and in particular $f \equiv 1$. 
 \item The back hole solutions which have a nontrivial $f(r)$ and end on a horizon in the IR.
\end{enumerate}
As usual the temperature is given by (the inverse of) the length of the compactified time direction, and is equal to the Hawking temperature of the black hole for the second type of solutions. Both of these geometries further split into two classes, one having zero and the other having nonzero tachyon. Therefore there can be up to four qualitatively different competing saddle points.

The finite temperature phase diagram has been studied in~\cite{alho}. The following structure was found at zero quark mass:
\begin{itemize}
 \item For $x_c \le x <\xBZ$, the finite temperature phase is the tachyonless black hole. A tachyonless thermal gas solution which has the same $r$-dependence as the zero temperature solution also exists at all temperatures but is subdominant. Therefore if the system is first prepared at zero temperature, any amount of heating makes the system jump to a different phase immediately. At temperatures which are much smaller than the characteristic scale of the RG flow between the two fixed points, the finite temperature backgrounds are obtained by deforming the zero temperature backgrounds, which are asymptotically AdS$_5$, only very close to the IR end. Consequently, small temperature thermodynamics is that of AdS, $p \propto T^4$, and the zero temperature transition is of 4th order.
 \item For $0<x<x_c$, the low temperature phase is the tachyonic thermal gas phase which is smoothly connected to the zero temperature solution and breaks chiral symmetry. The high temperature phase is the tachyonless black hole phase. There is always a first order (``deconfinement'') transition separating the phases, but it is also possible that an intermediate, chirally broken tachyonic black hole phase appears. If this is the case, chiral symmetry is restored at a separate second order transition, which has higher critical temperature than the deconfinement transition. As $x \to x_c$ from below, the critical temperatures go to zero following Miransky scaling.
\end{itemize}

At finite quark mass, all phases are tachyonic due to the UV boundary conditions, and chiral symmetry is broken. At low temperatures the dominant vacuum is the thermal gas phase. When the system is heated it undergoes a first order transition to the black hole phase, which is interpreted as the deconfinement transition (since there is no transition linked to chiral symmetry breaking). The phase structure is therefore similar for all $0<x<\xBZ$. The dependence of the critical temperature on $m_q$ and $x$ will be discussed in Sec.~\ref{sec:ft}. In addition to the phase transition, we find several crossovers which are linked to changes in the zero temperature geometry.

\section{Energy scales of (holographic) QCD in the Veneziano limit} \label{sec:scaling}

In this section we shall discuss the dependence of observables (at zero temperature) on the quark mass and $x=N_f/N_c$ for holographic models of QCD in the Veneziano limit and in general, i.e., not necessarily only for V-QCD. The results will not be proven rigorously, but we will sketch how they arise from rather natural assumptions. Comparison to quantitative results from V-QCD will be carried out in Sec.~\ref{sec:massscales}.

It is useful to discuss separately the regions with small (possibly zero) and large quark mass.
\begin{itemize}
 \item At small quark mass, we shall demonstrate that under certain natural assumptions, a universal picture arises from holography, which is not dependent on the details of the model. That is, the results of this section will apply to the V-QCD models with quite generic choices of potentials but also to holographic models with more general actions, if they satisfy some natural requirements which will be given below.
 \item At large quark mass, the $m_q$-dependence of energy scales and some other observables (such as meson mass gaps) can be found via arguments based on QCD. Large quark mass probes the UV structure in holography, where predictions may not be reliable. We shall see that the results from holography are model dependent in this region, and discuss what is needed to match with the known results in QCD. 
\end{itemize}
In this section we will concentrate on the behavior of the various energy scales for a generic holographic model. In the remaining of this article we will analyze concrete observables such as bound state masses, the chiral condensate, and the S-parameter for V-QCD. Also the behavior of these observables at small quark mass is to large extent independent of the model details, but it is just much easier to work with the explicitly fixed V-QCD action. 

We will exclude the BZ limit (i.e., the limit $x \to \xBZ$) from our analysis for the moment. In this region the theory is fully under perturbative control and holographic approach may not be useful.

First we need to give rough definitions for various energy scales at zero temperature, assuming a generic holographic model. There must be a field dual to the $\bar q q$ operator, which we call the tachyon, and the geometry must be asymptotically AdS in the UV. We take the UV boundary to lie at $r=0$ as above. 
Because we want to keep the discussion generic, the definitions for the energy scales will be rather sketchy. Precise definitions in the case of V-QCD will be given in Sec.~\ref{sec:massscales}.
\begin{itemize}
 \item $\LUV$ is the scale of the UV RG flow in QCD. In holography, it can be identified as the scale of the UV expansions of the vacuum solution. Since V-QCD also implements the UV RG flow of the coupling constant through the flow of the dilaton field, it will be most natural to define $\LUV$ in terms of the UV expansion of the dilaton (see~\eqref{laUV} in the next section). In the generic treatment of this section (which assumes small quark mass, i.e., $m_q/\LUV \ll 1$), this scale is most conveniently
 defined in terms of the UV behavior of the tachyon instead: it is identified as the (inverse of the) boundary of the interval where the standard UV expression~\eqref{tauruns} for the tachyon holds as a good approximation. 
 \item $\LIR$ is the soft IR scale which governs the IR expansions (or is set by an IR cutoff), in close analogy to the definition of $\LUV$ in the UV. Precise definition for V-QCD will be given in~\eqref{laIR} and~\eqref{laIRFP} in Sec.~\ref{sec:massscales}.
 \item The quark mass $m_q$ is defined as the source for the tachyon.
 \item $\L_\t$ is the scale of chiral symmetry breaking (whenever it is broken). In holography it can be identified as the energy scale (inverse of $r$) where the tachyon grows large (or becomes $\mathcal{O}(1)$, more precisely). As we shall see, its dependence on $m_q$ is similar to that of the constituent quark mass in nonrelativistic quark models.  
\end{itemize}
Notice that we only discuss the dynamical scales appearing through the background solutions, and for example the UV AdS radius is assumed to be fixed. More precisely, the ratios of the various energy scales are determined by the dynamics of the model, whereas one of the energy scales (most conveniently $\LUV$) can be taken as a fixed reference scale.

\subsection{Small quark mass}

By small quark mass we mean here that $m_q/\LUV \ll 1$, and $m_q$ may also be zero. In this case our results will be essentially universal, i.e., independent of the details of the holographic model. Naturally, several assumptions must be made on the model, in order to ensure that the physics resembles that of QCD in the Veneziano limit. 
Most importantly, there must be an IRFP and a conformal window, and the conformal transition at $x=x_c$ is assumed to be a BKT transition, which is naturally implemented in holography by the tachyon hitting its BF bound at the IRFP~\cite{son,Jensen:2010ga} (implemented through a geometry which is asymptotically AdS in the IR).
These assumptions are met by the V-QCD models with potentials which fulfill reasonable constraints in the UV and in the IR, and do not have any peculiar structure at intermediate energy scales~\cite{jk,Arean:2013tja}. But these assumptions can also hold more generally in holographic theories the actions of which cannot be written in the V-QCD form. Indeed similar results which we will
present here have been found in related top-down~\cite{kutasov} bottom-up~\cite{kutasovdbi,Alvares:2012kr,Evans:2014nfa} models. For simplicity we also assume that the upper edge of the conformal window lies at the QCD value, $x=11/2 \equiv x_\mathrm{BZ}$.

The model assumptions can be made explicit in terms of the tachyon background in various ranges of the bulk coordinate $r$. We will only need the solutions for small tachyon, so that the tachyon EoM is approximately linear (see Appendix~\ref{app:scaling} for more detailed derivation of these solutions). In the UV (where $r \to 0$), we assume the standard behavior determined by the dimension of the quark mass:
\be \label{tauruns}
 \tau \simeq m_q r + \sigma r^3\ , \qquad \left(r \ll \frac{1}{\LUV}\right) \ ,
\ee
where $\sigma$ is proportional to the chiral condensate.
Here logarithmic running of the quark mass (and the condensate) could also be included, but it is not important because it will not affect the leading scaling behavior.
In the vicinity of the fixed point when the tachyon is small and the BF bound is satisfied, corresponding to $x_c \le  x < \xBZ$ in QCD, we find that
\be \label{taufp}
 \tau \simeq C_m (r \LUV)^{\Delta_*} + C_\s (r \LUV)^{4-\Delta_*}\ , \qquad \left(\frac{1}{\LUV} \ll r \ll \frac{1}{\Lambda_\t}\right)\ ,
\ee
where the precise $x$-dependence of the anomalous dimension of the quark mass $\Delta_*$ depends on the model.
When the BF bound is violated, and $x_c-x$ is small 
\be \label{tauwalks}
 \tau \simeq C_w \left(r \LUV\right)^2 \sin\left[\n \log( r \LUV) +\phi\right]\ , \qquad  \left(\frac{1}{\LUV} \ll r \ll \frac{1}{\Lambda_\t}\right)\ ,
\ee
where $\n = \mathrm{Im}\Delta_* \simeq \pi\sqrt{(x_c-x)}/K$ as $x \to x_c$ from below. The coefficient $K$ depends on the model, and it satisfies
\be
 K = \frac{\pi}{\sqrt{\frac{d}{dx}\left[\Delta_*(\Delta_*-4)\right]_{x=x_c}}} \ .
\ee

Now it is straightforward to fix the integration constants ($\sigma$, $C_m$, $C_\s$, $C_w$ and $\phi$) and compute the mass dependence of the various energy scales by using the following recipes:
\begin{itemize}
 \item Both the normalizable and nonnormalizable terms of the tachyon solution are separately continuous\footnote{Near the IRFP in~\eqref{tauwalks} the normalizable and nonnormalizable terms cannot be separated, but the strategy is then interpreted as requiring the continuity of the tachyon and its first derivative.}. Therefore an approximate tachyon solution for all $r \ll 1/\Lambda_\t$ can be found by gluing together the results from the different regimes~\eqref{tauruns},~\eqref{taufp}, and~\eqref{tauwalks}.  Because the bulk mass of the tachyon varies smoothly, the terms of the tachyon can only jump by a factor \order{1} as we move from one regime to another.
 \item The IR behavior of the tachyon in the regime $r \gg 1/\Lambda_\t$ is nontrivial and qualitatively different from the formulas given above. Therefore, the IR boundary conditions obtained by requiring continuity at  $r\simeq 1/\Lambda_\t$ are taken to be generic, i.e., it is assumed that the conditions are not fine tuned to pick any specific solution for $r \ll 1/\Lambda_\t$.
\end{itemize}

\subsubsection{Zero quark mass}

Let us first recall how the energy scales depend on $x$ at zero quark mass. The results for $x<x_c$ are obtained by using the above assumptions and recipes. Therefore they are also independent of the details of the model. The computation can be found in Sec.~10 of~\cite{jk} and the scaling results will also be reproduced by the analysis of Sec.~\ref{sec:condensate} below, where the chiral condensate is studied in detail  (see also~\cite{kutasov}). We will only list the results here.   

When $x<x_c$, $\LIR \sim \Lt$ (i.e., the tachyon becomes sizeable where the asymptotic IR geometry starts), and $\Lt$ is not defined in the conformal window ($x_c \le x <\xBZ$). 
Depending on the value of $x=N_f/N_c$, there are three regions with qualitatively different behavior:
\begin{itemize} 
 \item In the QCD regime, meaning that $0<x<x_c$ and $x_c-x\gtrsim 1$, 
there is only one scale, $\LUV \sim \LIR$, which corresponds to $\Lambda_\mathrm{QCD}$. 
 \item In the walking regime, which is found when $x<x_c$ and $x_c-x \ll 1$, the IR and UV scales are related through Miransky scaling:
\be \label{Mscal}
 \frac{\LUV}{\LIR} \sim \exp\left[\frac{K}{\sqrt{x_c-x}}\right] \ .
\ee
The chiral condensate satisfies
\be
 \frac{\sigma}{\LUV^3} \sim \frac{\LIR^2}{\LUV^2} \sim  \exp\left[-\frac{2 K}{\sqrt{x_c-x}}\right] \ .
\ee
\item In the conformal window ($x\ge x_c$) the tachyon is zero as chiral symmetry is intact. In this case one expects that the scales of the UV and the IR expansions are the same, $\LUV\sim\LIR$ (as there is no obvious mechanism which would separate the scales).
\end{itemize}

\begin{figure}[!tb]
\begin{center}
\includegraphics[width=0.69\textwidth]{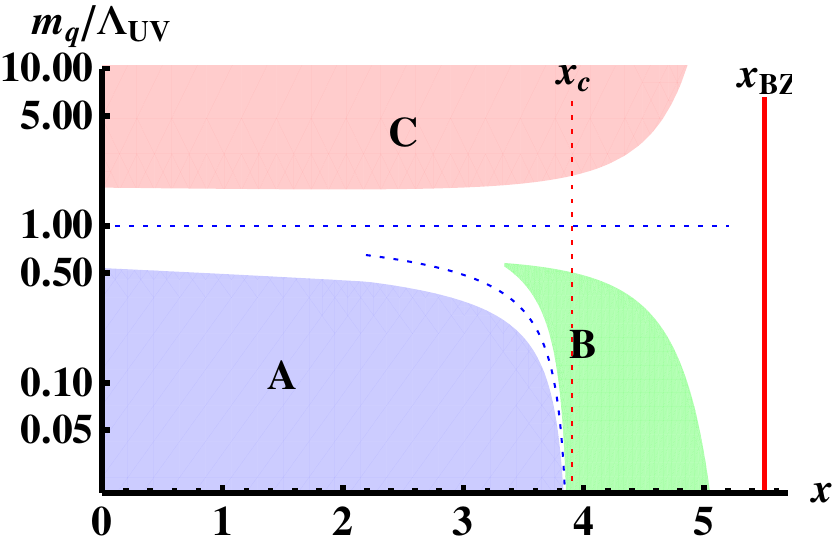}
\end{center}
\caption{A schematic diagram of the different scaling regions as functions of $x$ and $m_q$.}
\label{fig:scalingregions}\end{figure}

\subsubsection{Small but finite quark mass}

Let us then generalize these results to nonzero quark mass, first assuming a small mass ($m_q/\LUV \ll 1$). The detailed analysis uses the strategy formulated above, and can be found in Appendix~\ref{app:scaling}. We will only discuss the results here.

One can identify, for $m_q/\LUV\ll 1$, two regimes on the $(x,m_q)$-plane, shown schematically in Fig.~\ref{fig:scalingregions}:
\begin{itemize}
 \item[A] The regime where the quark mass is a small perturbation. This is possible for $0<x<x_c$ as the $m_q=0$ solution is continuously connected to the solution with finite $m_q$. In order to determine the extent of regime~A, we will perturb the solution at vanishing quark mass by adding a small $m_q$. It is a small perturbation so long as the nonnormalizable term remains small for $r \ll 1/\LUV$. This is equivalent to $m_q \ll |\sigma(m_q=0)/\LUV^2|$, or 
\be
 \frac{m_q}{\LUV} \ll 1 
\ee 
in the QCD regime, and
\be \label{regAcondw}
 \frac{m_q}{\LUV} \ll \exp\left[-\frac{2K}{\sqrt{x_c-x}}\right]
\ee
in the walking regime. Apart from the perturbation caused by the quark mass, the energy scales behave as in the $m_q=0$ case discussed above. In particular, the Miransky scaling law~\eqref{Mscal} applies for $x_c-x \ll 1$.
 \item[B] The ``scaling'' regime which involves walking of the coupling, and the amount of walking is determined by the quark mass. Continuity of the tachyon implies for $x\le x_c$ that (see Appendix~\ref{app:scaling} for details)
\be \label{regimeBlowx}
 \frac{m_q}{\LUV} \sim \frac{\LIR^2}{\LUV^2} \sim \frac{\sigma}{\LUV^3}\ ,  \qquad \left(x\le x_c\quad \mathrm{and} \quad \exp\left[-\frac{2K}{\sqrt{x_c-x}}\right] \ll \frac{m_q}{\LUV} \ll 1\right) \ ,
\ee
whereas in the conformal window
\be \label{regimeBhighx}
\!\!\!\!\!\!\! \begin{array}{rcl} 
\frac{m_q}{\LUV} &\sim& \left(\frac{\LIR}{\LUV}\right)^{\Delta_*} \\ \qquad \frac{\sigma}{\LUV^3} &\sim& \left(\frac{\LIR}{\LUV}\right)^{4-\Delta_*} \sim \left(\frac{m_q}{\LUV}\right)^{\frac{4-\Delta_*}{\Delta_*}}\end{array} \qquad \left(x_c \le x < \xBZ \quad \mathrm{and} \quad  \frac{m_q}{\LUV} \ll 1\right) \ .
\ee
Notice that~\eqref{regimeBlowx} is obtained from~\eqref{regimeBhighx} by setting $\Delta_* = 2$, which is indeed the value of the anomalous dimension as $x \to x_c$ from above, and the scaling behavior is therefore continuous and both expressions are valid at $x=x_c$. Moreover,~\eqref{regimeBhighx} gives what is termed the ``hyperscaling'' relation for the chiral condensate (see, e.g., \cite{DelDebbio:2010ze}) and often written in terms of the anomalous dimensions $\gamma_* = \Delta_*-1$,
\be
 \frac{\sigma}{\LUV^3} \sim \left(\frac{m_q}{\LUV}\right)^{\frac{3-\gamma_*}{1+\gamma_*}} \ .
\ee
Such scalings have been studied recently in a specific holographic model~\cite{Evans:2014nfa}. The behavior of $\Lambda_\t$ is the same as for $m_q=0$ and in the regime~A, i.e.,
\be
 \Lambda_\t \sim \LIR\ .
\ee
\end{itemize}

Notice that we have not discussed the $m_q$-dependence of bound state masses and decay constants, because they are difficult to analyze without specifying the details of the model. In the regime~A and at small $x$, however, it is clear that the lowest bound state masses (except for the light pions) and decay constants must be \order{\LIR} because this is the only available scale (apart from the small perturbation due to the quark mass). The pions are expected to obey the GOR relation since this arises in holography quite in general.
In the regime~B, as well as in the regime~A when $x_c-x \ll 1$, the most natural expectation is that the masses and decay constants are still given by the soft IR scale \order{\LIR}. This will be demonstrated for V-QCD below.

\subsection{Large quark mass}

Let us then discuss the limit $m_q/\LUV \gg 1$ (regime~C in Fig.~\ref{fig:scalingregions}). 
Since the quark mass is large, the tachyon grows large very close to the UV boundary where the dilaton is still small. Therefore regime~C probes the limit of small dilaton and large tachyon in the holographic model. Due to the smallness of the dilaton  it is not obvious that the holographic description is reliable for all observables in this regime.  But in analogy to how the UV structure of V-QCD is fixed in order to guarantee correct boundary conditions for the IR dynamics (as explained in Sec.~\ref{sec:vqcd}), it is important that the holographic model is as close to QCD as possible also at large quark mass, in order to have the best possible boundary conditions for the physics at small and \order{1} quark masses. We will therefore analyze what are the possibilities in this limit.

Deep in the UV, for $r \ll 1/m_q$, the tachyon will have the standard form of~\eqref{tauruns} which already implies scaling results for $\sigma$ and $\Lambda_\t$, with similar assumptions as above. At $r \sim 1/m_q$ it will grow large, after which nonlinear effects are important and a generic solution cannot be written down. In particular, the RG flow is not expected to approach the IRFP at any point, and the IR structure of the tachyon solution shown above is not relevant. 

Instead, additional scaling results can be derived assuming that the holographic model implements some key features of QCD (as will be the case in V-QCD). These are the RG flow of the 't Hooft coupling in QCD, which also gives the proper definition of the UV scale $\LUV$ in regime~C, and the decoupling of the massive quarks at energy scales much smaller than $m_q$. Such decoupling is automatically implemented in V-QCD by the Sen-like exponential tachyon potential of the flavor action, as we will explain in Sec.~\ref{sec:massscales}. 
Due to the decoupling, the RG flow is that of full QCD for $r \ll 1/m_q$ and that of YM theory for $r \gg 1/m_q$.
Continuity of the 't Hooft coupling (rather than the tachyon) is required. 

The collected results for the regime~C in Fig.~\ref{fig:scalingregions} are the following (details can again be found in Appendix~\ref{app:scaling}):
\begin{itemize}
 \item[C] The regime of large quark mass ($m_q/\LUV \gg 1$). The form of the tachyon solution implies that
 \be \label{largemqtauscales}
  \sigma \sim m_q^3 \qquad \mathrm{and} \qquad \Lambda_\t \sim m_q\ .
 \ee
 Here $\sigma$ will be connected to the properly renormalized chiral condensate (see Appendix~\ref{app:freeencoll} for details). The renormalization also involves scheme dependence which is important at large quark mass (see~\cite{ikp} for a discussion in the context of holography). The condensate is proportional to $m_q^3$ as in~\eqref{largemqtauscales} for generic schemes. 
 Further, analysis of the RG flow leads to
\be \label{largemqscalingt}
 \frac{\LUV}{\LIR} \sim \left(\frac{m_q}{\LUV}\right)^{b_0/b_0^\mathrm{YM}-1}\ , \qquad  \frac{m_q}{\LIR} \sim \left(\frac{m_q}{\LUV}\right)^{b_0/b_0^\mathrm{YM}}
\ee 
where $b_0$ ($b_0^\mathrm{YM}$) is the leading coefficient of the beta function for QCD in the Veneziano limit (YM theory at large $N_c$). That is,
\be
 \frac{b_0}{b_0^\mathrm{YM}} = 1-\frac{2x}{11} \ .
\ee
\end{itemize}

Let us then briefly comment on the size of the bound state masses in regime~C. First, as the quarks are decoupled for energy scales smaller than the quark mass, the glueballs are expected to decouple from the mesons, and have a  mass gap of \order{\LIR}. Recall that the meson states in QCD become nonrelativistic at large $m_q$, and therefore their mass gap is $\sim 2 m_q$, and the mass splitting of the low-lying states is much smaller than the gap. 
The example of V-QCD, which we shall discuss below, shows that obtaining such a mass gap in holography is nontrivial (see the analysis for V-QCD in Appendix~\ref{app:masses}), and the gap can be either \order{m_q} or \order{\LIR} for actions which produce reasonable results at small $m_q$.
If an IR cutoff is placed at the point where the tachyon grows large (as in the dynamic AdS/QCD models~\cite{Alvares:2012kr}), mass gap $\propto m_q$ is obtained. In the presence of such a cutoff, $\Lambda_\t \sim m_q$ is the only scale in the system. But in this case the splitting between the bound state masses is also expected to be \order{m_q}.

Finally let us comment on the structure of Fig.~\ref{fig:scalingregions} near the BZ region. Namely, the boundary of the regime~C bends toward higher $m_q/\LUV$ as $x$ grows. Similarly the upper boundary of the regime~B bends down in the BZ region. This is a generic feature due to RG flow which becomes slower and slower in the BZ limit $x \to \xBZ$. Due to slowness of the flow the separation of the energy scales $m_q$ and $\LUV$ needs to be larger for the scaling results to apply. This phenomenon is discussed in more detail in Appendix~\ref{app:scaling}.

Actually the BZ regime could be identified as an additional scaling regime. Here this is not done, however, since from the point of view of holography this regime is not the most interesting one. The coupling constant is restricted from above by its small value at the IRFP, the theory is perturbatively soluble, and therefore holographic description is not expected to be useful.

\section{Scaling of bound state masses in V-QCD} \label{sec:massscales}

While we discussed above the generic behavior of (holographic) QCD in the Veneziano limit, we shall now derive explicit predictions for the V-QCD models defined in Sec.~\ref{sec:vqcd}, and demonstrate that they agree with the results of the previous section.

\subsection{Energy scales in V-QCD}

The various energy scales can be defined explicitly in terms of the background solutions. For definiteness we will write down the definitions here. We will use the UV and IR expansions which can be found in Appendix~D of~\cite{jk} and in Appendix~D of~\cite{Arean:2013tja}. The UV expansions can also be found in Appendix~\ref{app:freeen}. 
\begin{itemize}
 \item First, $\LUV$ is the scale of the UV expansions. The precise definition is most conveniently given in terms of the UV expansion of the dilaton $\l$:  
\be \label{laUV}
 \l = -\frac{1}{b_0 \log (r\LUV)} -
\frac{  8 b_1 \log\left[-\log(r \LUV)\right]}{9 b_0^2\log(r \LUV)^2}+{\cal
O}\left(\frac{1}{\log(r\LUV)^3}\right) \ .
\ee
Recall that the coefficients of the potentials in the V-QCD action were matched to those of the QCD beta function in the UV. We used this mapping to write the coefficients in the expansion~\eqref{laUV} in terms of the coefficients $b_i$ of the QCD beta function in the Veneziano limit,
\be
 \beta(\l) \equiv \frac{d\l}{d\log \m} = -b_0 \l^2 +b_1 \l^3 + \cdots
\ee 
\item The IR scale $\LIR$ can be defined analogously in term of the IR expansions. For the standard geometry in V-QCD with an IR singularity the definition of $\LIR=1/R$ is given in limit $r \to \infty$ as~\cite{jk} 
\be \label{laIR}
 \log \l = \frac{3}{2} \frac{r^2}{R^2} +\morder{r^0} = \frac{3}{2} \LIR^2r^2 +\morder{r^0}  \ .
\ee
In the presence of an IRFP the definition is modified to
\be \label{laIRFP}
 \l \simeq \l_* -\left(\frac{r}{R}\right)^{-\delta} =\l_* -\left(r \LIR\right)^{-\delta}
\ee 
where $\delta = \Delta_{FF}-4$ is the anomalous dimension of the $\mathrm{Tr} F^2$ operator at the fixed point.
\item The quark mass is determined\footnote{It is well known that the quark mass can be defined only up to a constant. This constant is set to one for notational simplicity.} in terms of the UV asymptotics of the tachyon ($r\to 0$),
\be
 \frac{\tau(r)}{\ell} = m_q r (-\log(r\LUV))^{-\rho}\left[1+\morder{\frac{1}{\log(r\LUV)}}\right]\ ,
\ee 
where $\ell$ is the UV AdS radius and $\rho=\gamma_0/\beta_0$ is the ratio of the leading coefficients $\beta_0$ and $\gamma_0$ of the beta function and the anomalous dimension of the quark mass in QCD, respectively. Notice that in V-QCD, the logarithmic running of the quark mass is therefore included, and is matched to agree with that of QCD.
\item The scale $\L_\t$ where the tachyon grows large is defined simply by\footnote{For potentials~II slightly modified definition is used: the number on the right hand side is set to $1/10$. This is necessary because the transition to the IR region where nonlinear terms in the tachyon are important turns out to happen when the tachyon is still smaller than one, and in the nonlinear region the tachyon grows relatively slow. Therefore using exactly~\eqref{Lambdataudef} would lead to an energy scale which does not precisely reflect the change in the dynamics.}
\be \label{Lambdataudef}
 \left.\frac{\tau}{\ell}\right|_{A=\log \Lambda_\t} =  1 \ ,
\ee
where $\t$ and $A$ are to be evaluated on the background solution. 
\end{itemize}

The background EoMs for V-QCD are invariant under the transformation
\be
 r \to \L r \ , \qquad A \to A - \log\L \ .
\ee
This transformation changes all energy scales defined above by the same number, and reflects the choice of the units of energy on the field theory side. Consequently, only the ratios of the above energy scales are  a priori well-defined. Notice also that $\LIR$ and $\Lambda_\t$ do not have direct counterparts in field theory, whereas $\LUV$ is indeed mapped to the scale of the RG flow in field theory and $m_q$ is identified (up to a constant) as the quark mass on the field theory side. Therefore the values of $\LUV$ and $m_q$ (up to the ambiguities mentioned above) can be inferred from QCD data only, whereas the determination of $\LIR$ and $\Lambda_\t$ also requires pinning down the holographic action.

There is a small issue with the above definitions which will be visible in the explicit results below. Namely, the definition of  $\LUV$ which is natural at generic values of $x$ is not optimal at high $x$ and in particular in the BZ limit. In this limit 
one would expect $\LUV$ to match the scale of the UV RG flow, but this turns out not be be the case. 
Also at zero quark mass there should be only a single scale and therefore $\LUV$ should equal $\LIR$, but as it turns out, actually $\LUV$ becomes exponentially suppressed with respect to $\LIR$. Because the RG flow of the coupling is controlled by the two-loop beta function in the BZ limit, a single scale $\widetilde \L$ can be defined which has better behavior in this limit (see Appendix~B of~\cite{Kiritsis:2013xj}): 
\be
 \left(\frac{b_0}{b_1 \l}-1\right) \  \exp\left(\frac{b_0}{b_1 \l}\right)\simeq  \left(\widetilde \L r\right)^{-\frac{b_0^2}{b_1}}
\ee
This scale is related to $\LUV$ in the BZ limit by
\be
 \widetilde \L \simeq \left(\frac{b_1}{b_0^2}\right)^\frac{b_1}{b_0^2} \LUV \ ,
\ee
where $b_1/b_0^2 \sim (\xBZ-x)^{-2}$ grows large in the BZ limit (see Appendix~\ref{app:scaling} for some more details). Notice that $\LUV$ will anyhow be used as a reference scale in the numerical analysis below -- since the BZ region is not very interesting from holographic viewpoint, we have chosen to use the same definition of $\LUV$ as earlier literature, even though it has unnatural behavior in this region.

\subsection{Flavor nonsinglet masses and decay constants: scaling results}

Before going to the numerical results let us argue how the scaling laws for the masses and decay constants can be derived in V-QCD. The meson masses in each sector (for vectors, axial vectors, scalars, and pseudoscalars) may be defined into two classes: the flavor singlet and nonsinglet states. The former are singlets under the vectorial $SU(N_f)$ transformation, whereas the latter are the other fluctuation modes, which correspond to quark bilinear operators involving the Hermitean traceless generators 
$t_a$ of $SU(N_f)$. We restrict analytic considerations to the nonsinglet states, because the singlet mesons mix with the glueball states, which would lead to complications~\cite{Arean:2013tja,glue}. Only the main points are summarized here and the details can be found in Appendix~\ref{app:masses}. 

The fluctuation equations for the scalar nonsinglet mesons can transformed in the Schr\"odinger form as detailed in Appendices~A and~B of~\cite{Arean:2013tja}. The masses for each sector are then given as eigenvalues of a Schr\"odinger equation with a certain potential term and the Schr\"odinger coordinate running from $u=0$ (UV) to $u=\infty$ (IR). In order to find the behavior of the mass gaps and splittings one then needs to study the Schr\"odinger potentials $V_S(u)$. 

Let us first take finite but small quark mass ($m_q/\LUV \ll 1)$, so that the regimes~A and~B are covered. We use here the V-QCD action, but results in this region are not sensitive to the details of the action. The Schr\"odinger potential for $u \ll 1/\LIR$ is given by the background with small or walking dilaton and small tachyon, so that the geometry is close to AdS, and consequently $V_S(u) \sim \mathrm{const}/u^2$. For $u \gg 1/\LIR$ the diverging tachyon creates the confining potential ($V_s(u) \sim \LIR^4 u^2$ if the excitation spectrum is linear, $m_n^2 \sim n$ with $n$ being the excitation number). Therefore the only relevant scale is 
$\LIR$ and the (lowest lying) meson masses as well as the 
mass 
splittings between the modes are given by this scale:
\be \label{massscalingregsAB}
 m_n \sim \LIR \ ,
\ee 
which is also consistent with the results from Dyson-Schwinger and Bethe-Salpeter approaches in regime~A~\cite{Harada:2003dc}. In units of $\LUV$, by using the scaling laws from Sec.~\ref{sec:scaling} we therefore obtain  the same results as at vanishing quark mass~\cite{letter,Arean:2013tja} within regime~A: $m_n \sim \LUV$ in the QCD regime and
the masses go to zero obeying Miransky scaling, $m_n/\LUV \sim \exp(-K/\sqrt{x_c-x})$, in the walking regime. In regime~B, we find the ``hyperscaling relations''~\cite{DelDebbio:2010ze,Evans:2014nfa} for the low-lying meson masses:
\begin{align}
 \frac{m_n}{\LUV} &\sim \sqrt{\frac{m_q}{\LUV}} \ ,&  \  &\left(x\le x_c\quad \mathrm{and} \quad \exp\left[-\frac{2K}{\sqrt{x_c-x}}\right] \ll \frac{m_q}{\LUV} \ll 1\right) \ ,& \nn\\
\frac{m_n}{\LUV} &\sim \left(\frac{m_q}{\LUV}\right)^{1/\Delta_*}\ ,& \  &\left(x_c \le x < \xBZ \quad \mathrm{and} \quad  \frac{m_q}{\LUV} \ll 1\right) \ .&
\end{align}

The sole exception to these scaling results is the pion mode (for $x<x_c$), which is massless at $m_q=0$ and obeys the GOR relation in regime~A:
\be
 m_\pi^2 f_\pi^2 \sim m_q \sigma \ ,
\ee
so that it is lighter than the other meson states.
The GOR relation will be discussed in more detail in Sec.~\ref{sec:condensate}. At the level of the Schr\"odinger formalism the absence of the mass gap in the pseudoscalar sector is reflected in the negativity of the Schr\"odinger potential in the UV region. In regime~B, the pseudoscalar masses will also obey the scaling~\eqref{massscalingregsAB}.

The various decay constants are slightly more difficult to analyze. In Appendix~\ref{app:masses} it is argued that they are similarly of the order of $\LIR$ when the quark mass is small.

At large quark mass (regime~C) the meson masses and decay constants depends more on the details of the action. Actually only the form of the flavor action $S_f$ in the limit of large tachyon and small dilaton is relevant, as is shown in Appendix~\ref{app:masses}. In this limit the functions $a$, $\kappa$, and $V_{f0}$ become almost constants and the action of~\eqref{generalact} takes the form of the standard DBI action with the exponential Sen potential.

The exponential dependence  $V_f \propto  \exp(-a(\l) \t^2)$ of the flavor potential on the (squared) tachyon naturally implements the decoupling of the massive quarks. For large quark mass the tachyon is roughly proportional to $m_q r$ in the UV (for $r \ll 1/m_q$). When $r$ grows larger than $1/m_q$, that is, at energies lower than $m_q$ on the field theory side, the tachyon grows sizeable (see Appendix~\ref{app:largemq} for details) and the exponential factor in $V_f$ decays rapidly. Consequently, the flavor part of the V-QCD action becomes suppressed with respect to the glue part, and therefore the dynamics at energies below $m_q$ is governed by the gluons as expected. The decoupling of flavors will also be well visible in the numerical results for the scalar singlet states below which involve both glueball and $\bar qq$ components.

Interestingly, as shown in Appendix~\ref{app:masses}, the choice $V_f(\l,\t) = V_{f0}(\l) \exp(-a_0\t^2)$ of potentials~I, where the function $a(\l)$ is set to a constant value $a_0$, is a special case. This choice was motivated by the asymptotics of the meson trajectories at zero quark mass~\cite{jk,Arean:2013tja}, and noticeably $a(\l)$ is also constant in tachyon potentials obtained from string theory~\cite{Kraus:2000nj,Bigazzi:2005md,ckp}. In Appendix~\ref{app:masses} we show that this choice is essentially the only one which produces physically reasonable mass gap and splitting, if the value of $a_0$ is slightly modified from the value of potentials~I which reproduces the correct dimension of the quark mass and condensate in the UV.

We will anyhow discuss the results for potentials~I without this modification of $a_0$, because such a modification was not introduced for our numerical studies. It is found that the mass gap for the (flavor nonsinglet) mesons in all sectors (vectors, axials, scalars and pseudoscalars) is given by
\be \label{massscalingregC}
 m_\mathrm{gap} \sim m_q^\xi\ ; \qquad \xi = \frac{\left.3\ell^2\right|_{x=0}}{4 \ell^2} = \frac{3}{4} \left(1-\frac{x W_0}{12}\right)\ ,
\ee
where $\ell$ is the UV AdS radius, and $\xi=1$ would be required to match with the field theory result for nonrelativistic bound states\footnote{One can have both $\xi=1$ and the correct UV dimension for the quark mass if extra terms (e.g. $\propto (1 +\# \t^2)$) are added in the tachyon potential.}. 
We stress that this formula was obtained by fixing the value of $a_0$ to produce the UV dimension  of the quark mass and the condensate, as was done for the potentials~I used in the numerics. The last expression in~\eqref{massscalingregC} suggests that $\xi=1$ could be obtained by tuning the value of $W_0$. This is, however, problematic because the contribution involving $W_0$ should vanish in the probe limit $x \to 0$, and also because $W_0$ would negative, which causes $V_f$ to have a node~\cite{jk} at which the tachyon background equation becomes singular. 

The mass splitting is suppressed with respect to the gap as is the case in real quarkonia. For potentials~I it scales as the inverse of the mass gap. 
The decay constants of the lowest meson states are exponentially suppressed in the quark mass
\be
 f_n \sim \exp(-\# m_q^{2\xi})
\ee
and they are therefore decoupled. This is found for all mesons with masses below \order{m_q}.  The decay constants of the states with masses $\sim m_q$ are of the order of $\LIR$.

\subsection{Numerical results}

Computing the energy scales and masses numerically is straightforward (but tedious) after the background has been constructed numerically~\cite{jk,Arean:2013tja}. The potentials~I (given explicitly in Appendix~\ref{app:numerics}) are used here unless stated otherwise. We choose three reference values $x=1$, $4$, and $4.5$, which lie in the QCD regime, walking regime, and conformal window, respectively, and plot the observables as functions of the quark mass. We also show plots where $m_q/\LUV$ is fixed to $10^{-6}$ and $x$ is varied over the whole parameter space. These choices cover the most interesting structures of Fig.~\ref{fig:scalingregions}. 

Notice that when plotting dimensionful parameters as a function of $x$ we are comparing different theories and a choice for the reference scale must be made. 
A natural choice would be $\Lambda_\mathrm{UV}$, which roughly
corresponds to the scale where the 't Hooft coupling takes some fixed
tiny value very close to the UV boundary. In many of the plots below, however, 
$\Lambda_\mathrm{IR}$ is chosen as the reference scale instead, simply because this makes the plots more easily readable.

\begin{figure}[!tb]
\begin{center}
\includegraphics[width=0.49\textwidth]{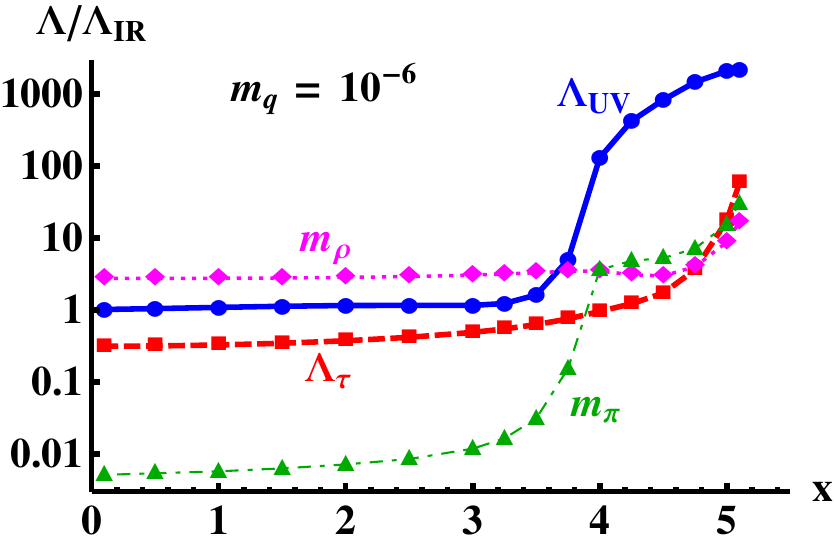}%
\includegraphics[width=0.49\textwidth]{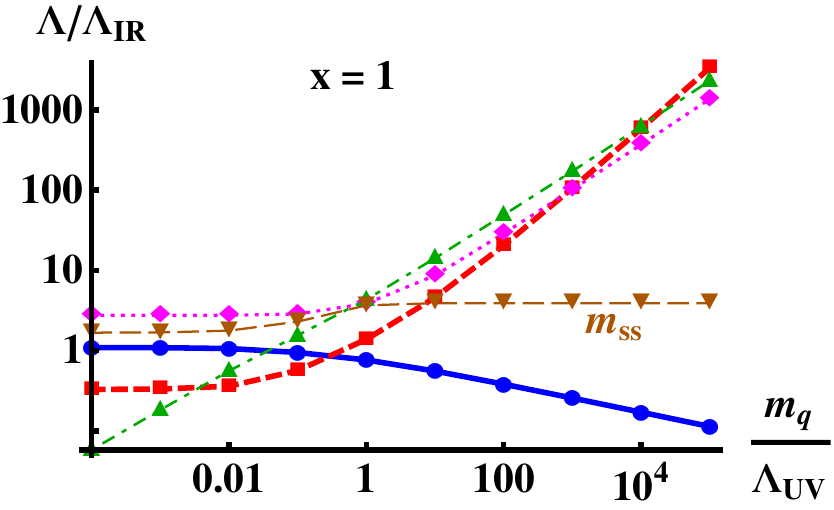}
\includegraphics[width=0.49\textwidth]{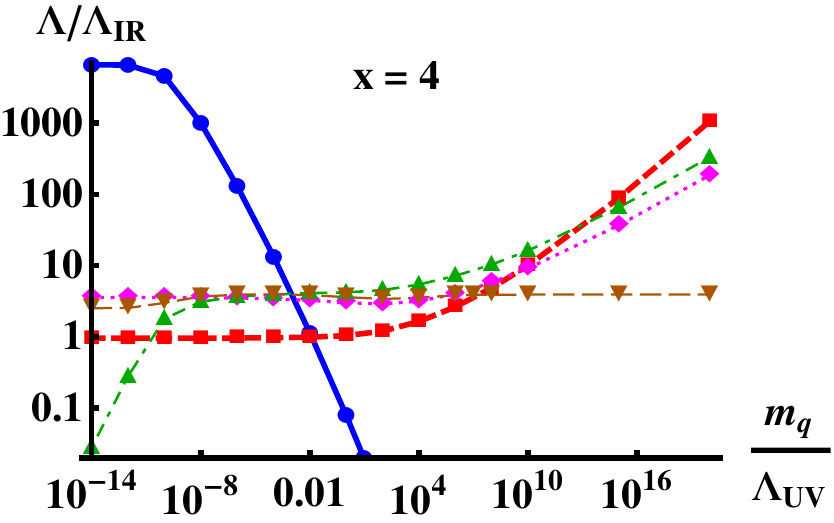}%
\includegraphics[width=0.49\textwidth]{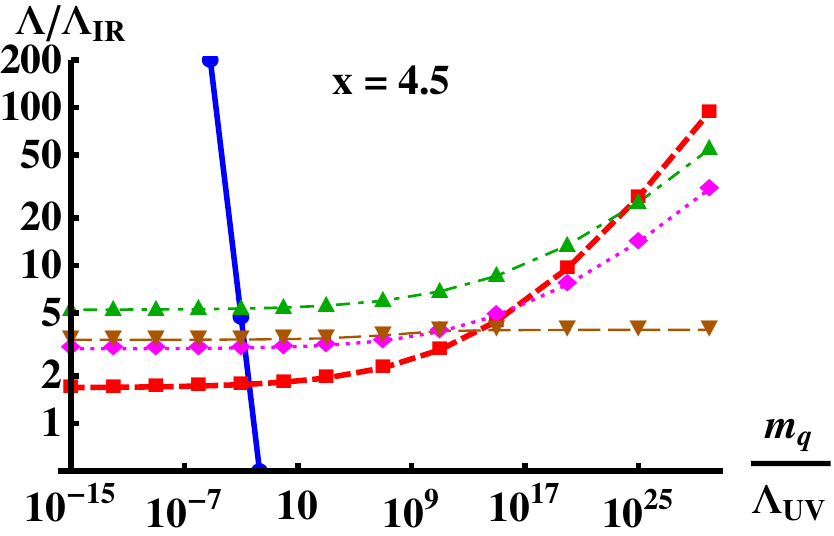}
\end{center}
\caption{Dependence of the energy scales on $x$ and $m_q$. Also the masses of the rho meson $m_\rho$, the lowest singlet scalar $m_\mathrm{ss}$, and the pions $m_\pi$ are shown for reference. The blue solid curves is $\LUV$, the red dashed curve is $\Lambda_\t$, the thin dotted magenta curve is the rho mass, the thin dotdashed green curve is the pions mass, and the thin long-dashed brown curve is the mass of the lowest singlet scalar in each of the plots. See text for more details.}
\label{fig:Escales}\end{figure}

\subsubsection{Energy scales}

In Fig.~\ref{fig:Escales} the dependence of the energy scales on $x$ and $m_q$ is demonstrated. We have chosen to show the scales $\LUV$ and $\Lambda_\t$ in the units of $\LIR$ since this makes the details visible. Some of the bound state masses are also shown as thin lines. 

The top-left plot shows the dependence of the scales on $x$ at a tiny quark mass ($m_q/\LUV = 10^{-6}$). The data extend only up to $x=5.1$ because in the BZ region it is difficult to do reliable numerics. The top-right plot shows the mass dependence in the running regime ($x=1$). The bottom-left plot is in the walking regime ($x=4$, which is close to $x_c\simeq 4.0830$). The bottom-right plot is in the conformal window ($x=4.5$).

The thick solid blue curves show the ratio $\LUV/\LIR$ in each plot. In the top-left plot the crossover from the QCD-like regime to conformal window is clearly visible: As $x \to x_c$ from below, the ratio first grows according to the Miransky scaling law~\eqref{Mscal} in regime~A until the condition~\eqref{regAcondw} no longer holds. Thereafter the ratio saturates to roughly $\sqrt{\LUV/m_q}$ in regime~B as predicted by~\eqref{regimeBlowx}, and then one moves out of regime~B at even higher $x$. 

The dependence of  $\LUV/\LIR$ on $m_q$ also follows the predicted scaling laws. In the regime~A (low $m_q/\LUV$ in top-right and bottom-left plots) the ratio is constant as the quark mass is a small perturbation. In the regime~C, i.e., at large $m_q/\LUV$, the ratio follows the power law of~\eqref{largemqscalingt} as best seen in the top-right plot. In the regime~B the ratio obeys the other power law of~\eqref{regimeBlowx} and~\eqref{regimeBhighx} as best seen at intermediate $m_q$ in the bottom-left plot. Notice that much of the solid blue curve was left out in the bottom row plots in order to make the details of the other curves better visible.

The thick dashed red curves show the ratio $\Lambda_\t/\LIR$. At small quark mass, including regimes~A and~B, the ratio is close to one as expected (except in the BZ limit). For the explanation of the divergence of $\Lambda_\t/\LIR$ in the BZ limit see Appendix~\ref{app:scaling}, Eq.~\eqref{LIRmqBZ}.  At large quark mass (regime~C), we find $\Lambda_\t \sim m_q$ as predicted in~\eqref{largemqtauscales}. 
At high $x$ (plots in the bottom row) the convergence toward this scaling law is quite slow due to the slow running of the quark mass. It could be demonstrated by continuing the plots up to much larger $m_q/\LUV$, but we have chosen not to do so in order to show the other details in the plots more clearly. Actually all variables vary slower and slower as functions of $m_q/\LUV$ when $x$ is grows, and therefore we have substantially increased the range of $m_q/\LUV$ in the plots with higher $x$, but this is still not enough to demonstrate the large $m_q$ scaling convincingly. The change in the $m_q$-dependence is due to the RG flow (see the end of Appendix~\ref{app:scaling} for some more details).

The dependence of $\sigma$ on $m_q$ will be discussed in Sec.~\ref{sec:condensate}.

\begin{figure}[!tb]
\begin{center}
\includegraphics[width=0.49\textwidth]{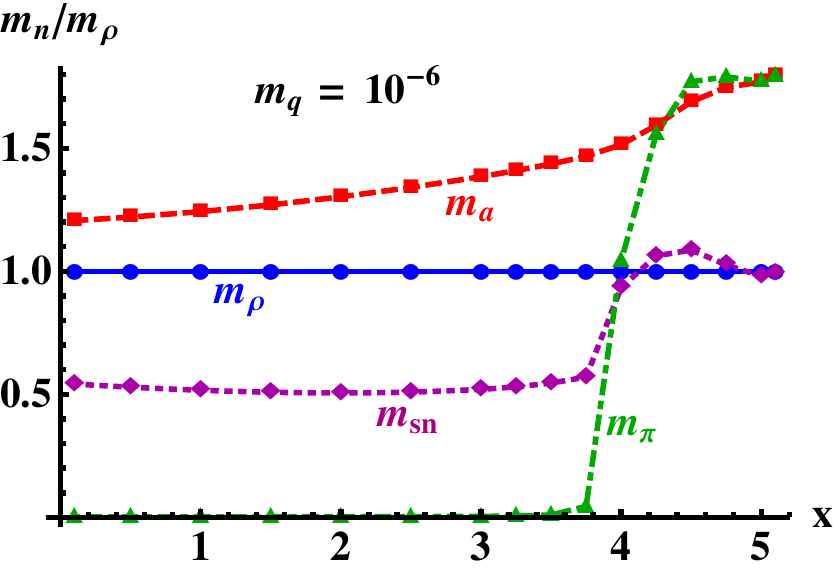}%
\includegraphics[width=0.49\textwidth]{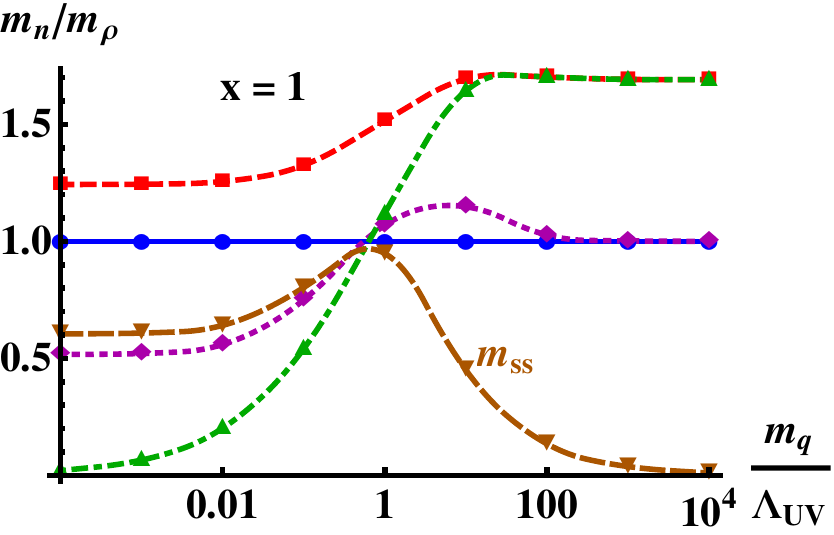}
\includegraphics[width=0.49\textwidth]{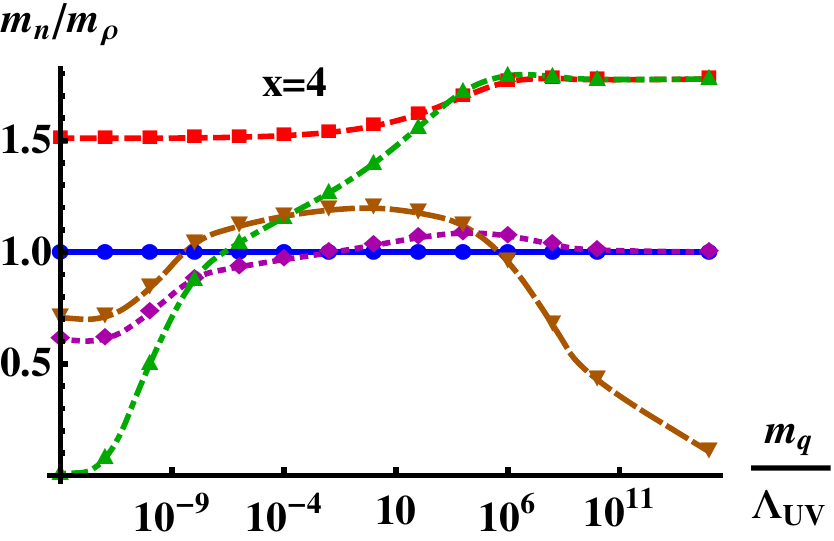}%
\includegraphics[width=0.49\textwidth]{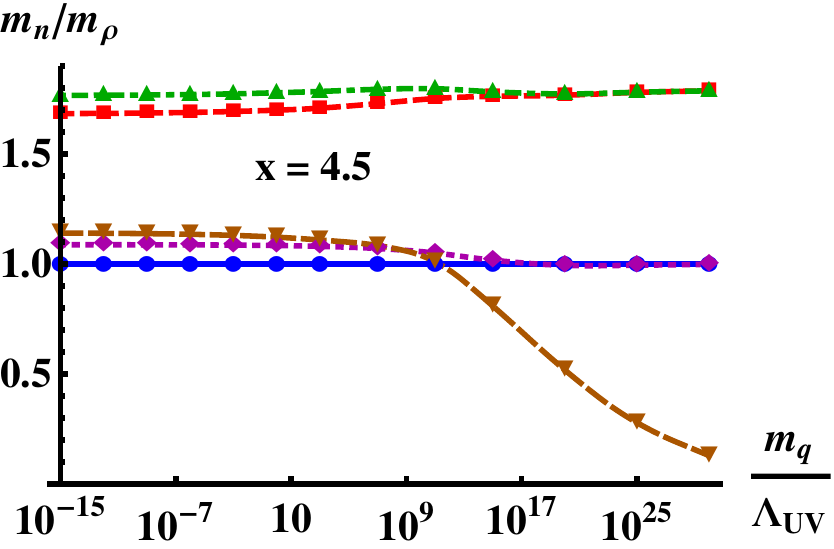}
\end{center}
\caption{Dependence of the mass ratios between the lowest excitations of each tower on $x$ and $m_q$. The choices of $x$ and $m_q/\LUV$ are as in Fig.~\protect\ref{fig:Escales}. Blue solid curve is the lowest vector ($\rho$ meson) mass, red dashed curve is the nonsinglet axial vector mass, dotted magenta curve is the lowest nonsinglet scalar mass, and the dotdashed green curve is the nonsinglet pseudoscalar (pion) mass in each plot. In addition, the long-dashed brown curve is the lowest singlet scalar mass. }
\label{fig:mratios}\end{figure}

\begin{figure}[!tb]
\begin{center}
\includegraphics[width=0.49\textwidth]{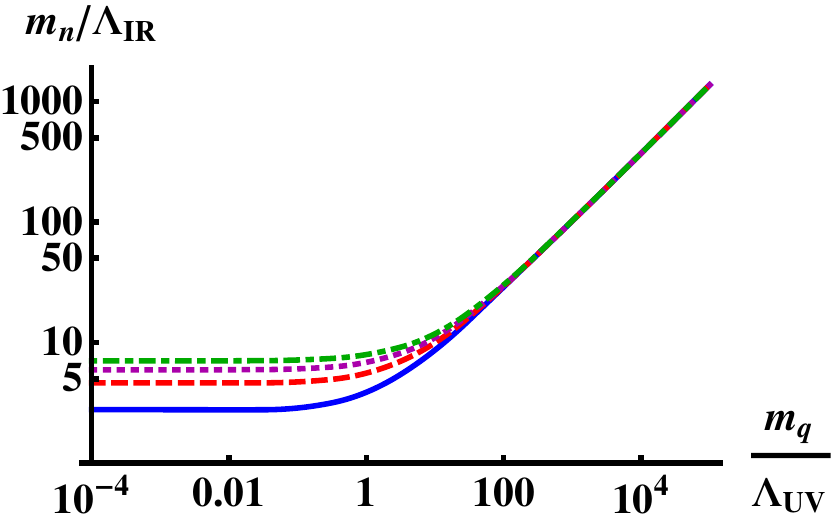}%
\includegraphics[width=0.49\textwidth]{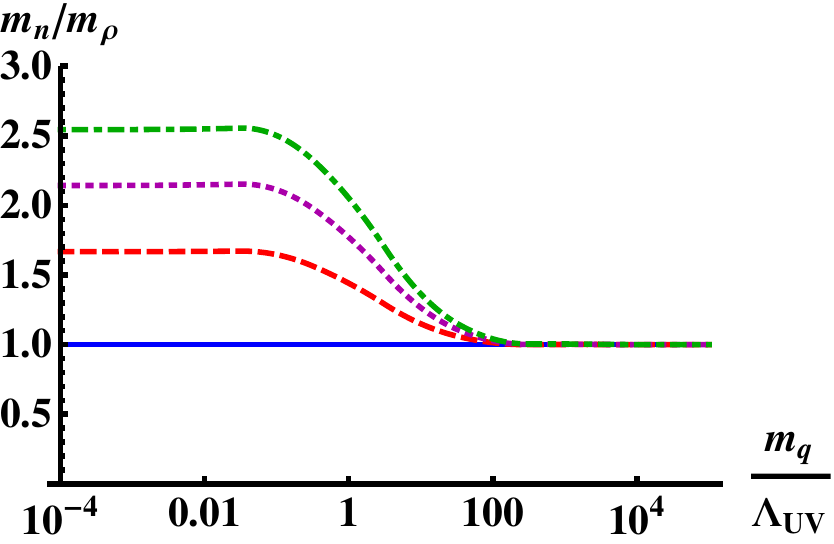}
\end{center}
\caption{Masses of the lowest four vector states as a function of $m_q$ at $x=1$. Left: masses in units of $\Lambda_\mathrm{IR}$. Right: masses normalized to the mass of the lowest vector meson (the $\rho$-meson).}
\label{fig:vmasses}\end{figure}

\subsubsection{Flavor nonsinglet masses and decay constants}

The flavor nonsinglet spectra can be computed as explained in~\cite{Arean:2013tja}. The results for different values of $x$ and $m_q$ are shown in Figs.~\ref{fig:Escales},~\ref{fig:mratios}, and~\ref{fig:vmasses}.

First, the $\rho$ and $\pi$ masses are shown in Fig.~\ref{fig:Escales} with thin dotted magenta and thin dotdashed green curves, respectively. As expected, the $\rho$ mass is \order{\LIR} at small quark mass (regimes~A and~B) and obeys the power law~\eqref{massscalingregC} in regime~C. The pion mass is close to the $\rho$ mass in regimes~B and~C, but in regime~A it obeys the GOR relation instead, as best visible from the top-right and bottom-left plots at small $m_q$. In the top-left plot, $m_\pi/\LIR$ is \order{\sqrt{m_q/\LUV}} at small $x$ but then increases with $x$ in the walking regime (actually obeying the Miransky scaling law) until the ratio is \order{1} at the point of the crossover near $x=x_c$.

Fig.~\ref{fig:mratios} shows the masses of the lowest meson states (i.e., the mass gaps) in each sector normalized to the $\rho$ mass. The choices of $x$ and $m_q$ are the same as in Fig.~\ref{fig:Escales}. The masses of the lowest vector, axial, scalar, and pseudoscalar states are given by the solid blue, dashed red, dotted magenta, and dotdashed green curves, respectively (the brown curves give the scalar singlet mass gap which will be discussed below). These ratios are mostly constant and close to one as predicted by the above scaling arguments. The exception is the pion mass (dotdashed green curves) which obeys the GOR relation in regime~A. Notice that in regime~C all meson mass gaps should approach the same number (roughly $2 m_q$) as expected for nonrelativistic bound states, but we find instead that the axial and pseudoscalar gaps are larger than those of the vectors and scalars. The reasons for this are analyzed in Appendix~\ref{app:masses}. Notice also that the lowest scalar states are 
lighter than vectors even at small values of $x$, which seems to be in conflict with QCD. 
Such details are, however, sensitive to the choice of the potentials in the V-QCD action, and can be changed by tuning the potentials.

Finally we plot the masses of the four lowest vector states in Fig.~\ref{fig:vmasses} in order to demonstrate the dependence of the mass splittings in the spectra on $m_q$. The masses are given in units of $\LIR$ (left hand plot) and normalized to the lowest mass, i.e., the $\rho$ mass (right hand plot). The splittings decrease with increasing $m_q$ in regime~C which is in qualitative agreement with the bound states becoming nonrelativistic, but the power laws are not exactly correct (see Appendix~\ref{app:masses} for details).

\begin{figure}[!tb]
\begin{center}
\includegraphics[width=0.49\textwidth]{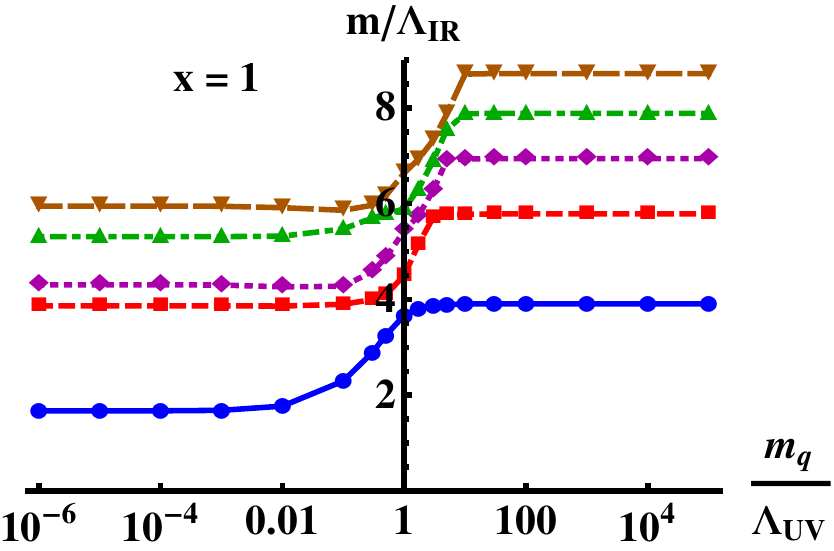}%
\includegraphics[width=0.49\textwidth]{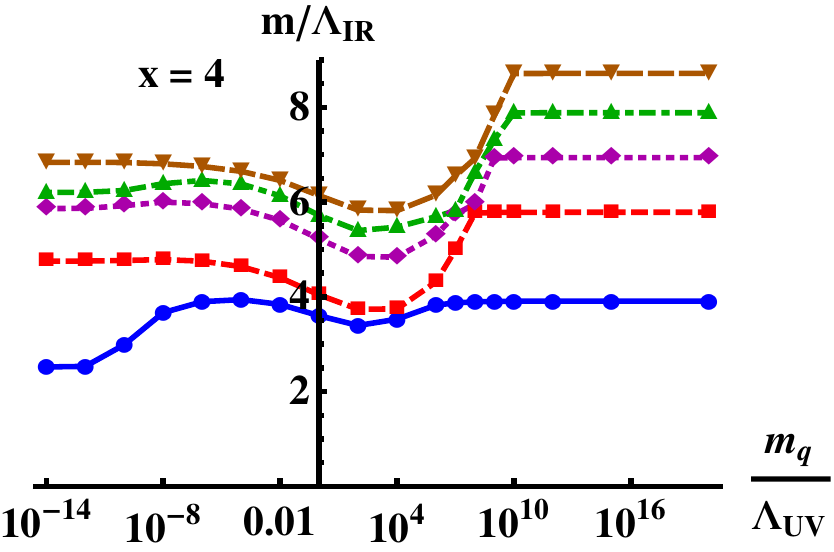}
\includegraphics[width=0.49\textwidth]{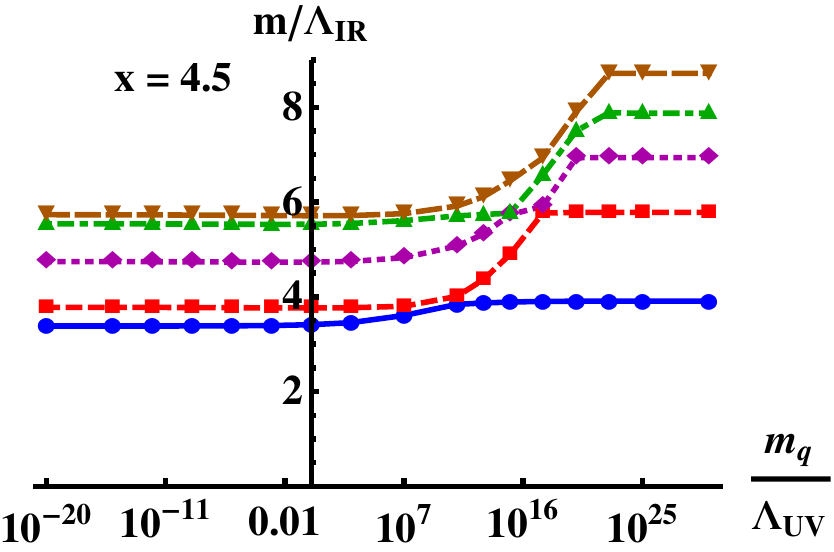}
\end{center}
\caption{The masses of the five lowest scalar singlet states as functions of $m_q$ for various choices of $x$.}
\label{fig:ssmasses}\end{figure}

\subsubsection{Scalar singlet masses} \label{sec:singlets}

The singlet sector is qualitatively different from the nonsinglet sector because it also contains glueball states which mix nontrivially with the singlet meson states. Such mixing takes place in the scalar and pseudoscalar sectors. The scalar sector will be discussed in detail here while the pseudoscalar sector will be analyzed in a future publication~\cite{cpodd}. 

Before going to the numerical results, let us discuss the generic features of the spectrum. The singlet mesons masses are expected to show similar\footnote{There is an exception in the singlet pseudoscalar sector: the lowest state, the $\eta'$ meson, is anomalously light at small $x$ in the Veneziano limit~\cite{Witten:1978bc,vu1}. This is also reproduced by V-QCD~\cite{cpodd}.} $m_q$ and $x$ dependence as the nonsinglet mesons above, given by~\eqref{massscalingregsAB} and~\eqref{massscalingregC}. In the singlet case there is, however, nontrivial mixing of the mesons with the glueball states, the masses of which should be independent of $m_q$ and therefore always characterized by $\LIR$. In particular in regime~C the masses of the mesons become much larger than the glueballs which suggest that the meson and glueball states decouple.

Let us then demonstrate these features numerically for the scalar singlet sector in V-QCD. The mass gap of the scalar singlets is shown by the long-dashed brown curves in Figs.~\ref{fig:Escales} and~\ref{fig:mratios}. The mass gap is that of the glueballs and therefore \order{\LIR} for all values of $m_q$ and $x$ as seen from Fig.~\ref{fig:Escales}. The fact that the lowest glueball mass is suppressed with respect to the rho mass $m_\rho$ in regime~C can also be seen in Fig.~\ref{fig:mratios}: the ratios $m_\mathrm{ss}/m_\rho$ given by the brown curves, where $m_\mathrm{ss}$ is the mass gap for the scalar singlet states, decrease with $m_q$ at large $m_q$.

The decoupling of the meson and glueball states is most clearly demonstrated by Fig.~\ref{fig:ssmasses}, where the masses of the five lowest scalar singlet states are plotted as a function of $m_q$ in the QCD regime ($x=1$, top-left), in the walking regime ($x=4$, top-right) and in the conformal window ($x=4.5$, bottom). When $m_q/\LUV \ll 1$ the spectrum has both glueball and meson states which are nontrivially mixed. As $m_q$ increases, the meson masses increase while the glueball masses stay constant, which leads to the crossing structure seen in the plots. This would be expected to happen at $m_q/\LUV \sim 1$, but as was discussed above, $\LUV$ is not exactly the scale of the UV RG flow when $x$ is large, and therefore the crossing structure shifts to higher values of $m_q$ as $x$ increases. At large $m_q/\LUV$ only the glueballs are left. Their decoupling from the mesons is demonstrated by the fact that the limiting values of their masses as $m_q \to \infty$ are independent of $x$, and in fact it can be 
checked that they match with the glueball masses obtained in the YM limit ($x \to 0$) of V-QCD.

It was shown in~\cite{letter,Arean:2013tja} that V-QCD does not have a light ``technidilaton'' mode~\cite{walk2} (which would be the lightest scalar singlet state) as $x \to x_c$. Both the singlet and nonsinglet scalars fluctuations do have critical~\cite{son} behavior in the near conformal region (see Appendix~I in~\cite{Arean:2013tja}) and the Schr\"odinger potential for the nonsinglet scalars is negative (as was also found in~\cite{Alanen:2011hh}), but it was shown numerically that this is not enough for a technidilaton to appear. The negative result is also seen in Figs.~\ref{fig:mratios} and~\ref{fig:ssmasses}: the lowest singlet scalar does not become light with respect to the other states in the walking regime ($x=4$). We do see, however, that it is lighter in regime~A than in regime~B.

\begin{figure}[!tb]
\begin{center}
\includegraphics[width=0.49\textwidth]{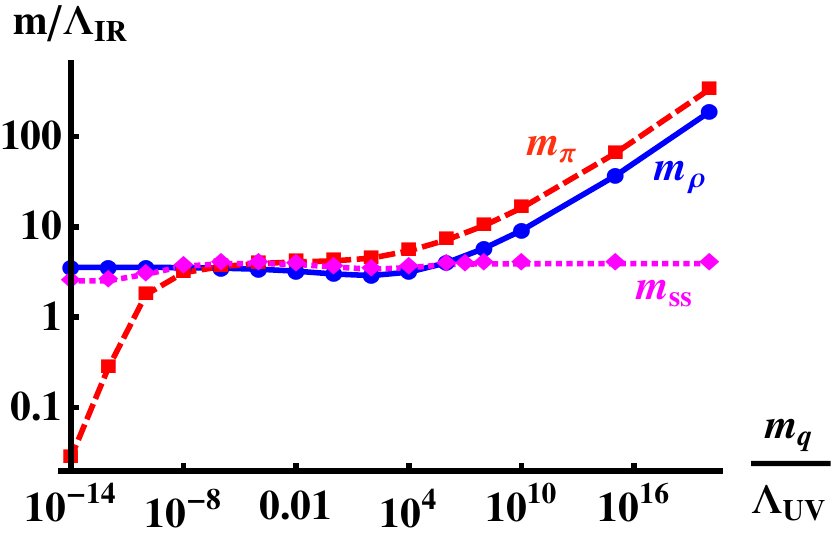}
\includegraphics[width=0.49\textwidth]{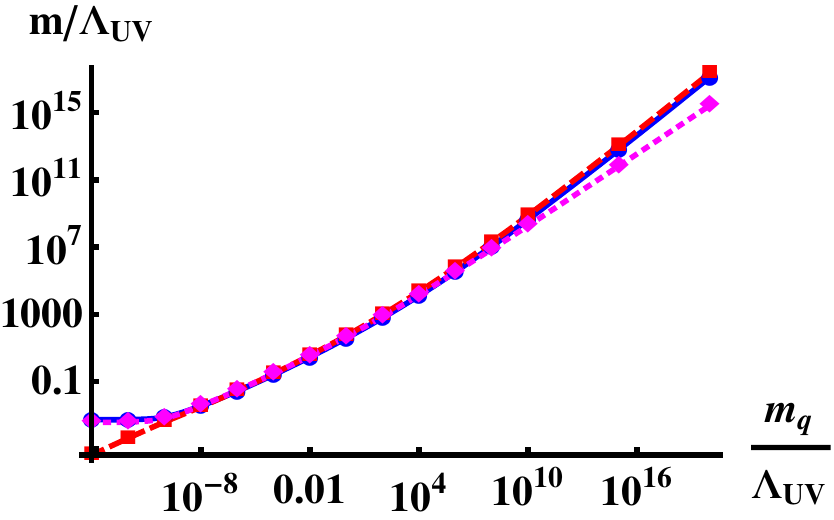}
\end{center}
\caption{The masses of the lowest vector (rho meson), pseudoscalar (pion), and singlet scalar states as a function of $m_q$ and in units of $\LIR$ (left) and $\LUV$ (right). The rho meson, pion, and singlet scalar masses are shown as the solid blue, dashed red, and doted magenta curves, respectively.}
\label{fig:massesonly}\end{figure}

\subsubsection{Masses near the conformal transition}

To conclude this section, let us add a few comments on the scaling of the bound state masses near the conformal transition.  We plot the masses of the rho meson, the pion, and the lowest scalar singlet state in Fig.~\ref{fig:massesonly}. The left hand plot shows the masses in units of $\LIR$ and the right hand plot shows the masses in units of $\LUV$. Notice that in units of $\LIR$ the pion mass deviates from the masses of the other states at extremely small $m_q$ (regime~A) where it obeys the GOR relation, but in units of $\LUV$ the pion mass obeys the same power law $\sim \sqrt{m_q/\LUV}$ in regimes~A and B. 

\begin{figure}[!tb]
\begin{center}
\includegraphics[width=0.69\textwidth]{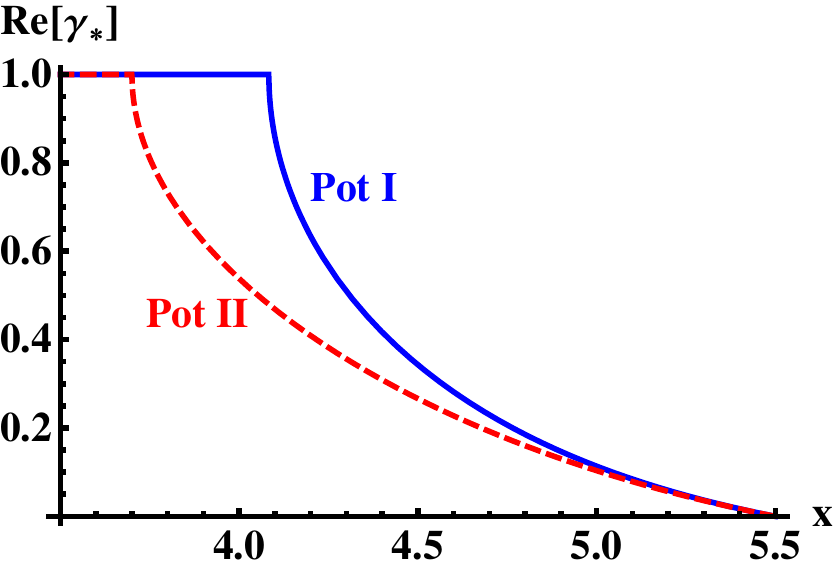}
\end{center}
\caption{The (real part of the) anomalous dimension at the fixed point as a function of $x$ for potentials I with $W_0=3/11$ (solid blue curve) and for potentials II with SB normalized $W_0$ (dashed red curve). The kink lies at $x=x_c$ for both potentials.}
\label{fig:gammastar}\end{figure}

Knowledge of the dependence of the meson masses on $m_q$ near the walking regime is important for the lattice studies which take place at finite quark mass, and aim to locate the conformal transition at $m_q=0$~\cite{Lombardo:2014mda,Lombardo:2014pda}. As we have pointed out above, the regime~B extends even to $x<x_c$ (see Fig.~\ref{fig:scalingregions}), and the crossover between regimes~A and~B moves to lower $x$ as $m_q$ is increased. This suggests that studies at finite $m_q$ lead to an underestimate for $x_c$. One should recall, however, that the scaling in regime~B involves $\Delta_* =\gamma_*+1$ which depends strongly on $x$. 
We show $\mathrm{Re} \gamma_*$, which controls the scaling exponents of the masses, as a function of $x$ for both potentials~I and~II in Fig.~\ref{fig:gammastar}. The kinks in the plots are located exactly at $x=x_c$, and $\gamma_*$ drops rapidly right above the kinks. This supports the idea that $x=x_c$ can be located by extracting $\gamma_*$ from the meson masses on the lattice. 
In fact, recent lattice results for $\gamma_*$ in the conformal window report very low values~\cite{Cheng:2013eu,Lombardo:2014pda} that are in apparent contradiction of the curves in Fig.~\ref{fig:gammastar}.  
Recall however that the model has not been tuned to fit any QCD data yet\footnote{One should also keep in mind that we are working in a bottom-up model, which is defined in the Veneziano limit whereas the lattice data was computed at finite $N_f$ and $N_c$.}. It is actually not difficult to construct potentials for which $\gamma_*$ drops much more rapidly when $x>x_c$.

\section{Quark mass and the chiral condensate} 
\label{sec:condensate}

Let us then discuss the mass dependence of the chiral condensate. Recall that the tachyon solution in the UV reads
\be \label{mqsigmadef}
\begin{split}
 \frac{\tau(r)}{\ell} &= 
m_q r \left(-\log r\LUV\right)^{-\rho}\left[1+\morder{\frac{1}{\log r\LUV}}\right]  \\
 & \ \ \ + \sigma r^3 \left(-\log r\LUV\right)^{\rho}\left[1+\morder{\frac{1}{\log r\LUV}}\right]
\end{split}
\ee
where $\sigma$ can be identified as the chiral condensate\footnote{When $m_q$ is finite, the UV expansions in practice only define $\sigma$ up to a linear term in $m_q$. This issue will be discussed below and in Appendix~\ref{app:qbarqnorm}.} (the exact identification is studied in Appendix~\ref{app:qbarqnorm}), and $\rho$ can be expressed in terms of the leading coefficients of the beta and gamma functions as $\rho=\gamma_0/b_0 = 9/(22-4 x)$.

Notice that the analysis of previous sections was restricted to the standard, dominant vacuum. It is known~\cite{jk,Arean:2013tja}, however, that there are subdominant ``Efimov'' vacua in the QCD and walking regimes ($x < x_c$) which quite in general appear in connection to the BKT transition (see, e.g., ~\cite{Iqbal:2011in,kutasov,kutasovdbi}). These vacua are mapped to different values of the chiral condensate on the $(m_q,\sigma)$-plane: all possible regular vacua form a spiral structure, which will be called the ``Efimov spiral'' below. The results for the spiral structure will be used to analyze four-fermion deformations of QCD in Sec.~\ref{sec:4f}.

\subsection{Efimov spirals}

Let us first review the structure of the subdominant vacua. Including the solutions with finite quark mass~\cite{jk}:
\begin{itemize}
 \item When $x_c\le x<\xBZ$, only one vacuum exists, even at finite quark mass.
 \item When $0<x<x_c$ and the quark mass is zero, there is an infinite tower of (unstable) Efimov vacua in addition to the standard, dominant solution\footnote{It is assumed here for simplicity that there is an IRFP for all positive values of $x$, and the BF bound is violated at the fixed point all the way down to $x=0$.}.
 \item When $0<x<x_c$ and the quark mass is nonzero, there is an even number (possibly zero) of Efimov vacua. The number of vacua increases with decreasing quark mass for fixed $x$.
\end{itemize}

The infinite tower of Efimov vacua, which appears at zero quark mass, admits a natural enumeration $n=1,2,3,\ldots$ where $n$ is the number of tachyon nodes of the background solution as we shall demonstrate below (see also Sec.~10 and Appendix~H in~\cite{jk}). A generic feature of these backgrounds is, that they ``walk'' more than the dominant, standard vacuum, so that the scales $\Lambda_\mathrm{UV}$ and $\Lambda_\mathrm{IR}$ become well separated for all $0<x<x_c$ when $n$ is large enough. It is possible to show that
\be \label{scalescalEf1}
  \frac{\Lambda_\mathrm{UV}}{\Lambda_\mathrm{IR}} \sim  \exp\left(\frac{\pi n}{\n}\right)\ ,\qquad (n\to\infty)\ ,
\ee
for any $0<x<x_c$. The coefficient $\n$ will be given below in Eq.~\eqref{nudef} (see also Appendix~F in \cite{jk}). In the walking regime, one finds that
\be \label{scalescalEf2}
 \frac{\Lambda_\mathrm{UV}}{\Lambda_\mathrm{IR}} \sim  \exp\left(\frac{K(n+1)}{\sqrt{x_c-x}}\right) \ , \qquad (x \to x_c{}^-)
\ee
for any value of $n$. In particular, $n=0$ corresponds to the standard solution discussed in the previous sections, and the relation~\eqref{scalescalEf2} gives the standard Miransky scaling, whereas for $n>0$ the scaling is even faster. We also found a similar scaling result for the free energies of the solutions as $x \to x_c$ in \cite{jk}, therefore proving that the Efimov vacua are indeed subdominant in this limit, and verified this numerically for all $0<x<x_c$. In~\cite{Arean:2013tja} it was shown that the Efimov vacua are perturbatively unstable (again analytically as $x \to x_c{}^-$, and numerically for all $0<x<x_c$).

When the BF bound is violated at the IRFP, the quark mass and the condensate are known to show an oscillating behavior for the (chirally broken) backgrounds where the coupling flows very close to the fixed point~\cite{jk}. Let us first discuss how these oscillations arise from the tachyon EoM. 

First, take a background at zero tachyon which reaches the fixed point as $r \to \infty$. Then consider turning on an ``infinitesimal'' tachyon. It satisfies the linearized tachyon EoM 
\be
 \tau'' - \frac{3}{r} \tau' + \frac{2  \ell_*^2 a(\l_*)}{r^2\kappa(\l_*)} \tau = 0 \ , \qquad (r \to \infty) \ ,
\ee
where $\l_*$ is the value of the coupling at the fixed point and $\ell_*$ is the IR AdS radius.
Inserting the Ansatz $\tau\sim r^{\Delta_*}$ yields
\be
 \Delta_*(4-\Delta_*) = \frac{2  \ell_*^2 a(\l_*)}{\kappa(\l_*)} = \frac{24 a(\l_*)}{V_\mathrm{eff}(\l_*) \kappa(\l_*)}
\ee
where
\be 
 V_\mathrm{eff}(\l) = V_g(\l)- x V_{f0}(\l)
\ee
in terms of the dilaton and tachyon potentials. The BF bound is thus given by
\be
 \frac{24 a(\l_*)}{V_\mathrm{eff}(\l_*) \kappa(\l_*)} \le 4 \ .
\ee
When the BF bound is violated we denote
\be \label{nudef}
 \n = \mathrm{Im} \Delta_* = \sqrt{\frac{24 a(\l_*)}{V_\mathrm{eff}(\l_*) \kappa(\l_*)}-4} \ .
\ee
In this case the asymptotic infinitesimal tachyon solution is oscillatory,
\be \label{inftauIR}
 \tau \sim  r^2 \sin(\n\log r + \phi)\ , \qquad (r \to \infty) \ .
\ee

Let us consider next specific tachyon solutions $\tau_m$ and $\tau_\sigma$, where either an infinitesimal quark mass $m_q$ or a condensate $\sigma$ is turned on in the UV, respectively. These solutions are conveniently expressed in units of $\LUV$, and will have the asymptotics of~\eqref{inftauIR}. We denote
\begin{align}
\label{taumdef}
 \frac{\tau_m}{\ell} &\simeq \frac{m_q}{\LUV} K_m \left(r \LUV\right)^2\ \sin\left[\n\log\left(r\LUV\right) + \phi_m \right]\ ,& \quad &\left(r \gg \frac{1}{\LUV}\right) & \\
\label{tausdef}
 \frac{\tau_\sigma}{\ell} &\simeq \frac{\sigma}{\LUV^3} K_\sigma \left(r \LUV\right)^2\ \sin\left[\n\log\left(r\LUV\right) + \phi_\sigma \right]\ ,& \quad &\left(r \gg \frac{1}{\LUV}\right) &
\end{align}
where the coefficients $K_i$ and $\phi_i$ cannot be computed analytically but it is easy to extract them from numerical solutions. One can require that $K_i>0$ and $-\pi/2 \le \phi_i<3\pi/2$.

We are, however, interested in the solutions where the tachyon is small and finite. In this case the tachyon will eventually grow large when $r \sim 1/\L_\t \sim 1/\LIR$, and drive the system away from the fixed point. The above formulas~\eqref{taumdef} and~\eqref{tausdef} then hold as approximations for $1/\LUV \ll r \ll 1/\LIR$. 
The solution for $r \gtrsim 1/\LIR$ depend on the details of the model 
in the IR. However, in order to satisfy the boundary conditions imposed by the good IR singularity, the tachyon must have certain fixed normalization and phase when expressed in IR units:
\be \label{tauIR}
 \frac{\tau}{\ell} \simeq K_\mathrm{IR} \left(r \LIR\right)^2\ \sin\left[\n\log\left(r\LIR\right) + \phi_\mathrm{IR} \right] \ , \qquad \left(\frac{1}{\LUV} \ll r \ll \frac{1}{\LIR}\right) \ .
\ee
In general the IR asymptotics of the background depends on one parameter (e.g., $T_0$ for potentials~I~\cite{jk}) but as the fixed point is approached, the dependence on this parameter appears only through the ratio $\LUV/\LIR$ whereas $K_\mathrm{IR}$ and $\phi_\mathrm{IR}$ take fixed values (see Appendix~I in~\cite{Arean:2013tja}).

As the fixed point is approached, the result~\eqref{tauIR} must match with the sum of~\eqref{taumdef} and~\eqref{tausdef}. Therefore one finds
\bea
 &&\frac{m_q}{\LUV}  K_m \ \sin\left[\n\log\left(r\LUV\right) + \phi_m \right] + \frac{\sigma}{\LUV^3}  K_\sigma\ \sin\left[\n\log\left(r\LUV\right) + \phi_\sigma \right] \\\nn
&=&   K_\mathrm{IR}  \left(\frac{\LIR}{\LUV} \right)^2\ \sin\left[\n\log\left(r\LUV\right) + \n\log\frac{\LIR}{\LUV}+ \phi_\mathrm{IR} \right]\ , \qquad  \left(\frac{1}{\LUV} \ll r \ll \frac{1}{\LIR}\right) \ .
\eea
From here one can solve
\be \label{spiraleqs}
\begin{split}
 \frac{m_q}{\LUV} &= \frac{K_\mathrm{IR} }{K_m}\,\frac{\sin\left(\phi_\mathrm{IR}-\phi_\sigma-\n \vs  \right)}{ \sin\left(\phi_m-\phi_\sigma\right)} \, e^{-2 \vs} \\
 \frac{\sigma}{\LUV^3} &= \frac{K_\mathrm{IR} }{K_\sigma }\,\frac{\sin\left(\phi_\mathrm{IR}-\phi_m-\n \vs  \right)}{\sin\left(\phi_\sigma-\phi_m\right)}\, e^{-2 \vs} 
\end{split}
\ee
where
\be
 \vs = \log\frac{\LUV}{\LIR}\ .
\ee
As $\vs$ varies the equations~\eqref{spiraleqs} define a spiral on the $(m_q,\sigma)$-plane, which has been studied recently at finite chemical potential in a different context (see~\cite{Iqbal:2011in}). Notice that $\n$ does not need to be small.  It can be verified numerically that the handedness of the spiral is such that its phase increases with increasing $\vs$ (counter clockwise direction) on the $(m_q,\sigma)$-plane. This means that
\be \label{handedness}
 \sin(\phi_m-\phi_\sigma) > 0 \ .
\ee
As we shall show in Appendix~\ref{app:qbarqnorm}, this is also required in order for the standard solution (with nonzero and nodeless tachyon) to be dominant (for $x<x_c$ so that the BF bound is violated and the spiral exists).

Finally, let us point out some properties of the spiral as $x \to x_c$ from below. Notice from~\eqref{nudef} that $\nu = \morder{\sqrt{x_c-x}}$. The approximations~\eqref{taumdef} and~\eqref{tausdef} are valid for $1/\LUV \ll r \ll 1/\LIR$, but they should join smoothly with the UV asymptotics of the tachyon in~\eqref{mqsigmadef}. At $r \sim 1/\LUV$ we find that
\be
 \frac{\t_m}{\ell} (r \LUV)^{-2} \sim \frac{m_q}{\LUV}\ K_m \sin\phi_m \ , \qquad 
 r \frac{d}{dr}\left[\frac{\t_m}{\ell} (r \LUV)^{-2}\right] \sim \frac{m_q}{\LUV}\ K_m \n \cos\phi_m \ .
\ee
These estimates, and similar estimates for the solution~\eqref{tausdef}, can be matched with the UV asymptotic formulas if
\be
 \frac{1}{K_m} \sim\frac{1}{K_\s} \sim \sin\phi_m \sim \sin\phi_\s \sim \sqrt{x_c-x} \ , \qquad (x \to x_c{}^-) \ .
\ee
An analogous argument shows that for~\eqref{tauIR} to satisfy generic IR boundary conditions,
\be
 \frac{1}{K_\mathrm{IR}} \sim \sin\phi_\mathrm{IR} \sim \sqrt{x_c-x} \ , \qquad (x \to x_c{}^-) \ .
\ee
We have found numerically that (with the convention $K_i>0$ and $-\pi/2 \le \phi_i<3\pi/2$)
\be
 \phi_m = \morder{\sqrt{x_c-x}} = \phi_\s \ , \qquad \phi_\mathrm{IR}-\pi = \morder{\sqrt{x_c-x}} \ ,
\ee
(and $\phi_m-\phi_\s>0$ such that~\eqref{handedness} holds) for all potentials which we have studied. Then the first node of the mass in Eqs.~\eqref{spiraleqs} in their regime of validity $u \ll 1$ occurs at
\be \label{firstnode}
 \n u = \n \log \frac{\LUV}{\LIR} = \phi_\mathrm{IR}-\phi_\s = \pi + \morder{\sqrt{x_c-x}} \ .
\ee
This node is identified as the ``standard'' solution, where the tachyon has no nodes (in particular no nodes appear in the regime of validity of~\eqref{taumdef}, \eqref{tausdef}, and~\eqref{tauIR}). Notice that solving $\LUV/\LIR$ from~\eqref{firstnode} results in Miransky scaling\footnote{In principle it could be possible to satisfy the boundary conditions with $\phi_\mathrm{IR} = \morder{\sqrt{x_c-x}}$. We speculate that this happens in the model of~\cite{Evans:2013vca} where no Miransky scaling was found.} of~\eqref{Mscal}. Nodes at larger values of $u$, i.e., $\n u \simeq (n+1) \pi$, with $n=1,2,\ldots$ are identified as the Efimov solutions, where the tachyon has $n$ nodes.

\begin{figure}[!tb]
\begin{center}
\includegraphics[width=0.49\textwidth]{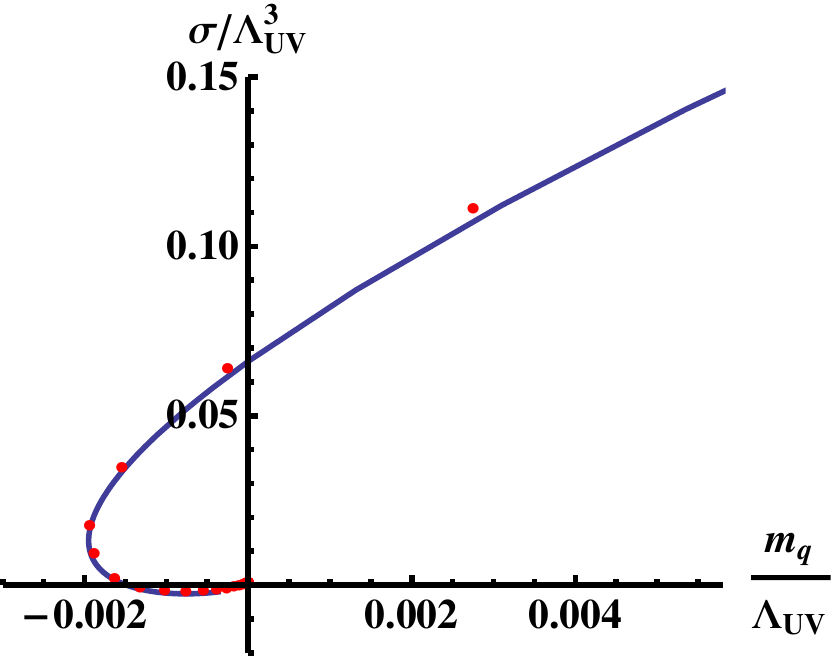}
\includegraphics[width=0.49\textwidth]{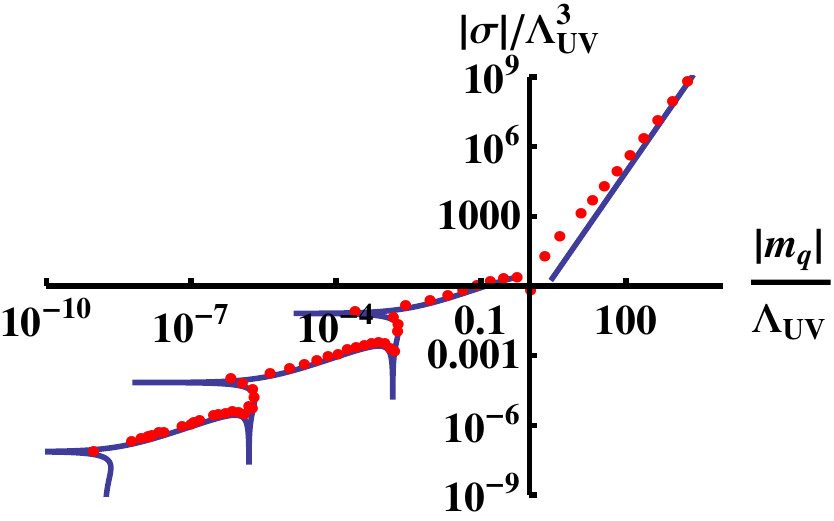}
\end{center}
\caption{The dependence of $\sigma$ on the quarks mass for potentials~II with SB normalized $W_0$ at $x=2$.  Left: the spiral of physical solutions on the $(m_q,\sigma)$-plane. Right: the same data after taking logarithms of both $\sigma$ and the quark mass. The red dots are numerical data while the blue curves are analytic fits. See text for details.}
\label{fig:spiral}\end{figure}

\subsection{Numerical results}

As an example, we have computed $m_q$ and $\sigma$ numerically for potentials~II with SB normalization\footnote{It is much easier to extract the Efimov spiral numerically for potentials~II than for potentials~I.} for $W_0$ at $x=2$. The results are shown in Fig.~\ref{fig:spiral}. The blue dots are our data and the red curves are given by Eqs.~\eqref{spiraleqs}. The blue line on the right hand side is a power-law fit $\sigma \simeq - C m_q^3$. The spiral structure is not well visible on the left hand plot because the distance of the curve from the origin decreases exponentially with increasing $u$ in~\eqref{spiraleqs}. In order to make the spiral structure visible, we have plotted the logarithms of (the absolute values of) $\sigma$ and $m_q$ on the right hand side of Fig.~\ref{fig:spiral}. Since the action is symmetric under $\tau \to -\t$ there is actually also another spiral which not shown in the left hand plot but can be obtained simply by a rotation of 180 degrees around the origin.

Notice that when $m_q \ne 0$ it is difficult to define $\sigma$ unambiguously in practical calculations, because the vacuum expectation value (vev) solution of the tachyon cannot be separated from the subleading terms of the source solution. 
This means that we also have to specify more carefully how  $K_m$ and $\phi_m$ in~\eqref{taumdef} are defined in our numerical analysis:
we pick a reference solution which has $\sigma=0$ by definition, and consequently defines $K_m$ and $\phi_m$ (see also Appendix~\ref{app:qbarqnorm}). It is natural to expect that the $\sigma=0$ solution has a quark mass of \order{\LUV} in the QCD regime. Therefore the reference solution was fixed to be the standard, dominant vacuum solution with $m_q/\LUV =1$. This kind choice is important in order to avoid unnatural fine tuning effects.

The various coefficients in the solutions in~\eqref{spiraleqs} were determined as follows. Three solutions were chosen for the background for which $\sigma$ and $m_q$ are very small such that~\eqref{inftauIR} holds as a good approximation for at least two periods of oscillation. The first (second) solution was tuned to have approximately zero $\sigma$ ($m_q$) and was used to fit the constants $K_i$, $\phi_i$ in~\eqref{taumdef} [in~\eqref{tausdef}]. The third solution was used to fit the constants $K_\mathrm{IR}$ and  $\phi_\mathrm{IR}$ in~\eqref{tauIR}.

\begin{figure}[!tb]
\begin{center}
\includegraphics[width=0.49\textwidth]{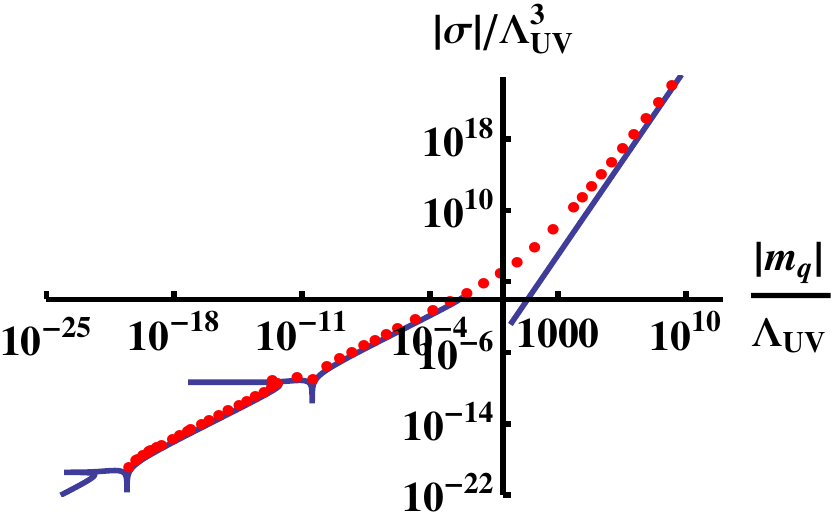}%
\includegraphics[width=0.49\textwidth]{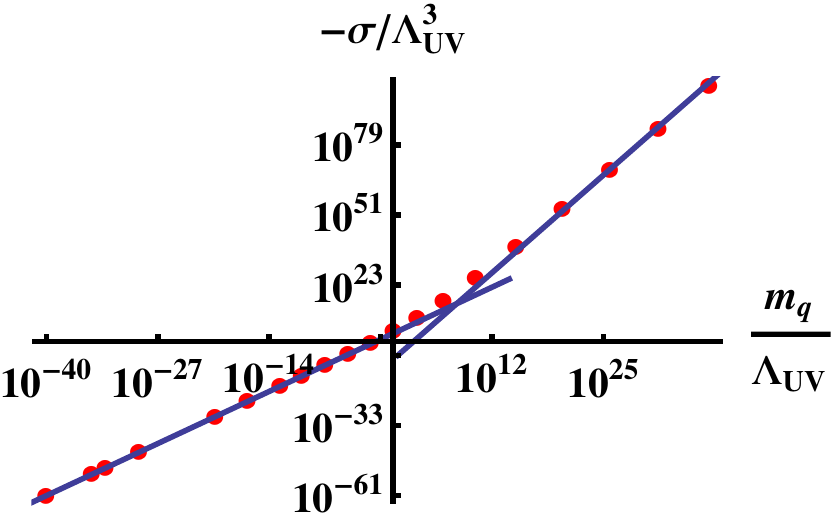}
\end{center}
\caption{The dependence of $\sigma$ on the quarks mass for potentials~II with SB normalized $W_0$. Left: results in the walking regime, $x=3.6$. Right: results in the conformal window, $x=4$. The red dots are numerical data while the blue curves are fits based on Eqs.~\protect\eqref{spiraleqs}. See text for details.}
\label{fig:sigmacw}\end{figure}

In the walking regime, extra care is needed in the choice of the reference solution having $\sigma=0$. The chiral condensate is expected to have a node around such values of the quark mass where the normalizable and nonnormalizable terms in the tachyon solutions are nontrivially coupled. This happens for the standard vacuum at the crossover point between the regimes~A and~B, i.e., for
for 
\be
 \frac{m_q}{\LUV} \sim \exp\left[-\frac{2 K}{\sqrt{x_c-x}}\right] \ .
\ee

The spiral in the walking regime ($x=3.6$, close to $x_c \simeq 3.7001$) is show on the log-log scale in Fig.~\ref{fig:sigmacw} (left). By studying numerically various quantities (for example the S-parameter, plotted below in Fig.~\ref{fig:massdepwalk}) it is found that the choice $m_q/\LUV = 3 \times 10^{-11}$ lies at the crossover and is therefore chosen as the reference point with $\sigma=0$. 

As the critical value $x_c$ is approached, the Efimov spiral becomes ``squeezed'', as seen by comparing plots in Fig.~\ref{fig:spiral} (right) and~\ref{fig:sigmacw} (left). In the walking regime, the consecutive solutions with vanishing $m_q$ and $\sigma$ are close so that the spiral approaches a straight line in the log-log scale. This reflects our analytic results for the spiral~\eqref{spiraleqs} derived above in the limit $x \to x_c{}^-$: in particular $\phi_m-\phi_\sigma \sim \sqrt{x_c-x} \to 0$.

We have also repeated the analysis in the conformal window. In this case the scaling of~\eqref{regimeBhighx} is expected to hold for small quark mass and that $\sigma \sim m_q^3$ for large quark mass. The results for potentials~II with SB normalization for $W_0$ at $x=4$ are shown in Fig.~\ref{fig:sigmacw} (right). The red dots are the data while the blue curves are given by the power laws mentioned above.

Again we need to specify how $\sigma$ is extracted in the numerical analysis because the tachyon vev solution cannot be separated from the subleading terms of the source solution. In the conformal window, however, the solution is simple because the values of $\sigma$ grow fast with increasing quark mass. Therefore one can choose any reference solution ``with $\sigma=0$'' at very small $m_q$, and the results are essentially independent of the choice.

Notice that the plots in the walking regime (left) and conformal window (right) do not seem too different. Indeed the curves undergo a smooth transition at $x=x_c$. The coefficients of the power laws are continuous at the transition, and the nodes where $\sigma=0$ or $m_q=0$ approach the origin of the spiral very fast as $x \to x_c$ from below. Actually the rate of the approach is given by the Miransky scaling law.

\subsection{Gell-Mann-Oakes-Renner relation} \label{sec:gmor}

The GOR relation can be obtained as usual by combining the results from two computations. The first result is the expression for the pion mass at small $m_q$, which is obtained by analyzing the fluctuation equations at small $m_q$ as done in Appendices~\ref{app:gor} and~\ref{app:fpi}. One finds that\footnote{The pion decay constant $f_\pi$ will be discussed in detail below in Sec.~\ref{sec:S}.}
\be \label{intIres}
 m_\pi^2 f_\pi^2 = 2 M^3 N_f N_c \ell^5 W_0 \kappa_0 m_q \s \left[1+\morder{\frac{m_q\LUV^2}{\s}}\right]
\ee
where $\kappa_0=\kappa(\l=0)$ and $W_0 = V_{f0}(\l = 0)$.  The second result is the relation between $\sigma$ and the chiral condensate as $m_q \to 0$, which is obtained by deriving the renormalized on-shell action (i.e., the vacuum energy $\mE$) with respect to $m_q$ (see Appendix~\ref{app:qbarqnorm}):
\be \label{qbarqnorm}
 \langle \bar q q \rangle = \frac{\pa \mE}{\pa m_q} =  -2 M^3 N_f N_c \ell^5 W_0 \kappa_0 \s\left[1+\morder{\frac{m_q\LUV^2}{\s}}\right] \ .
\ee
The combination is the GOR relation:
\be \label{GORtext}
 m_\pi^2 f_\pi^2 =  -m_q \langle \bar q q \rangle \left[1+\morder{\frac{m_q\LUV^2}{\s}}\right] \ .
\ee
It can be checked that the proportionality coefficient (here minus one) is correct for our normalization of $f_\pi^2$, which differs by the factor $N_f/2$ from the standard normalization in chiral perturbation theory (see, e.g.,~\cite{Bijnens:2009qm}).

\begin{figure}[!tb]
\begin{center}
\includegraphics[width=0.49\textwidth]{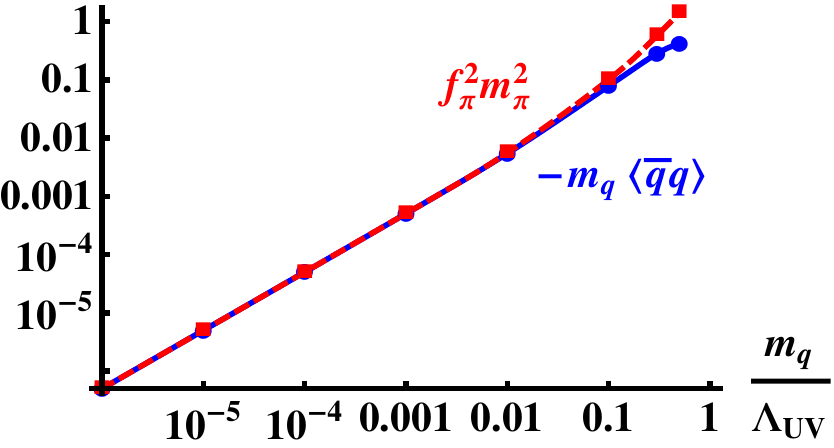}
\includegraphics[width=0.49\textwidth]{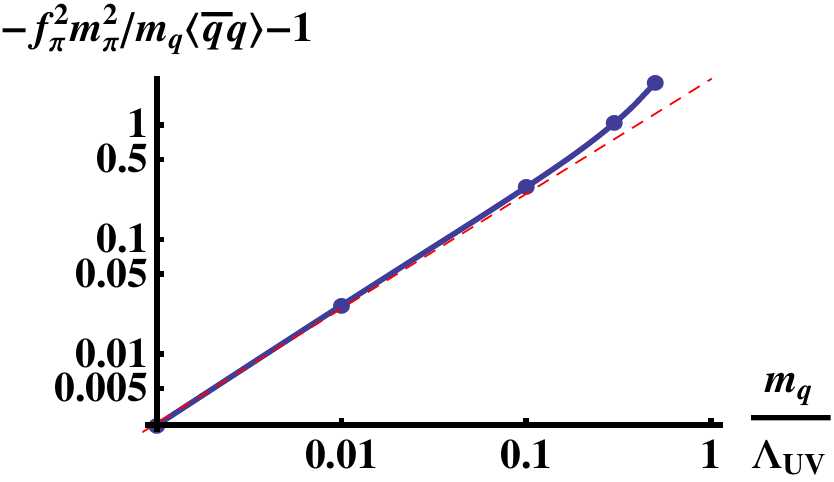}
\end{center}
\caption{Numerical test of the GOR relation. Left: Both sides of the GOR relation plotted against the quark mass. The blue curve is $m_q\langle \bar q q\rangle$ (in units of $\LUV^4$), while the red dashed curve is $f_\pi^2 m_\pi^2$. Right: The leading correction to the relation. The blue curve is the numerical data, and the thin dashed red line is a linear fit ($2.5\ m_q/\LUV$). Potentials I with $x=1$ and $W_0=3/11$ were used.}
\label{fig:GOR}\end{figure}

Notice that
\begin{enumerate}
 \item One might expect that the logarithmic terms which appear in the UV expansion of the fields would result in correction that are only suppressed by $1/\log m_q$ in~\eqref{intIres}. Remarkably, the logarithmic terms completely cancel, as they should, since the  QCD the relation holds up to linear corrections in the quark mass. Why this happens is shown in Appendix~\ref{app:gor}. The cancellation of these corrections is also consistent with the combination $m_q \langle\bar qq\rangle$ being RG invariant~\cite{Kiritsis:2014kua}. 
 \item The derivative in~\eqref{qbarqnorm} is nontrivial, since changing the quark mass also affects the geometry even at $m_q=0$ due to the full backreaction between the flavor and glue sectors, which may add contributions to the derivative (see Appendix~\ref{app:qbarqnorm}).
 \item The relation is valid only in regime~A: the correction terms in~\eqref{GORtext} become large at the crossover between regimes~A and~B.
\end{enumerate}

We have tested the GOR relation numerically for potentials~I at $x=1$. Both sides\footnote{To be precise, we are checking equation~\eqref{intIres} -- the normalization factor of~\eqref{qbarqnorm} was assumed in the computation and subleading corrections were dropped.} of the relation are plotted as functions of the quark mass in Fig.~\ref{fig:GOR} (left) and good agreement is found. The subleading correction in Fig.~\ref{fig:GOR} (right) is clearly linear in the quark mass as it should.

\section{Four-fermion operators in V-QCD} \label{sec:4f}

After we have constructed the solutions on the $(m_q,\sigma)$-plane it is straightforward to analyze the effect of multitrace deformations following the recipe of~\cite{Witten:2001ua}. In the presence of such deformations, the coefficient of the source term, which was denoted by $m_q$ above, is no longer trivially related to the quark mass. Let us therefore denote this coefficient by $\alpha_m$ in this section. Similarly, the coefficient of the normalizable term is denoted by $\beta_m$.

One can now study the following extra terms on the field theory Lagrangian:
\be \label{Wdef}
 \frac{1}{N_fN_c } \widetilde W = -m_q \int d^4 x \mO + \sum_{n=2}^{n_\mathrm{max}} \frac{(-1)^ng_n}{n} \int d^4 x  \mO^n \ ,
\ee
where $\mO = \bar q q/N_fN_c$.
We first replace $\mO$ by $-c_\sigma \beta_m$ in $\widetilde W$, where $c_\s = 2 M^3 W_0 \kappa_0 \ell^5$ is the proportionality coefficient between $\s$ and $\langle \mO \rangle$ in the absence of the multi-trace deformations\footnote{For simplicity we will omit here the difficulty of defining $\beta_m$ in practice: we take it to be directly proportional to $\langle \bar q q\rangle$ (so that $k_\t=0$ in~\eqref{qqbarmqfin} of Appendix~\ref{app:qbarqnorm}).}. 
The result is the functional  
\be \label{Wbulk}
 \frac{1}{N_fN_c } W[\b_m] = m_q c_\s \int d^4 x \b_m + \sum_{n=2}^{n_\mathrm{max}} \frac{g_n c_\s^n}{n} \int d^4 x  \b_m^n \ .
\ee
Then the boundary conditions are obtained 
by setting the source term $\alpha_m$ to the value~\cite{Witten:2001ua} (see also~\cite{Mueck:2002gm,Minces:2002wp})
\be \label{alphacond}
 \alpha_m = \frac{1}{N_f N_c c_\sigma} \frac{\delta W}{\delta \beta_m(x)} = m_q +\sum_{n=2}^{n_\mathrm{max}} g_n c_\sigma^{n-1}\beta_m^{n-1} \ ,
\ee
and the vev is given by
\be \label{betacond}
 \beta_m = \sigma \equiv -\frac{1}{c_\s} \langle \mO\rangle  = -\frac{1}{N_fN_cc_\s}\langle \bar q q\rangle  \ .
\ee

The possible vacua can then be identified by overlapping this condition with the curves of regular solutions on the $(\alpha_m,\beta_m)$-plane (shown above in Figs.~\ref{fig:spiral} and~\ref{fig:sigmacw}).

\begin{figure}[!tb]
\begin{center}
\includegraphics[height=0.4\textwidth]{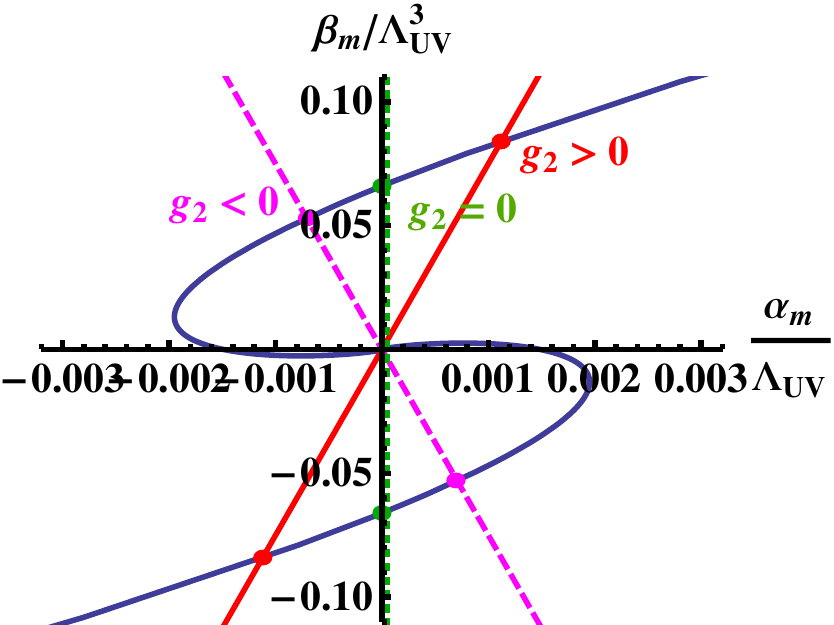}%
\hspace{5mm}\includegraphics[height=0.4\textwidth]{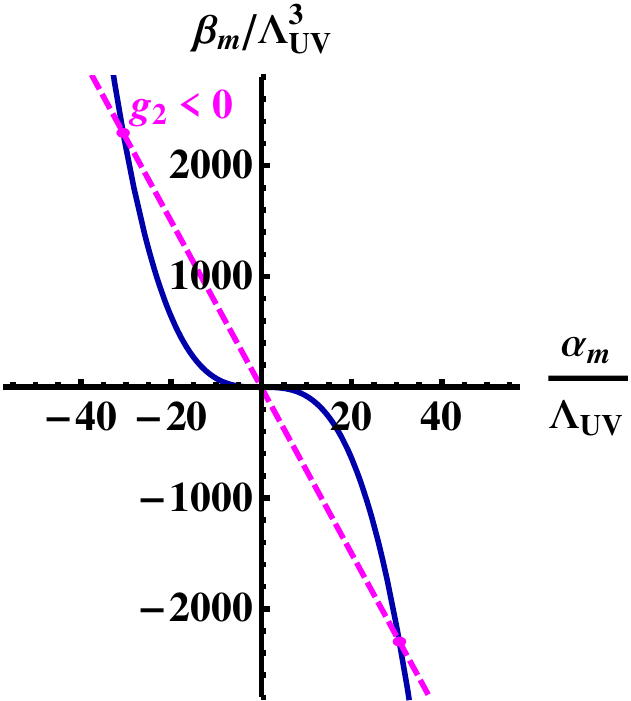}
\end{center}
\caption{Solutions in the presence of a double trace deformation in the QCD regime.}
\label{fig:doubletr}\end{figure}

We are most interested in the case of double trace deformation, $g_2\ne 0$ with other couplings equal to zero, since this operator becomes marginal at the critical point $x=x_c$. Let us also set $m_q=0$. In this case
\be \label{g2constr}
 \alpha_m = g_2 c_\s \beta_m \ .
\ee
The overlap plot is shown for the phase with the Efimov spiral ($0<x<x_c$) 
in Fig.~\ref{fig:doubletr}. Notice that the second branch of the spiral, obtained by reflection about the origin,
which was omitted in earlier plots (for example Fig.~\ref{fig:spiral}) was now also included\footnote{The solutions on the second branch are nontrivially related to those on the first branch only if operators with odd $n$ have been included.}.  
For this special case where $g_2$ is the only nonzero coupling, as the mapping $(\alpha_m,\beta_m) \mapsto (m_q,\s)$ in~\eqref{alphacond} and~\eqref{betacond} implies, the change of the UV boundary conditions with respect to the standard case $g_2=0$ corresponds to changing the vertical axis to the line defined by~\eqref{g2constr} (examples are the red and dashed magenta lines in Fig.~\ref{fig:doubletr}) while the horizontal axis is kept fixed.

Notice that the solutions with $g_2=0$ lie on the vertical axis, and are denoted by the green line in Fig.~\ref{fig:doubletr}. The green dot shows the (stable) standard solution, but there is also an infinite tower of additional intersection points near the origin, which are not visible as the spiral converges very fast towards zero. These intersection points give the Efimov solutions.

In order to draw the phase diagram at nonzero $g_2$, we need to solve the free energy of each solution and find the dominant vacuum. We can start from the identity
\be \label{Eder}
  \frac{\pa\mathcal{E}}{\pa m_q} = \langle \bar q q \rangle = - N_f N_c c_\s \s \ .
\ee
In Appendix~\ref{app:qbarqnorm} we show that the conditions~\eqref{alphacond} and~\eqref{betacond} are indeed consistent with~\eqref{Eder} and that the higher order expectation values satisfy
\be \label{Egder}
  \frac{n (-1)^{n+1}}{N_fN_c} \frac{\pa\mathcal{E}}{\pa g_n} = \langle \mO^n \rangle = \langle \mO \rangle^n
\ee
in agreement with the large $N$ factorization of expectation values.

Recall from Sec.~\ref{sec:condensate} that the Efimov spiral can be parametrized as 
\be \label{spiraleqsab}
\begin{split}
 \frac{\alpha_m}{\LUV} &= \frac{K_\mathrm{IR} }{K_m}\,\frac{\sin\left(\phi_\mathrm{IR}-\phi_\sigma-\n \vs  \right)}{ \sin\left(\phi_m-\phi_\sigma\right)} \, e^{-2 \vs} \\
 \frac{\beta_m}{\LUV^3} &= \frac{K_\mathrm{IR} }{K_\sigma }\,\frac{\sin\left(\phi_\mathrm{IR}-\phi_m-\n \vs  \right)}{\sin\left(\phi_\sigma-\phi_m\right)}\, e^{-2 \vs} 
\end{split}
\ee
asymptotically at small values of the variable
\be
 \vs = \log\frac{\LUV}{\LIR} \ .
\ee
Notice that if only $g_2$ is nonzero, $m_q$ and $\s$ still satisfy~\eqref{spiraleqsab} (with $\alpha_m$ ($\beta_m$) replaced by $m_q$ ($\s$), respectively), if the coefficients $K_i$ and $\phi_i$ are redefined. This is seen by inserting the conditions~\eqref{alphacond} and~\eqref{betacond}, and is consistent with the change of boundary conditions simply corresponding to a new choice of axes in Fig.~\ref{fig:doubletr}.

By inserting this parametrization in~\eqref{Eder} and integrating along the Efimov spiral we find that for the solutions with $m_q=0$ (see Appendix~\ref{app:qbarqnorm})
\be
 \frac{1}{N_f N_c} \left(\mathcal{E}-\mathcal{E}_0\right) = - \frac{\n c_\s K_\mathrm{IR}^2 e^{-4 u}}{8 K_m K_\s \sin(\phi_m-\phi_\s)} \ ,
\ee
where $\mathcal{E}_0$ is the free energy of the solution with $\alpha_m = 0 = \beta_m$. Since $\sin(\phi_m-\phi_\s)>0$, the dominant solution from those in the range of validity of~\eqref{spiraleqsab} is that with the largest value of $u$. In the walking regime this can be seen explicitly. Namely, we argued in Sec.~\ref{sec:condensate} that in the walking regime $\n \simeq \pi\sqrt{(x_c-x)}/K \to 0$, and that the solutions are found at $\n \vs \simeq (n+1) \pi$, with $n=0,1,2,\ldots$. Therefore
\be
 \frac{1}{N_f N_c} \left(\mathcal{E}-\mathcal{E}_0\right) \simeq - \frac{\n c_\s K_\mathrm{IR}^2}{8 K_m K_\s \sin(\phi_m-\phi_\s)} \exp\left(-\frac{4 K(n+1)}{\sqrt{x_c-x}}\right) \ , \qquad (x_c-x \ll 1) \ ,
\ee
which agrees with the scaling of the free energy found in~\cite{jk} (see Sec.~10 and Appendix~H there).

Also quite in general the solution with largest $|\sigma|$ is the dominant vacuum. From~\eqref{Eder} we see that the energy density is given in terms of the (oriented) area between the spiral and the horizontal axis. For clockwise oriented spirals the minimum energy is reached at the solutions furthest away from the origin.

\begin{figure}[!tb]
\begin{center}
\includegraphics[width=0.6\textwidth]{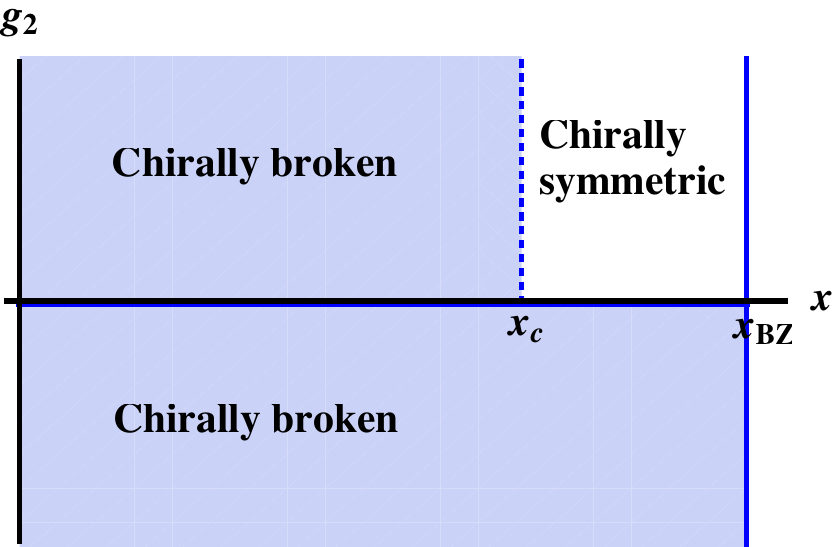}
\end{center}
\caption{The phase diagram of V-QCD in the presence of a four-fermion deformation (at zero quark mass and temperature). The blue horizontal line is a discontinuity at $g_2 = 0^-$. The dotted vertical line presents a BKT transition.}
\label{fig:4fphases}\end{figure}

It is then straightforward to construct the phase diagram, which is shown in Fig.~\ref{fig:4fphases} for the case of $m_q=0$ and nontrivial $g_2$. 
For $x<x_c$ the diagram can  be found simply by analyzing the solutions in Fig.~\ref{fig:doubletr}. When $g_2>0$ the configuration is qualitatively similar to that at $g_2=0$: the dominant vacuum for $x_c<x$ is the vacuum with the standard tachyon solution having no nodes. 
When $g_2<0$, 
the dominant solution is that of Fig.~\ref{fig:doubletr} (right) having sizeable $|\s|$. As this solution is absent for $g_2 \ge 0$, there is a discontinuity (a ``zeroth order'' transition) at $g_2 = 0 -$. It is smoothly connected, though, to the dominant solution of $g_2 > 0$ through the limit $g_2 \to \pm \infty$ (where the constraint~\eqref{g2constr} gives a horizontal line in the plots of Fig.~\ref{fig:doubletr}). There is also a subdominant solution shown in Fig.~\ref{fig:doubletr} (left), namely the continuation of the standard, dominant solution at $g_2=0$ to negative $g_2$. We will argue below that this solution is metastable for small $|g_2|$ and unstable for large $|g_2|$.

Notice that the chiral condensate is a scheme dependent quantity, and the scheme dependence is important, in particular, at large $m_q$ (see~\cite{ikp} for a discussion of the scheme dependence in the context of holography). Therefore the dominant solution $g_2<0$ appears scheme dependent at least for small $|g_2|$. We have, however, found the behavior of generic schemes $\s \propto m_q^3$ with a negative proportionality constant for all choices of potentials we have tried, and independently of the precise definition of $\s$. This is enough to guarantee that the phase diagram is that of Fig.~\ref{fig:4fphases}. Actually, as stressed in Appendix~\ref{app:masses}, the results at very large $m_q$ are essentially independent of the choices of the various functions in the V-QCD action.

In the conformal window, the situation is even simpler. For $g_2 \ge 0$ there is only the solution with zero quark mass and the condensate. When $g_2<0$, there is again a solution with large $|\s|$, analogous to that shown in Fig.~\ref{fig:doubletr} (right). When $g_2>0$, the transition at $x=x_c$ is similar as at $g_2=0$, i.e., a BKT transition, since changing the value of $g_2$ only amounts to changing the axes in Fig.~\ref{fig:doubletr} without affecting the structure of the spiral.

One can also show that when a finite $m_q$ is turned on, the BKT transition and the chirally symmetric phase disappear, but the discontinuity at $g_2= 0^-$ remains.

\subsection{Perturbative stability}

Finally perturbative stability of the solutions with modified boundary conditions could be analyzed following~\cite{Arean:2013tja}. Here we will only discuss which of the solutions are expected to be unstable, and will not prove the stability of any solution. Naturally, the modification of the boundary conditions for the background also implies that the boundary conditions of the fluctuations are similarly changed as the fluctuations must preserve the physical value of $m_q$.

Let us first analyze any solutions with walking, i.e., $\vs \gg 1$. Recall that the standard solution has been shown to be stable, while the Efimov solutions are unstable when $g_2=0$~\cite{Arean:2013tja}. The instability appeared in the scalar flavor singlet and nonsinglet sectors. 
In order to study stability of the other solutions, one should look at the scalar fluctuation equations. In the walking regime, they admit simple solutions in the UV and in the vicinity of the (approximate IR) fixed point. In fact, as argued in Appendix~I of~\cite{Arean:2013tja}, the fluctuation equations (for sufficiently small mass of the fluctuation) take the same form as the EoMs for the background. This is true both in the flavor singlet and in the nonsinglet sectors. Therefore, the fluctuations in the vicinity of the (approximate) IRFP are analogous to~\eqref{tauwalks}:
\be \label{phiwalks}
 \psi_S(r) \simeq C_s r^2 \sin\left[\n\log(r\LUV) + \phi_s \right] \ , \qquad \left(\frac{1}{\LUV} \ll r \ll \frac{1}{\L_\t}\right) \ ,
\ee 
where $\psi_S$ is the radial wave function of any scalar fluctuation mode.
The solution necessarily has a node if this approximation is valid for more than half a period of the sine function. In terms of variable $\vs = \log(\LUV/\LIR)$ the node therefore appears for $\n \vs \gtrsim \pi$ (where we used the fact that $\LIR \sim \L_\tau$ whenever walking is present). 

Such a node of the wave function (say at zero momentum) implies an instability, for generic 
UV boundary conditions for the fluctuations. It is straightforward to prove this in the case of nonsinglet scalars for which the fluctuation equations can be cast into the Schr\"odinger form, and with the standard UV boundary conditions such that the fluctuation wave function is normalizable in the UV (see also the analysis in~\cite{Anguelova:2013tha}). We shall sketch the proof here. 

Denote the UV normalizable (but not necessarily IR normalizable) Schr\"odinger wave function by $\phi$ and the location of any of its nodes by $r_0$. By studying the variation of 
\be
 0 = \int_0^{r_0(m^2)} \phi(r) (-\phi''(r) + V_S(r)\phi(r) - m^2 \phi(r)) dr
\ee
with respect to the mass, we find that
\be \label{nodeevol}
 \frac{d r_0}{dm^2} = - \frac{1}{\phi'(r_0)^2} \int_0^{r_0} \phi^2(r) dr < 0 \ .
\ee
We see that all nodes move towards the IR as $m^2$ is lowered. But as $m^2 \to -\infty$ the solution for generic UV boundary conditions is $\phi \propto \exp(|m|r)$ and has no nodes for $r \ll 1/m$. Therefore all nodes must disappear either by moving to $r=\infty$ or by merging with other nodes. However, the Schr\"odinger equation does not admit solutions with a double node for regular potentials $V_S$, so merging of the nodes is not possible, and consequently all nodes must disappear by moving to the far IR. In particular, if there the wave function has a node when $m^2=0$, the node must move to $r = \infty$ at some negative value of $m^2$. At this value, $\phi$ is IR normalizable: otherwise it could not have a node in the far IR if $m^2$ is slightly perturbed. Tachyonic normalizable mode marks the presence of instability. Putting the above observations together, we conclude that a node of the $m^2=0$ wave function marks the presence of an instability.

The above proof does not apply directly to our case because the UV boundary conditions are modified, and consequently the integral in~\eqref{nodeevol} is divergent. The divergence can be regulated by introducing a UV cutoff at $r=\eps$, but this results in extra counterterms on the right hand side of~\eqref{nodeevol} and its negativity is no longer obvious.

The proof can be fixed, however, in our case (i.e., scalar fluctuations and node in the region of validity of~\eqref{phiwalks}) when $x \to x_c$ from below. This is because the node becomes well separated from the UV region: the location of the node satisfies $\log (r_0/\LUV)  \sim 1/\sqrt{x_c-x}$ as seen from~\eqref{phiwalks}, where $\n \sim \sqrt{x_c-x}$. For generic boundary conditions, the right hand side of~\eqref{nodeevol} is then dominated by the contributions to the integral near the node. The counterterms which cancel the divergence of the integral are essentially independent of $x$ for any reasonable boundary conditions, and therefore negligible. The rest of the proof remains unchanged.

Let us then study the solutions with $m_q=0$ as $x \to x_c$. Recall that the phase differences $\phi_\mathrm{IR}-\phi_\s-\pi$ and $\phi_\s-\phi_m$ will be $\propto \sqrt{x_c-x}$ in this limit as we argued in Sec.~\ref{sec:condensate}. The standard solution (the green dot on the vertical axis in Fig.~\ref{fig:doubletr}) is in the regime of validity of the approximations leading to~\eqref{spiraleqsab}. It is found at 
\be \label{standardvac}
 \n \vs = \phi_\mathrm{IR}-\phi_\sigma = \pi + \morder{\sqrt{x_c-x}} \ .
\ee
Based on the above analysis, the onset of the perturbative instability is also expected at $\n \vs - \pi = \morder{\sqrt{x_c-x}}$. As the standard solution is stable, the critical value of $\vs$ must be larger than that given in~\eqref{standardvac}. Inserting this in~\eqref{spiraleqsab}, the critical value of $g_2$ is given by the ratio $\a_m/(\b_m c_\s)$. The factors of $\sqrt{x_c-x}$ cancel leaving a \order{1/\LUV^2} number, which must be negative given the handedness of the spiral. The critical value of $g_2$ is therefore of the same order as the value corresponding to the dashed magenta line in Fig.~\ref{fig:doubletr}. The parts of the spiral which are closer to the origin from the critical points (near the intersection points with magenta dots) are unstable. 

The above analysis cannot be applied in the QCD regime where $x_c-x$ is not small. This regime could be analyzed numerically. The natural expectation is that the critical value is still negative and \order{1/\LUV^2}. This is supported by the fact that the Efimov vacua are unstable, which has been shown by a numerical computation~\cite{Arean:2013tja}.

To conclude this section, the phase diagram is that shown in Fig.~\ref{fig:4fphases}: For $g_2>0$, the diagram is identical to that of $g_2=0$. The result is somewhat different from the result obtained in the gauged Nambu-Jona-Lasinio model, where the four-fermion operator is also slightly different, and the lower bound of the conformal window appears to increase with $g_2$~\cite{Yamawaki:1996vr}. As any negative $g_2$ is turned on, the ``standard'' vacuum immediately becomes unstable and the dominant vacuum has much larger $|\s|$. For $x<x_c$ and $g_2<0$, the perturbation analysis suggests the the standard vacuum is metastable for small $|g_2|$ and becomes perturbatively unstable for larger $|g_2|$, with the critical value being \order{1/\LUV^2}.

\section{S-parameter and current-current correlators} \label{sec:S}

The vector-vector and axial-axial correlators have the structure
\begin{align}
 i \int d^4 x\ e^{-i q x} \langle 0|\ T\!\left\{ J_{\mu}^{a\,(V)}(x) J_{\n}^{b\,(V)}(0)\right\} |0\rangle &=- \frac{2\delta^{ab}}{N_f}
\left( q^2 \eta_{\m\n}-{q_{\m} q_{\n}}\right)\Pi_V (q^2) &\\
 i \int d^4 x\ e^{-i q x} \langle 0|\ T\!\left\{ J_{\mu}^{a\,(A)}(x) J_{\n}^{b\,(A)}(0)\right\} |0\rangle &=- \frac{2\delta^{ab}}{N_f}\Big[
\left( q^2 \eta_{\m\n}-{q_{\m} q_{\n}}\right)\Pi_A (q^2) \\\nn
 & \qquad \qquad \ \  + q_{\m} q_{\n}\Pi_L (q^2)\Big] \ , &
\end{align}
where 
\be
 J_{\mu}^{a\,(V)} = \bar q \gamma_\mu t^a q \ , \qquad J_{\mu}^{a\,(A)} = \bar q \gamma_\mu\gamma_5 t^a q \ .
\ee
and $t^a$ are the generators of $SU(N_f)$ with the normalization ${\mathbb Tr}\, t^a t^b =\delta^{ab}/2$.
The factors $2/N_f$ on the right hand side were added to ensure that $\Pi_V$ and $\Pi_A$ are proportional to $N_f$. The numerical factor was chosen such that these factor are equal to one when $N_f=2$, which is the smallest number for which the flavor non-singlet currents are defined. This will result in the standard normalization of the S-parameter. The sign convention is $\eta_{\m\n} = \mathrm{diag}(-,+,+,+)$. Notice that $\Pi_L$ vanishes for zero quark mass because then $\partial^\mu J_{\mu}^{a\,(A)} =0$.  

When the spectrum is discrete (i.e., $0<x<x_c$ or $m_q$ is finite), we may formally write the correlators as sums over the contributions from the meson states:
\bea \label{PiAsp}
 \Pi_A &=& \frac{S_0}{q^2} + \sum_{n=1}^\infty \frac{(f_{n}^{(A)})^2}{q^2+(m_{n}^{(A)})^2} \\
\label{PiVsp}
 \Pi_V &=& \sum_{n=1}^\infty \frac{(f_{n}^{(V)})^2}{q^2+(m_{n}^{(V)})^2} \\
 \label{PiLsp}
 \Pi_L &=& \frac{S_0}{q^2} - \sum_{n=1}^\infty \frac{(f_{n}^{(P)})^2}{q^2+(m_{n}^{(P)})^2} \ .
\eea
Depending on the choice of the holographic action, these series may not converge. This issue will be discussed below in the context of the S-parameter.
The residues $S_0$ of the ``spurious'' $q^2=0$ pole in~\eqref{PiAsp} and~\eqref{PiLsp} must identical for the pole to cancel in the full correlator. At zero quark mass also $S_0$ must be related to the pion decay constant, $S_0 = f_\pi^2 \equiv (f_1^{(P)})^2$, since $\Pi_L$ vanishes.

The difference of the vector-vector and axial-axial correlators involves the quantity
\be \label{Ddef}
 D(q^2) = q^2 \Pi_A(q^2)- q^2 \Pi_V(q^2) \ ,
\ee
which is nontrivial only when chiral symmetry is broken. The expansion of $D(q^2)$ at $q^2=0$ defines the S-parameter:
\be \label{Sdef}
 D(q^2) = S_0 - \frac{S}{4\pi} q^2 + \frac{S_2}{4\pi}q^4 + \cdots \ .
\ee
Here $S_0$ is the same coefficient which appears in~\eqref{PiAsp} and~\eqref{PiLsp}, and we will also study the higher order coefficient $S_2$ below.

\subsection{Correlators and the S-parameter in V-QCD}

Let us then recall how the correlators can be computed in V-QCD. As pointed out in Sec.~\ref{sec:vqcd}, the vector currents are dual to the gauge fields in the DBI action~\eqref{generalact}. Following the standard approach (see~\cite{Arean:2013tja} for additional details), we carry out the fluctuation analysis writing down an Ansatz which separates the spatial and radial dependence of the fluctuation modes in momentum space. The spatial (radial) wave functions of the vector, transverse axial, and longitudinal axial modes are denoted by $\mV$, $\mA$ and $\mP$ ($\psi_V$, $\psi_A$, and $\psi_L$). The radial wave functions are IR normalizable and satisfy the UV boundary conditions 
\be
 1=\psi_V(\eps,p^2)=\psi_A(\eps,p^2)=\psi_L(\eps,p^2)-\psi_P(\eps,p^2) \ .
\ee 
Here $\psi_P$ is the radial pion wave function. 

The terms of the on-shell V-QCD action which are quadratic in the vector fields can then be written as
\bea
 S_V &=& \frac{1}{4} M^3 N_c \int \frac{d^4 p}{(2\pi)^4} \mV_\m^{a}(p) P^{\m\n} \mV_\n^{a}(-p) \left.V_f e^A w^2 \partial_r \psi_V(r,p^2)\right|_{r=\eps}\ , \\
 S_A &=& \frac{1}{4} M^3 N_c \int \frac{d^4 p}{(2\pi)^4} \mA_\m^{a}(p) P^{\m\n} \mA_\n^{a}(-p) \left.V_f e^A w^2 \partial_r \psi_A(r,p^2)\right|_{r=\eps}\ , \\
 S_L &=& \frac{1}{4} M^3 N_c \int \frac{d^4 p}{(2\pi)^4} \mP^{a}(p) \mP^{a}(-p) \left.V_f e^A p^2 w^2 \partial_r \psi_L(r,p^2)\right|_{r=\eps} \ ,
\eea 
where
\be
 P^{\m\n} = \eta^{\m\n} - \frac{p^\m p^\n}{p^2}
\ee
projects to the transverse parts of the wave functions.  To compute the vector-vector correlators, we need the precise dictionary given in terms of the couplings to field theory currents:
\bea
  &&\int \frac{d^4p}{(2\pi)^4} J_\mu^{a\,(V)}(-p) \mV^{\m\, a}(p)\ , \\
  &&\int \frac{d^4p}{(2\pi)^4} J_\mu^{a\,(A)}(-p)\left[P^{\m\n}\mA_\n^{a}(p) - i p^\m \mP^a(p)\right] \ .
\eea
Applying the gauge/gravity correspondence with these couplings leads to the following expressions for the form factors:
\be \label{Pidef}
 q^2 \Pi_I(q^2) = -\frac{1}{4} M^3 N_f N_c \left.V_f e^A w^2  \partial_r \psi_I(r,q^2)\right|_{r=\eps} \ ,
\ee
where $I=V,A,L$. 

Notice that for $m_q>0$ (so that $\Pi_L$ is nonzero) the wave functions $\psi_A$ and $\psi_L$ satisfy the same fluctuation equations as $q^2 \to 0$. Consequently
\be
  \lim_{q\to 0} q^2 \Pi_A(q^2) = \lim_{q\to 0} q^2 \Pi_L(q^2) 
\ee
and this number equals $S_0$ of~\eqref{PiAsp} and~\eqref{PiLsp} which is therefore well defined. This equality ensures the cancellation of the ``pole'' at $q^2 = 0$ as pointed out above.

By inserting~\eqref{Pidef} in the definitions~\eqref{Ddef} and~\eqref{Sdef} we obtain for the S-parameter
\be \label{Sdiffform}
 S = \pi M^3 N_f N_c V_f e^A w^2 \left[ \frac{\partial^2}{\partial r\partial q^2} \psi_A(r,q^2)-\frac{\partial^2}{\partial r\partial q^2} \psi_V(r,q^2) \right]_{r=\eps,\,q^2=0} \ .
\ee
This formula is, however, not convenient for high precision numerical computations, because the subleading terms at $r=\eps$ are only suppressed by logarithms of $\eps$ in V-QCD, so that extremely small values of $\eps$ would be needed to obtain accurate results. By an analysis of the fluctuation equations, it is possible to derive more convenient integral representations for $S$ (see Appendix~\ref{app:fpi}). We find that
\be \label{Dint}
 D(q^2) = q^2 \Pi_A(q^2)- q^2 \Pi_V(q^2) =   M^3 N_f N_c \int_0^\infty du\  \psi_V V_f e^{3A} \kappa \t^2 \psi_A \ ,
\ee
where $u$ is the Schr\"odinger coordinate defined in Appendix~\ref{app:fpi}, and that
\begin{align}
 S &= \pi M^3 N_f N_c \int_0^\infty du \ V_f e^A w^2 \left[\psi_A^2-\psi_V^2\right]_{q^2=0} \\
&= \pi M^3 N_f N_c \int_0^\infty du \ V_f e^A w^2 \left[\psi_A^2-1\right]_{q^2=0} \ .
\label{Sparamint}
\end{align}
The formula for the S-parameter is well-known in the context of simple bottom-up models (see, e.g.,~\cite{Barbieri:2003pr}). In Appendix~\ref{app:fpi} we write it in a form which holds for very generic holographic models. We used the fact that
\be
 \left.\psi_V\right|_{q^2=0} = 1
\ee
in order to obtain the last expression~\eqref{Sparamint}.

\begin{figure}[!tb]
\begin{center}
\includegraphics[width=0.49\textwidth]{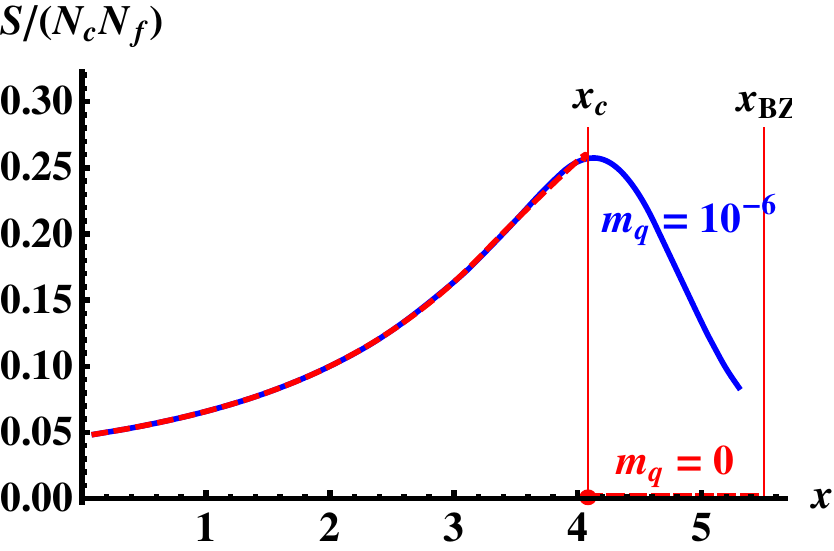}
\includegraphics[width=0.49\textwidth]{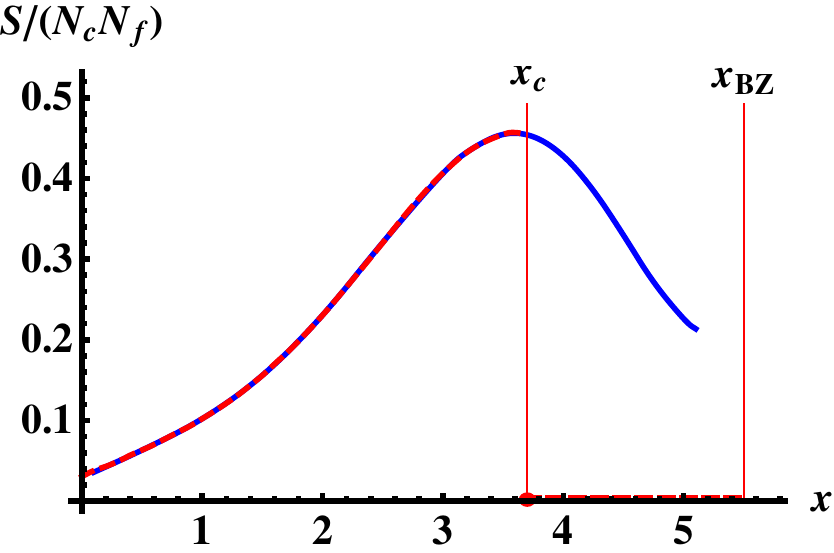}
\end{center}
\caption{The normalized S-parameter as a function of $x$ for $m_q/\LUV=10^{-6}$ (blue solid curves) and $m_q=0$ (red dashed curves). Left: potentials~I with $W_0=3/11$. Right: potentials~II with SB normalized $W_0$.}
\label{fig:Sparam}\end{figure}

\subsection{Numerical results for the S-parameter and $f_\pi$}

Using equations~\eqref{Pidef}, \eqref{Dint}, and~\eqref{Sparamint} it is straightforward to compute the S-parameter, the pion decay constant as well as the coefficients $S_0$ and $S_2$ of~\eqref{Sdef} numerically (see~\cite{Arean:2013tja} for additional details, and~\cite{Sparam,cobi,parnachev} for analysis of the S-parameter in other holographic models). Figure~\ref{fig:Sparam} shows the results for $S$ at $m_q=0$ (dashed red curves) and at $m_q/\LUV=10^{-6}$ (blue curves) for potentials~I (left) and~II (right). The numerical value of $M^3$ was fixed such that the asymptotics of the vector-vector correlator matches with perturbative QCD (see Eq. (C.10) in~\cite{Arean:2013tja}).

The most striking feature in these plots is the discontinuity of the S-parameter in the conformal window. When $x \ge x_c$, the S-parameter immediately jumps from zero to a \order{N_f N_c} number when any finite $m_q$ is turned on. The mechanism which leads to this discontinuity will be discussed in detail below (from the holographic viewpoint), but it appears rather natural: the S-parameter is \order{N_fN_c} whenever the geometry has the IR singularity, and vanishes only for zero quark mass in the conformal window where there is an IRFP instead. The result is also consistent with the analysis based on field theory at qualitative level~\cite{sannino,DelDebbio:2010ze}: the S-parameter is finite except for exactly zero mass in the conformal window.

There is, however, one striking difference~\cite{letter,Arean:2013tja} with respect to previous results: the S-parameter increases with $x$ in regime~A, whereas many earlier analyses~\cite{SinDS,walkingS,as,Sparam} suggest that the S-parameter is suppressed in the walking regime and may even vanish as $x \to x_c$. Recall, however, that the IR behavior of the potentials in the V-QCD action has not yet been fitted to QCD or lattice data, and such fits may affect the $x$-dependence of the S-parameter. By analyzing the form of the fluctuation wave functions (see Appendix~G in~\cite{Arean:2013tja}) in the integral formula~\eqref{Sparamint} one can indeed check that it is dominated in the IR (for small $m_q$ and for all values of $x$). This is consistent with the analysis of Appendix~I of~\cite{Arean:2013tja}, where the same is argued to hold for the meson masses and decay constants.

\begin{figure}[!tb]
\begin{center}
\includegraphics[width=0.49\textwidth]{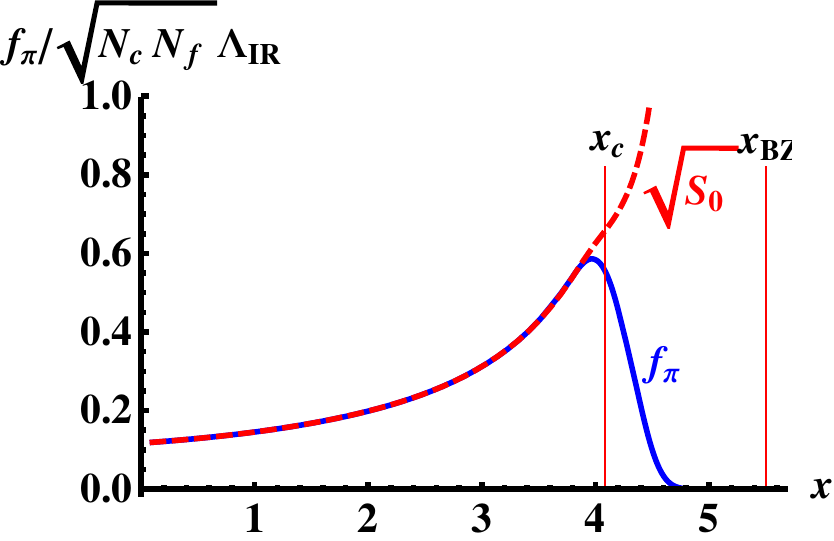}
\includegraphics[width=0.49\textwidth]{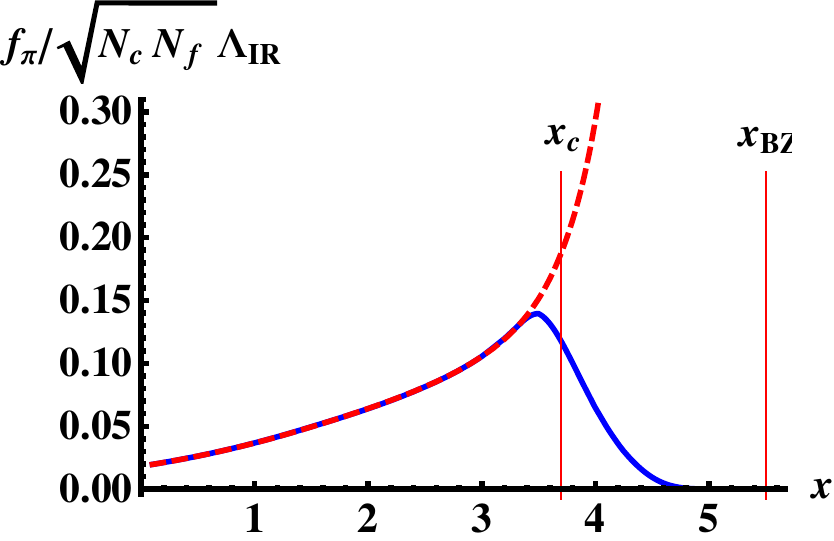}
\end{center}
\caption{The pion decay constant as a function of $x$ for $m_q/\LUV=10^{-6}$. Left: potentials~I with $W_0=3/11$. Right: potentials~II with SB normalized $W_0$. The (normalized and squared) pion decay constant is shown as the blue curves, and the values of the constant $\sqrt{S_0}$ are also shown as the red dashed curves for comparison. }
\label{fig:fpi}\end{figure}

We have also computed the pion decay constant which is defined in terms of the residue of $\Pi_L$ at the pion mass whenever $m_q \ne 0$ (see Eq.~\eqref{fpimqne0} in Appendix~\ref{app:fpi}). The results for both potentials are given in Fig.~\ref{fig:fpi}, and they are also compared\footnote{Notice that $S_0$ is UV divergent whenever the quark mass is finite, as can be seen by inserting the UV expansions of the wave functions~\cite{Arean:2013tja} in~\eqref{Pidef}, and needs to be renormalized. The divergence is $\propto m_q^2$ in agreement with the one-loop field theory computation of Appendix~\ref{app:SFT}. At small quark masses it is irrelevant how the renormalization is done because the difference between all reasonable renormalization schemes is negligible due to the smallness of the coefficient in the divergent term.} to the constant $S_0$. As expected the pion decay constant (blue curves) match with   $\sqrt{S_0}$ (dashed red curves) in the QCD regime. The ratio $f_\pi^2/S_0$ decreases fast with 
increasing $x$ for both 
potentials when $x\gtrsim x_c$, suggesting that the pion decouples.

\begin{figure}[!tb]
\begin{center}
\includegraphics[width=0.49\textwidth]{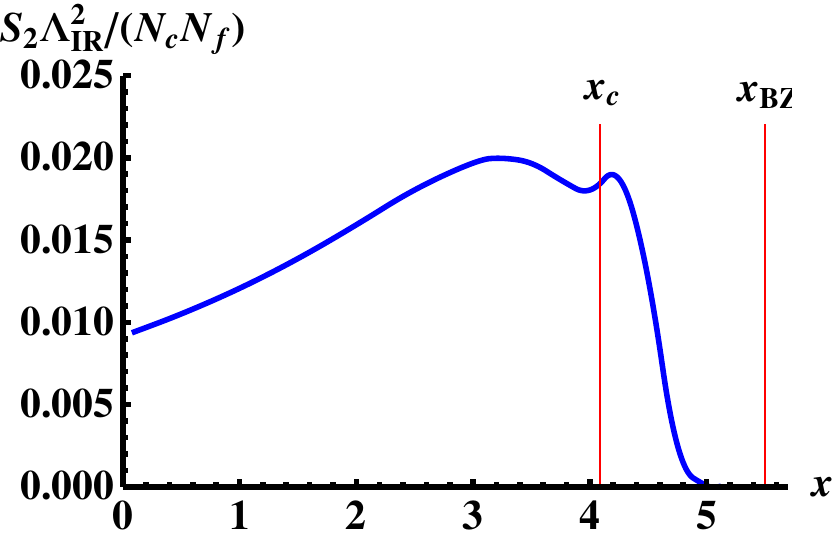}
\includegraphics[width=0.49\textwidth]{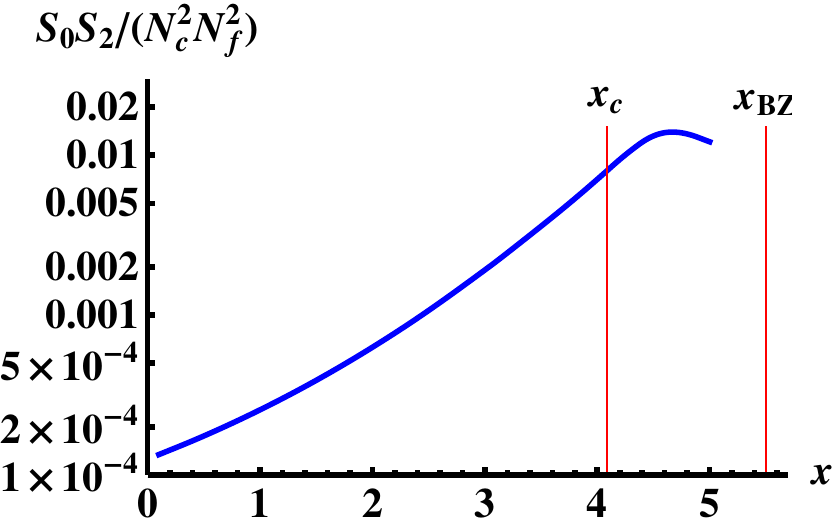}
\end{center}
\caption{The $x$ dependence of the higher order coefficient $S_2$ for $m_q/\LUV=10^{-6}$ and for potentials~I with $W_0=3/11$. Left: $S_2$ in units of $\L_\mathrm{IR}$. Right: The dimensionless product $S_0 S_2$.}
\label{fig:Sprime}\end{figure}

The higher order coefficient $S_2$ (in units of $\LIR$) is also shown for potentials~I in Fig.~\ref{fig:Sprime} (left) and the product $S_0S_2$ in Fig.~\ref{fig:Sprime} (right). The dependence on $x$ is similar as for the S-parameter when $x \lesssim x_c$.

We have not tried to analyze the various observables in the BZ limit $x \to \xBZ$ because this region is not the most interesting one from a holographic viewpoint. In general, however, the slow RG flow in the BZ limit causes that all ratios of energy scales to be typically $\propto \exp[\#(\xBZ-x)^{-2}]$. Therefore even ratios which are expected to be close to one for generic values of $x$ easily blow up in the BZ limit -- in this case $\#$ in the above relation is small, but it is difficult to define the scales such that it would be exactly zero. In view of this, it is not surprising that only the dimensionless quantities $S$ and $S_0S_2$ approach finite values in the BZ region.

\begin{figure}[!tb]
\begin{center}
\includegraphics[width=0.69\textwidth]{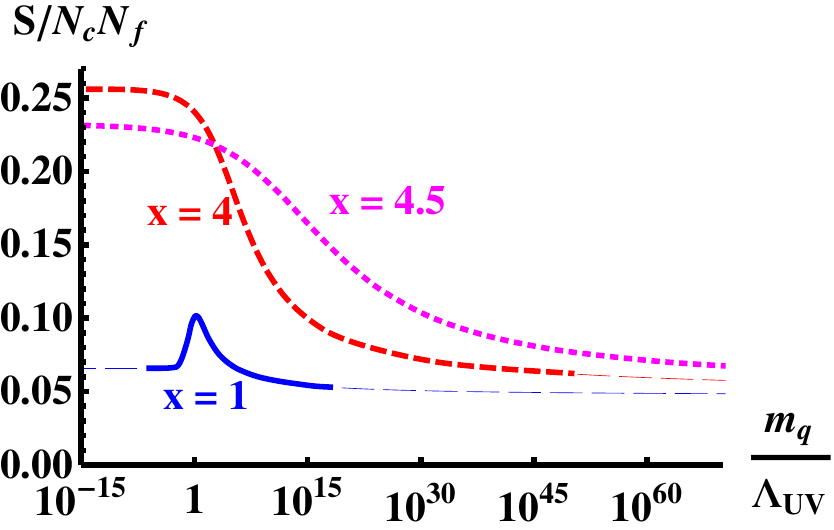}
\end{center}
\caption{The mass dependence of the S-parameter for $x=1$ (blue curve), $x=4$ (dashed red curve), and $x=4.5$ (dotted magenta curve). Potentials~I with $W_0=3/11$ were used. The thin lines are extrapolations given by fits to the asymptotic behavior.
}
\label{fig:massdep}\end{figure}

Let us then analyze the mass dependence of $f_\pi$ and $S$ in more detail. Fig.~\ref{fig:massdep} shows the mass dependence of the S-parameter in the QCD-regime ($x=1$, blue curve), in the walking regime ($x=4$, dashed red curve), and in the conformal window ($x=4.5$, dotted magenta curve). The dependence on $m_q$ is relatively mild for all values of $x=0$ (apart from the discontinuity at $m_q=0$ which is only present in the conformal window and the fact that $S$ varies slower as $x$ increases which is due to the RG flow as discussed at the end of Appendix~\ref{app:scaling}).  In particular the limiting value as $m_q \to \infty$ is independent of $x$. In this limit the S-parameter is expected to approach the value $N_c N_f/12\pi$ from perturbative QCD (see Appendix~\ref{app:SFT}). Even though V-QCD is not expected to reproduce perturbative results in general, the limiting value in V-QCD is numerically close to the QCD number $1/12 \pi \simeq 0.0265$.

\begin{figure}[!tb]
\begin{center}
\includegraphics[width=0.49\textwidth]{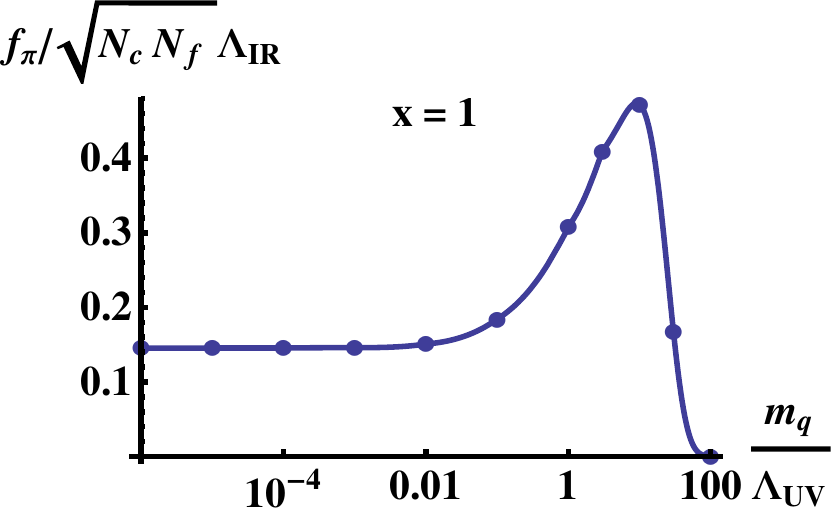}%
\includegraphics[width=0.49\textwidth]{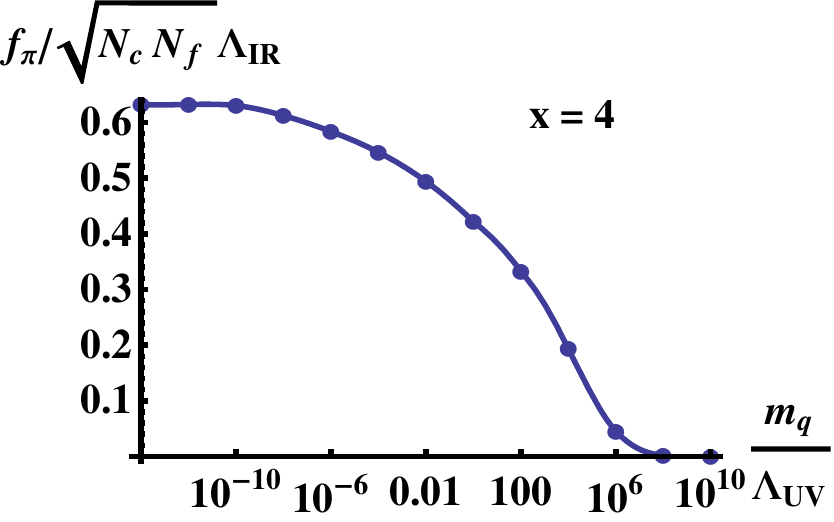}
\includegraphics[width=0.49\textwidth]{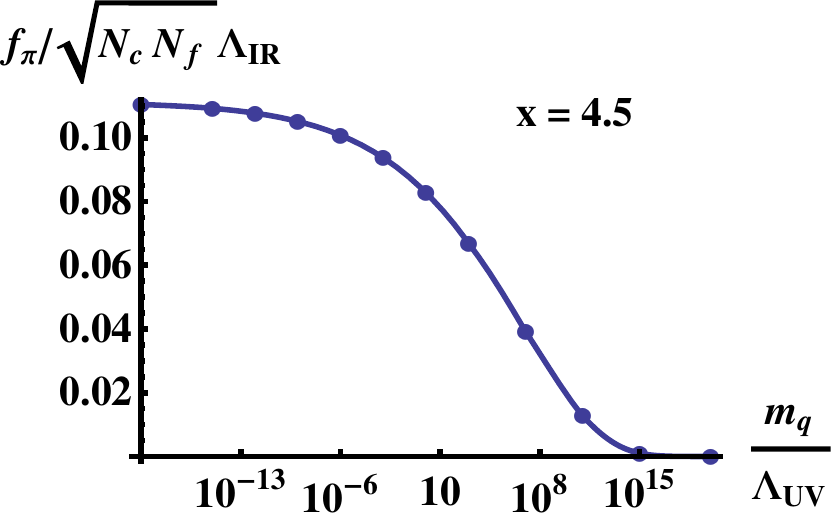}
\end{center}
\caption{The mass dependence of $f_\pi$ for $x=1$ (top left), $x=4$ (top right), and $x=4.5$ (bottom). Potentials~I with $W_0=3/11$ were used. 
}
\label{fig:fpimassdep}\end{figure}

The dependence of $f_\pi$ on $m_q$ is demonstrated in Fig.~\ref{fig:fpimassdep}. Again the different plots are in the QCD regime ($x=1$, top left plot), in the walking regime ($x=4$, top right plot), and in the conformal window ($x=4.5$, bottom plot). For small $m_q$ the dependence is weak, but when $m_q/\LUV \gg 1$ the decay constant vanishes very fast with increasing $m_q$. This signals the decoupling of the pion mode in regime~C, and is consistent with the findings of Appendix~\ref{app:masses} (see Eq.~\eqref{fnpotI}). 
Actually, all low-lying meson states are expected to decouple for potentials~I.

\begin{figure}[!tb]
\begin{center}
\includegraphics[width=0.49\textwidth]{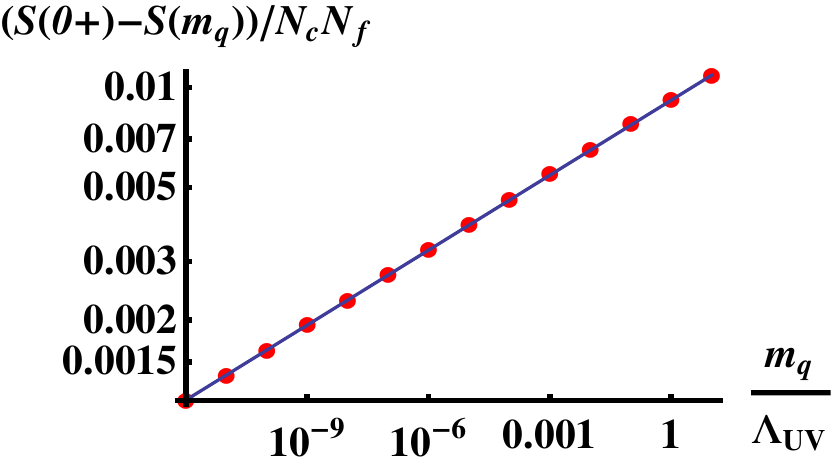}
\includegraphics[width=0.49\textwidth]{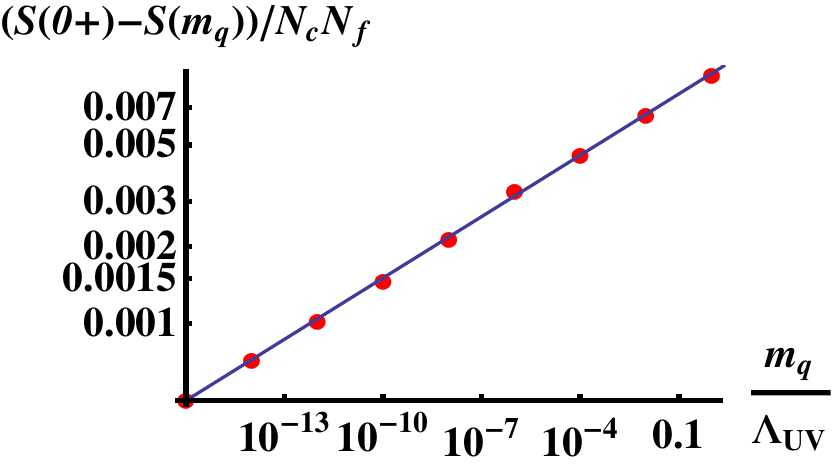}
\end{center}
\caption{The mass dependence of the S-parameter for $x=4.5$ in log-log scale. The red dots are the data and the blue lines are power-law fits $\propto m_q^{0.08}$. Left: potentials~I with $W_0=3/11$. Right: potentials~II with SB normalized $W_0$.}
\label{fig:massdeplog}\end{figure}

\subsection{Scaling of the S-parameter}

We have also analyzed the S-parameter in the limit $m_q \to 0$. Numerical results are shown in the conformal window (for $x=4.5$) in Fig.~\ref{fig:massdeplog}, where the red dots are our data and the lines are given by the functions
\be \label{Sfit}
 \frac{S(m_q)-S(0^+)}{N_cN_f} = \beta_1 m_q^{\beta_2} \ ,
\ee
where the parameters $\beta_{1,2}$, as well as $S(\eps)$, were fitted to the data. Here it is understood that
\be
 S(0^+) \equiv \lim_{m_q \to 0 +} S(m_q) 
\ee
which is a finite number whereas the S-parameter vanishes at zero quark mass: $S(0)=0$. Thus there indeed is a discontinuity at $m_q=0$. For both potentials $\beta_2 \simeq 0.08$ fits the data very well.

In order to understand the power law in~\eqref{Sfit}, it is useful to first discuss in more detail how the discontinuity at $m_q=0$ arises. The mechanism is the same which was studied in detail in the case of $m_q=0$ and $x \to x_c$ in section~6 and Appendix~I of~\cite{Arean:2013tja}. Here instead $x \ge x_c$ and $m_q \to 0$. In both cases the RG flow of the coupling approaches the fixed point ($\l=\l_*$) but misses it finally (due to the finite tachyon). For such flows, it is useful to divide the background to the UV and IR sections, having $\l<\l_*$ and $\l>\l_*$, respectively.  
Considering flows which get closer and closer to the fixed point, the S-parameter can be computed more and more precisely in terms of the IR section. The IR part takes a fixed shape in this limit, explaining the finite value of the S-parameter. For exactly zero quark mass the IR section of the background becomes disconnected from the UV section and is therefore not present in the physical vacuum solution, which now ends at the IRFP. This is reflected in the vanishing value of the S-parameter.

By using similar arguments, we can also sketch how the power law in the mass dependence arises. It is understood to be the leading ``perturbation'' of the IR background due to the fact that the fixed point was not reached exactly.  In the conformal window, the flow towards the fixed point is given by
\be
 \l \simeq \l_* - C_\l \left( r\LUV\right)^{-\delta}\ ,
\ee
where $\delta$ is related to the dimension of the $\mathrm{Tr} F^2$  operator at the fixed point. It can be computed as the derivative of the holographic beta function at the fixed point (when the tachyon is set to zero exactly)~\cite{jk,Gubser:2008yx}. One finds that
\be
 \Delta_{FF}-4 = \delta = \sqrt{4-\frac{9 V_2 \l_*^2}{V_0}}-2 \ ,
\ee
where $V_i$ are the coefficients of the Taylor expansion of $V_\mathrm{eff}$ at $\l=\l_*$:
\be
 V_\mathrm{eff} \equiv V_g - x V_{f0} = V_0 + V_2 (\l-\l_*)^2 + \cdots \ .
\ee
Notice that $V_2<0$. Flow toward the fixed point ends when $r\sim 1/\LIR$. The difference of $\l$ with respect to the fixed point value when this happens is given by
\be \label{massdepCW}
 \l_*-\l_\mathrm{IR} \equiv \l_*-\l(r=1/\LIR) \sim \left( \frac{\LIR}{\LUV}\right)^{\delta} \sim  \left( \frac{m_q}{\LUV}\right)^{\frac{\delta}{\Delta_*}} \ , \qquad \left(\frac{m_q}{\LUV} \ll 1\right)
\ee
where~\eqref{CWscaling} was used to obtain the last expression. This difference controls the deviation of the IR section of the background from its limiting shape as $m_q \to 0$ and correspondingly the deviation of the S-parameter from the limiting value $S(0^+)$. Indeed by using the explicit expressions for the potentials one obtains
\be
 \frac{\delta}{\Delta_*} \simeq 0.0780 \ , \qquad (x=4.5) \ ,
\ee
a value in a very good agreement with the fit from above, shown in Fig.~\ref{fig:massdeplog}. 

Notice that the difference $\l_*-\l_\mathrm{IR}$ not only controls the corrections to the S-parameter at small mass, but also to other quantities that can be defined in terms of the IR section of the background. Examples are decay constants and meson masses in IR units, which are therefore expected to have qualitatively similar $m_q$ dependence to the S-parameter at small masses\footnote{We have found numerically that ratios of masses or decay constants typically follow the scaling of~\eqref{massdepCW} more accurately than their values in IR units.}. Indeed, the power law of~\eqref{massdepCW} agrees with that found for the scaling corrections to generic correlators by analyzing the Wilsonian RG flow perturbatively in the vicinity of the fixed point~\cite{DelDebbio:2013qta}.

\begin{figure}[!tb]
\begin{center}
\includegraphics[width=0.49\textwidth]{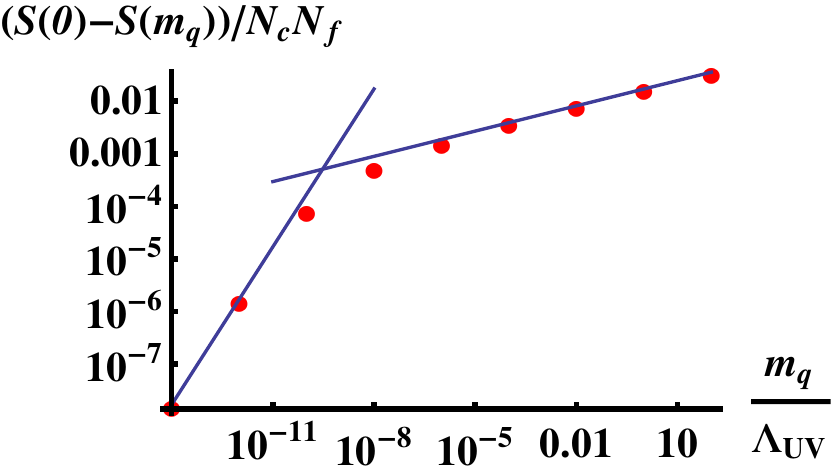}
\includegraphics[width=0.49\textwidth]{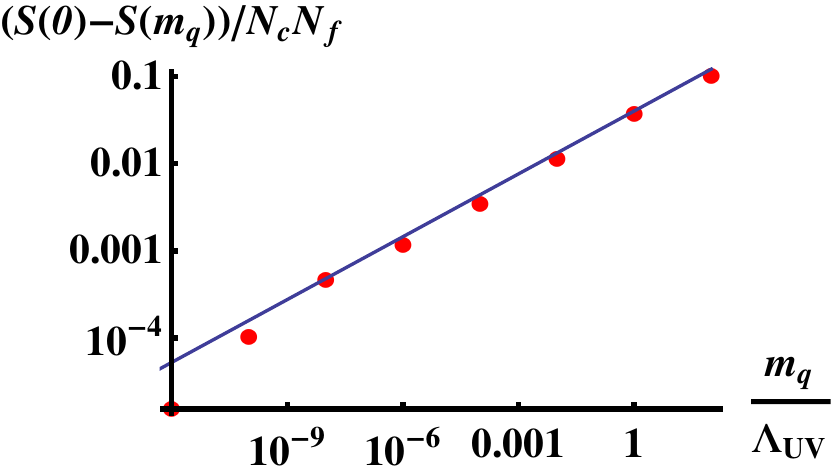}
\end{center}
\caption{The mass dependence of the S-parameter in the walking regime in log-log scale. The red dots are the data and the blue lines are power-law fits. Left: potentials~I with $W_0=3/11$ and with $x=4$. Right: potentials~II with SB normalized $W_0$ and $x=3.6$.}
\label{fig:massdepwalk}\end{figure}

The mass dependence of the S-parameter can be analyzed similarly in the walking regime ($x \to x_c{}^-$). The above calculation is approximately valid in regime B, i.e, when the quark mass controls the amount of walking. One can use~\eqref{regimeBlowx} together with~\eqref{massdepCW} to obtain
\be \label{massdepwalking}
 \l_*-\l_\mathrm{IR}  \sim  \left( \frac{m_q}{\LUV}\right)^{\frac{\delta}{2}} \ , \qquad \left(\exp\left[-\frac{2K}{\sqrt{x_c-x}}\right]\ll \frac{m_q}{\LUV} \ll 1\right) \ .
\ee
Therefore one should effectively take $\Delta_* \to 2$ in~\eqref{massdepCW}, meaning that the power is continuous over the conformal transition at $x=x_c$, as $\Delta_*=2$ at the transition. For even smaller $m_q$, i.e., in regime~A, the mass term of the tachyon can be treated as a linear perturbation to the whole background, and therefore the mass dependence of the S-parameter is linear.  
Both the scaling of~\eqref{massdepCW} and the linear dependence can be seen in the numerical results in the walking regime, see Fig.~\ref{fig:massdepwalk}.

\begin{figure}[!tb]
\begin{center}
\includegraphics[width=0.49\textwidth]{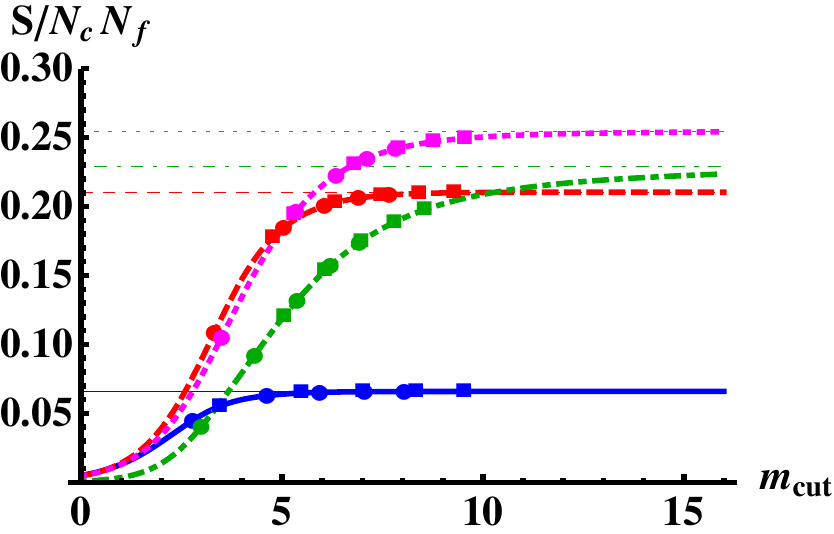}%
\includegraphics[width=0.49\textwidth]{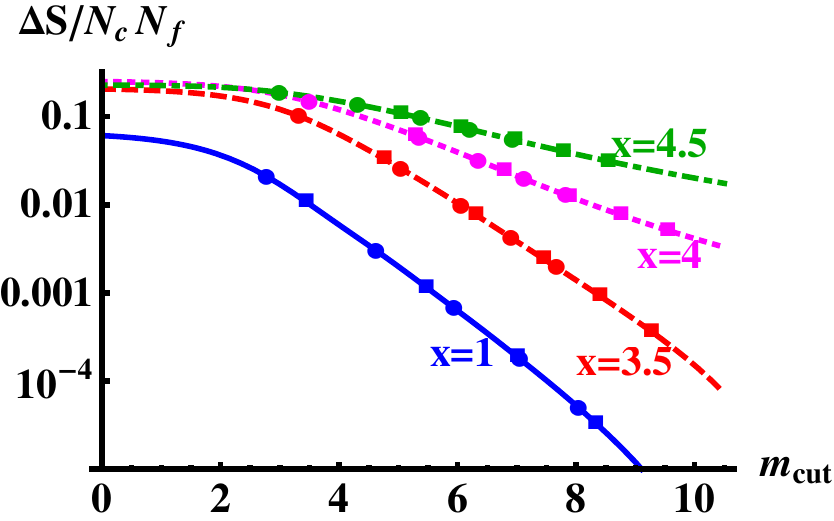}
\end{center}
\caption{S-parameter computed from the spectral representation compared to the exact value. Left: the dependence of the value of $S$ on the mass cutoff in the series. The thin horizontal lines are the limiting values which match with direct computation of $S$. Right: the difference between the number obtained from the series and the exact value as a function of the cutoff. The blue, dashed red, dotted magenta, and dotdashed green curves have $x=1$, $3.5$, $4$, and $4.5$, respectively, in both plots. The filled circles and boxes denote the masses of the lowest five vector and axial mesons, respectively.}
\label{fig:Sconv}\end{figure}

\subsection{Convergence of the spectral representation}

Finally we will analyze the spectral representation of the S-parameter in order to understand better why it increases with $x$ in the QCD and walking regimes.
By inserting~\eqref{PiAsp} and~\eqref{PiVsp} in the definition for $S$, one obtains
\be \label{Sseries}
 S = 4 \pi \sum_{n=1}^\infty\left(\frac{(f_{n}^{(V)})^2}{(m_{n}^{(V)})^2}-\frac{(f_{n}^{(A)})^2}{(m_{n}^{(A)})^2}\right) \ .
\ee
This series may, however, be ill defined. For example, the decay constant approach asymptotically constant values whereas the masses have linear trajectories $m_n^2 \sim n$ for potentials~I (see Appendices~E and~F in~\cite{Arean:2013tja}). One can check that~\eqref{Sseries} is convergent thanks to the asymptotic cancellation of the vector and axial terms, but it is not absolutely convergent, and therefore the result may depend on the ordering of the terms. The definition of~\eqref{Sseries}, where the states are ordered in terms of their excitation numbers $n$, would work for potentials with linear trajectories if the slopes of the vector and axial spectra are the same. This is not\footnote{The question whether the slopes should be the same in QCD is unsettled~\cite{Golterman:2001nk}, but usually they are assumed to be, which can be obtained for potentials~I by changing the IR asymptotics of $w(\l)$~\cite{Arean:2013tja}.} the case for potentials~I, and consequently~\eqref{Sseries} is incorrect. The solution 
to this issue is simple: the contributions should be ordered by the meson 
masses rather than by the excitation numbers. 

It is straightforward to verify numerically that~\eqref{Sseries} converges towards the S-parameter if the terms are ordered according to the masses. It turns out to be convenient to define a mass dependent cutoff,
\be \label{Scut}
 S(m_\mathrm{cut}) = 4 \pi \sum_{n=1}^\infty\left(\frac{(f_{n}^{(V)})^2}{(m_{n}^{(V)})^2}\, f_\mathrm{cut}(m_{n}^{(V)})-\frac{(f_{n}^{(A)})^2}{(m_{n}^{(A)})^2}\, f_\mathrm{cut}(m_{n}^{(A)})\right)
\ee
where
\be
 f_\mathrm{cut}(m) = \frac{1}{2}\left(1- \tanh \frac{m-m_\mathrm{cut}}{\delta m}\right) \ .
\ee
A smooth cutoff function was chosen instead of a step function because it improves convergence drastically. The convergence also means that the value of the S-parameter is determined through the dynamics in the deep IR, because the same holds for the masses and decay constants (the argument can be made precise in the walking regime, see Appendix~I in~\cite{Arean:2013tja}). 

The convergence of the regulated series~\eqref{Scut} towards the S-parameter is demonstrated in Fig.~\ref{fig:Sconv}. The resolution of the cutoff function was fixed to the mass difference of the two lowest vector states: $\delta m =m_2^{(V)}-m_1^{(V)}$. The speed of the convergence is best visible from the right hand plot, which shows
\be
 \Delta S =  S -   S(m_\mathrm{cut})
\ee
as a function of the cutoff. The convergence is exponential for all values of $x$, but becomes significantly slower as $x$ increases and one moves from the QCD regime to the conformal window. The slowness of the convergence is not due to changes in the spectra. To show this, the filled circles and boxes marking the masses of the lowest five vector and axial mesons, respectively, were added in each plot. It is seen that the spectrum changes relatively little with $x$ in unit of $\LIR$.

Finally let us try to extract the reason for the increase of $S$ with increasing $x$ in the region of low values of $x$ from the plots of Fig.~\ref{fig:Sconv}. Notice that the curves for $x=1$, $3.5$ and $4$ essentially overlap at small values of $m_\mathrm{cut}$ in the left hand plot. The curves for $x=1$ and $x=3.5$ deviate from that of $x=4$ as $m_\mathrm{cut}$ increases, and after deviating rapidly saturate to the final value of $S$. Therefore it appears that contributions to the S-parameter are roughly mass-independent up to a saturation scale, which increases with $x$. In order to have a good estimate for the S-parameter, a growing number of terms need to be included in the sum with increasing $x$, whereas the individual terms in the sum are of roughly constant size. Therefore the increase of $S$ with $x$ in the QCD and walking regimes can be seen to be due to slower convergence of the sum. In the conformal window, i.e., for the curve with $x=4.5$, something different happens. The convergence of the 
sum is even slower, but the contributions at fixed $m_\mathrm{cut}$ are suppressed, resulting in the decrease of the S-parameter 
with increasing $x$.

\section{Finite temperature phase diagram}
\label{sec:ft}

The finite temperature phase diagram has been studied in detail for IHQCD in~\cite{ihqcd2} and for V-QCD at zero quark mass and at small values of the quark mass in~\cite{alho}. Here this study is extended to large values of $m_q$ as well as very high values of $x$. The code for constructing solutions at finite temperature is available at~\cite{code}. Some extra tricks are necessary in order to obtain reliable results in the BZ region and at very large $m_q$ (see Appendix~\ref{app:numerics}). 

First recall the generic structure of the ($x,T$) phase diagram~\cite{alho}, which was already reviewed in Sec.~\ref{sec:vqcd}. At zero quark mass, there is a first order deconfinement transition in the QCD and walking regimes, but there is also the possibility (depending on the choice of potentials and the value of $x$) of a chiral symmetry restoration at a separate second order transition. In the conformal window, there is a continuous phase transition at zero temperature. 
When a finite quark mass is turned on, chiral symmetry is always broken, and the second order chiral transition will become a fast crossover when the quark mass is small, and completely disappear at larger quark masses.
The system is in a tachyonic thermal gas (TG) phase at small temperatures for all $0<x<\xBZ$. As the system is heated, there is a first order deconfinement transition to the high temperature phase, which is implemented through a transition from TG phase to the black hole (BH) phase in holography. 

The existence of the deconfinement transition requires an order parameter. While QCD at finite $N_c$ and $N_f$ has no order parameter related to deconfinement, the pressure acts as an effective order parameter at large $N_c$. This is clear in the 't Hooft limit, where the number of degrees of freedom (and consequently the pressure) is \order{N_c^0} in the low temperate phase and \order{N_c^2} in the high temperature phase. In the Veneziano limit the number of degrees of freedom is of the same order in both phases, but the phase transition may still be identified as a discontinuity of the pressure.
In fact, the pressure of the model is still exactly zero in the TG phase, because our approach does not capture the contributions corresponding to loops of pions (as well as mesons with higher masses) in this phase. Including these contributions in the model would affect the critical temperature, and potentially even alter the order of the transition~\cite{pionloop}.

\subsection{Scaling laws at finite temperature}

The critical temperature has nontrivial dependence on $m_q$, which can be analyzed analytically. Notice that the temperature brings in an additional energy scale with respect to the zero temperature solutions. Another difference is that the definition of the standard reference scale $\LIR$ cannot be extended to the BH solutions in a natural manner (because the geometry now ends at a horizon rather than an IR singularity). Therefore it is understood that $\LIR$ is defined below through the TG solution (or equivalently through the zero temperature solution at the same values of $m_q$ and $x$).

Let us first discuss the mass dependence of the critical temperature, which can be 
inferred by using the results from~\cite{alho} and from Sec.~\ref{sec:scaling}. The temperature of the black hole can be related to the metric through the formula
\be \label{Tdef}
 T = \frac{1}{4 \pi e^{3A_h}}\left(\int_0^{r_h}\frac{dr}{e^{3A(r)}}\right)^{-1} \ ,
\ee
where $A_h$ and $r_h$ are the values of the scale factor and the bulk coordinate at the horizon. 
In V-QCD models $T(r_h)$ has a nontrivial minimum (for tachyonic BHs) and the transition takes place at the scale of the minimum. 
Indeed, the entropy density 
\be \label{sdef}
 s_\mathrm{BH} = 4 \pi M^3 N_c^2 e^{3 A_h}
\ee
decreases monotonically (and fast) with $r_h$, as suggested by the UV and IR (zero temperature) expansions of $A(r)$ and as can be verified numerically. Further, the geometry of the BH solution approaches smoothly the TG solution as $r_h \to \infty$, and therefore $p_\mathrm{TG} = \lim_{r_h \to \infty} p_\mathrm{BH}$. Integrating $p_\mathrm{BH}'(r_h) = s(r_h) T'(r_h)$, a node $p_\mathrm{BH} = 0$, and consequently a first order phase transition, is found near the minimum of $T(r_h)$.

In conclusion, one should locate the minimum of $T(r_h)$  in order to determine the scaling of $T_c$. When the geometry is close to AdS,~\eqref{Tdef} implies that $T \sim 1/r_h$. This result holds both in the UV asymptotic region and when there is an approximate IRFP, i.e., walking. When the quark mass is large, there is also an approximately AdS region where the flavors are already decoupled but the dilaton $\l$ is still small. In summary,  
\be \label{TUV}
 T \sim 1/r_h\ ,\qquad  \left(r_h \ll \frac{1}{\LIR}\right) \ . 
\ee
For $r_h \gg 1/\LIR$ the temperature increases with $r_h$ as seen by studying the IR expansions (see~\cite{alho}). Therefore, the minimum of $T(r_h)$ takes place at $r_h\sim 1/\LIR$. By continuity,~\eqref{TUV} implies that
\be
  T_c \sim \LIR
\ee
for all $m_q>0$ and $0<x<\xBZ$.

The scaling results in units of $\LUV$ immediately follow by using the results from Sec.~\ref{sec:scaling}:
\begin{itemize}
 \item In regime A, $T_c \sim \LIR \sim \LUV$ for small values of $x$ and 
\be \label{TcinregA}
 \frac{T_c}{\LUV} \sim \frac{\LIR}{\LUV} \sim \exp\left[-\frac{K}{\sqrt{x_c-x}}\right]
\ee
 as $x \to x_c$ from below. The dependence of $T_c$ on $x$ at $m_q=0$~\cite{alho} is in qualitative agreement with analysis based on field theory (see, e.g.,~\cite{Braun:2009ns}), and the dependence on $m_q$ is expected to be a linear perturbation. 
\item In regime B,
\be
 \frac{T_c}{\LUV} \sim \sqrt{\frac{m_q}{\LUV}} 
\ee 
when $x \le x_c$ and
\be
 \frac{T_c}{\LUV} \sim \left(\frac{m_q}{\LUV}\right)^\frac{1}{\Delta_*} 
\ee 
when $x_c \le x <\xBZ$.
\item In regime C,
\be \label{TcinregC}
 \frac{T_c}{\LUV} \sim \left(\frac{m_q}{\LUV}\right)^{1-b_0/b_0^\mathrm{YM}} \ .
\ee
\end{itemize}

In addition to the phase transition, also various crossovers can be identified as the maxima of the interaction measure
\be
 \frac{\eps-3p}{T^4} = \frac{T s -4 p}{T^4} \ .
\ee
As it turns out, such crossovers reflect the different regions of the zero temperature geometry. This can be understood by approximating the horizon as a sharp cutoff added on the zero temperature background. Substituting an AdS metric in the formulas~\eqref{Tdef} and~\eqref{sdef}, one finds that the interaction measure vanishes. 

First, there is the crossover which marks the transition from the quasi-conformal or walking phase (with approximate IRFP) to the asymptotic UV phase \cite{alho,Tuominen:2012qu}. Such a crossover is found whenever there is walking, i.e., in regime B, and the $x \to x_c$ edge of regime A. The UV asymptotics is valid for $r \ll 1/\LUV$, and the flow from the UV fixed point to the IRFP is characterized by $\LUV$. Consequently, using~\eqref{Tdef} and~\eqref{TUV}, the crossover temperature is expected to be
\be \label{Tqclaw}
 T_\mathrm{co,qc} \sim \LUV
\ee
independently of $m_q$. 

Second, there is a crossover at large quark mass, corresponding to the transition from the region where the quarks are decoupled to the UV asymptotic region. The decoupling of the quarks takes place at $r \sim 1/m_q$ as pointed out in Sec.~\ref{sec:scaling}. Consequently, the crossover temperature is given in terms of the quark mass:
\be \label{Tmqlaw}
 T_\mathrm{co,{\it m_q}} \sim m_q \ .
\ee

\begin{figure}[!tb]
\begin{center}
\includegraphics[width=0.49\textwidth]{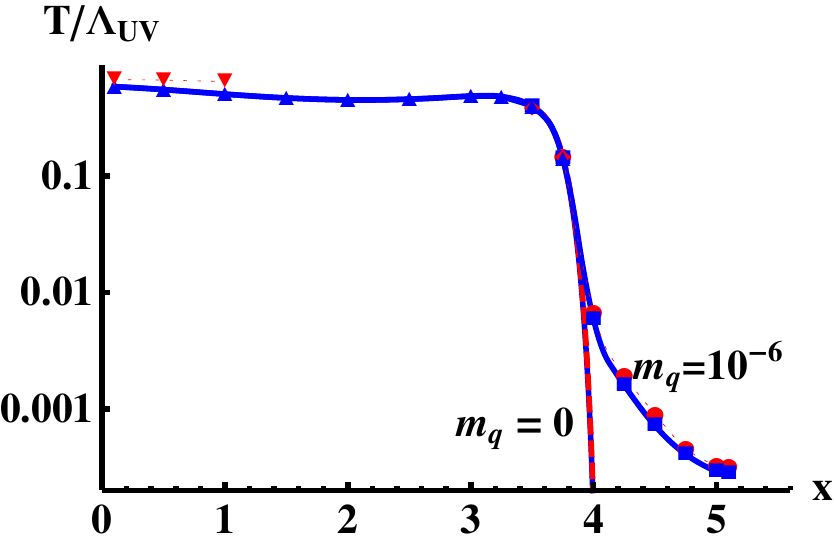}%
\includegraphics[width=0.49\textwidth]{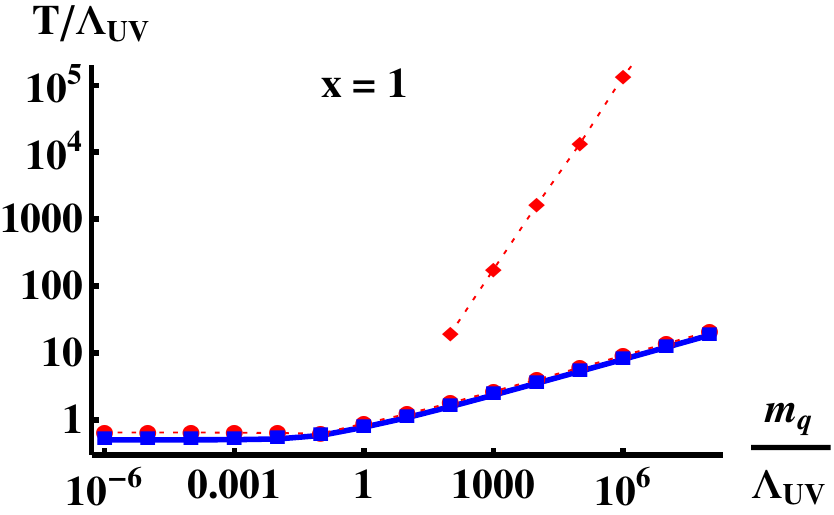}
\includegraphics[width=0.49\textwidth]{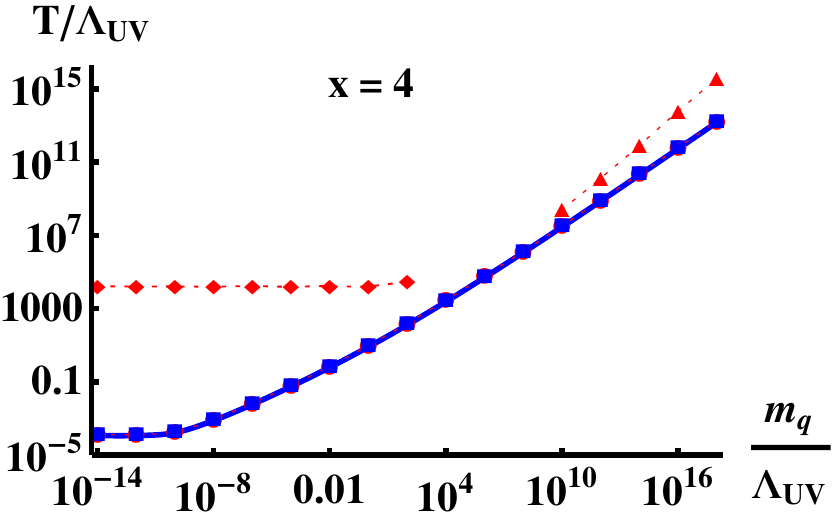}%
\includegraphics[width=0.49\textwidth]{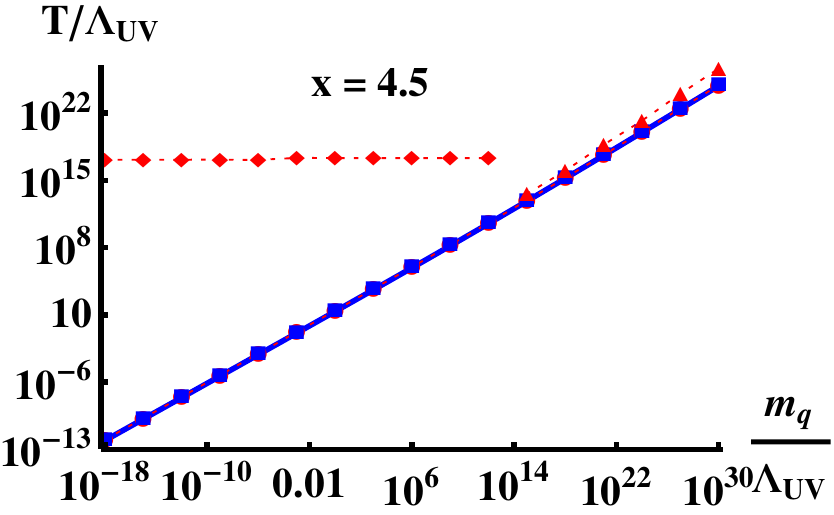}
\end{center}
\caption{The dependence of the critical transition temperatures (solid blue curves) and various crossover temperatures (thin dotted red curves) on $x$ and $m_q$. In the top left plot we also included the $m_q=0$ data, and the critical temperature of the second order chiral transition which is shown as the thick red dashed curve near $x=x_c$.}
\label{fig:FT}\end{figure}

When $T \ll m_q$ the quarks are effectively decoupled, and therefore the thermodynamics is the same as for pure YM (that is, the $x\to 0$ limit of V-QCD). Notice that $\LUV$, however, is defined in terms of the UV asymptotics, i.e., effectively at infinite energy, and different from that of YM even as $m_q \to \infty$: for YM, $\LUV \sim \LIR$ but in the limit of large $m_q$ these scales are related through~\eqref{largemqscalingt} instead. Consequently, in order for the thermodynamics to smoothly approach YM thermodynamics as $m_q \to \infty$, dimensional quantities should be expressed in units of $\LIR$ or $T_c$ rather than in units of $\LUV$.

\subsection{Numerical results}

Let us then illustrate the dependence of the various critical temperatures on $x$ and $m_q$ numerically. The basic features of the phase diagram at small quark mass are demonstrated in the top-left plot of Fig.~\ref{fig:FT}. The first order ``deconfinement'' transition temperatures for potentials~I at zero and at tiny ($10^{-6}$) quark mass are shown as functions of $x$ on the logarithmic scale. The first order transitions are shown as blue curves. The curves overlap at small $x$, but as the conformal transition is approached, the curves become separated. The lower curve (which overlaps with the red dashed curve and is therefore not well visible) is the transition temperature at $m_q=0$ which goes to zero with Miransky scaling as $x \to x_c$. For $x>x_c$ there is no transition when $m_q=0$. When a tiny quark mass is turned on the transition (upper blue curve) is present for all values of $x$. the critical temperature $T_c$ decreases with $x$ inside the conformal window\footnote{In~\cite{alho} also a region 
where $T_c$ increased with $x$ was seen at high $x\simeq 4.5$ (see Fig.~27 there), but this effect turns out to be due to the UV cutoff being too low in the 
numerics.}.

The dashed thick red curve is the second order chiral restoration transition, which also shows Miransky scaling as $x \to x_c$ and is absent for $x>x_c$. This transition only exists for $m_q=0$ in the walking regime.

The dependence of $T_c$ on $m_q$ is also demonstrated numerically for V-QCD in Fig.~\ref{fig:FT}. The top-right, bottom-left, and bottom-right plots are in the running regime ($x=1$), walking regime ($x=4$), and in the conformal window ($x=4.5$), respectively, and $T_c$ is given by the blue curve in each plot. The power laws are in agreement with the above formulas. For example, in the regime~C with large $m_q$ we find that the exponent of~\eqref{TcinregC} is $1-b_0/b_0^\mathrm{YM} = 2x/11 \simeq 0.182$ at $x=1$ and $0.727$ at $x=4$, which is consistent with the plots. 

The crossover between the quasiconformal and UV regions at $T=T_\mathrm{co,qc}$ is seen as the horizontal lines in the bottom row of Fig.~\ref{fig:FT}. The ratio $T_\mathrm{co,qc}/\LUV$ is constant as expected, but the value of the constant deviates significantly from its expected value, i.e., one. This happens because $\LUV$ deviates from the scale of the UV RG flow at large values of $x$, as was explained above in Sec.~\ref{sec:massscales}. The crossover due to the decoupling of the quarks at large $m_q$ at the temperature $T \sim m_q$ is best visible in the top-right plot of Fig.~\ref{fig:FT}.

In addition, in large part of the parameter space there is also a separate maximum of the interaction measure in connection to the first order transition. This kind of maxima have also been included as thin red curves in Fig.~\ref{fig:FT}, and can be found close to the blue curves denoting the transition temperatures.

\begin{figure}[!tb]
\begin{center}
\includegraphics[width=0.49\textwidth]{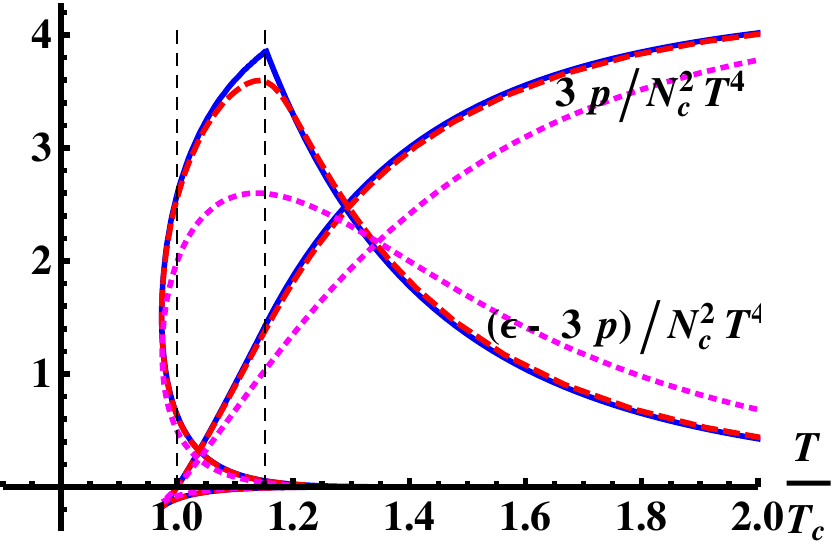}%
\hspace{2mm}\includegraphics[width=0.49\textwidth]{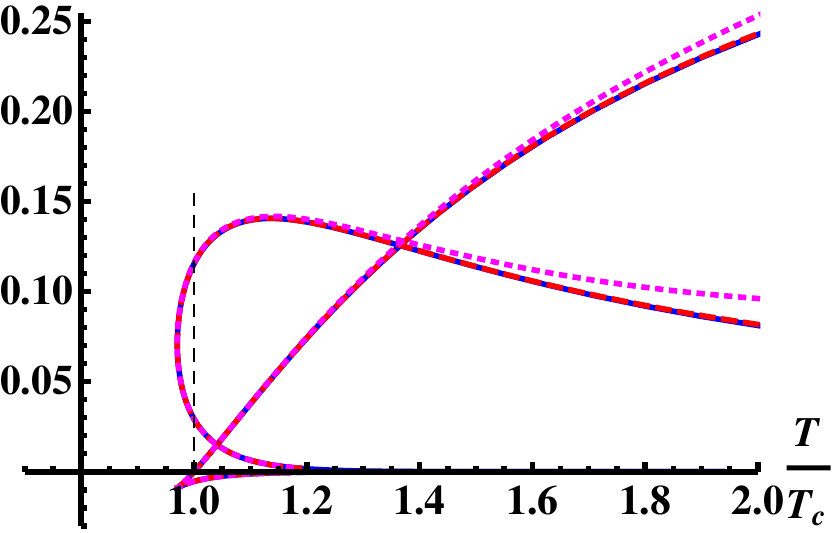}
\end{center}
\caption{Pressure and interaction measure as a function of temperature for small (left hand plot) and large (right hand plot) quark mass. On the left, the solid blue, dashed red, and dotted magenta curves have $m_q/\LUV=0$, $10^{-12}$, and $10^{-10}$, respectively. On the right, the solid blue curves are the YM thermodynamics in our model, while the dashed red and dotted magenta curves have $m_q/\LUV=10^{10}$ and $m_q=10^8$, respectively. We used potentials I with $W_0=3/11$ and $x=4$ (except for the YM curves which correspond to $x=0$).  }
\label{fig:thermo}\end{figure}

Recall that in regime~C the mass gap of the mesons does not have the expected behavior $m_\mathrm{gap} \simeq 2 m_q$ in V-QCD (if Sen-like tachyon potential is assumed). Nevertheless, the crossover due to the decoupling of the quarks takes place correctly at $T_\mathrm{co,{\it m_q}} \sim m_q$. This is seen as a consequence of the decoupling of the mesons having an unphysically low mass, which was discussed above in Sec.~\ref{sec:massscales}.

Finally let us study in some details of the thermodynamic functions at very small and very large $m_q$. We plot the (normalized) energy density, pressure, and interaction measure for potentials~I with $x=4$ in Fig.~\ref{fig:thermo} (left). The solid blue, dashed red, and dotted magenta curves have $m_q/\LUV = 0$,  $10^{-12}$, and $10^{-10}$, respectively. The value of $x=4$ was chosen such that the model shows the separate second order chiral restoration transition, which appears as a kink in the energy density and the interaction measure of the functions at zero quark mass. The vertical thin dashed black lines mark the locations of the phase transitions. As a tiny quark mass ($m_q/\LUV =10^{-12}$) is turned on, the second order transition turns into a crossover. Indeed the dashed curves follow closely the solid curves except for very close to the kink, where the curves have a smooth behavior instead. At $m_q/\LUV =10^{-10}$  a much larger deviation is seen already.

The approach to YM theory can be seen by studying the thermodynamics at very large quark mass. Thermodynamic functions at sizeable $m_q$ are shown for the same choice of potentials and compared to the results for the limit of YM theory ($x \to 0$) in Fig.~\ref{fig:thermo}. As the quark mass increases, the quarks are decoupled and the thermodynamics near $T=T_c$ is expected to converge to that of YM theory. Indeed, the (dashed red) curves for $m_q/\LUV = 10^{10}$ lie essentially on top of the (solid blue) curves of the YM thermodynamics\footnote{In order to show the convergence, the Planck mass $M$ was fixed to the value obtained in the YM limit  (by matching the pressure to the SB law at high temperatures) also at $x=4$. If we matched the pressure separately at $x=4$, there would be a small deviation, because a fixed value of $W_0$ was used rather than the ``SB normalized'' value~\cite{alho} so that $M$ would vary as a function of $x$.}. For smaller quark mass ($m_q/\LUV =10^8$, dotted magenta curves) a much 
large deviation is seen
already.

\section*{Acknowledgments}\label{ACKNOWL}

I would like to thank E.~Kiritsis for advice during this project and F.~Sannino for explanations of the mass dependence of the S-parameter. In addition, I would like to thank D.~Albrecht, T. Alho, F. Aprile, B. Gout\'eraux, P. Hoyer, T. Ishii, K.~Kajantie, W. Li, T. Morita, F. Nitti, M. Panero, C. Pica, C. Rosen, R. Shrock, K. Tuominen, and R.~Zwicky for interesting discussions, relevant comments, or correspondence.

This work was partially supported by European Union's Seventh Framework Programme
under grant agreements (FP7-REGPOT-2012-2013-1) No 316165, PIF-GA-
2011-300984, the ERC Advanced Grant BSMOXFORD 228169, the EU program
Thales MIS 375734. It was also co-financed by the European Union (European Social
Fund, ESF) and Greek national funds through the Operational Program ``Education
and Lifelong Learning'' of the National Strategic Reference Framework (NSRF)
under ``Funding of proposals that have received a positive evaluation in the 3rd and
4th Call of ERC Grant Schemes'', as well as under the action ``ARISTEIA''.  
I also thank the ESF network Holograv for partial support, and acknowledge the use of the computational resources of CCTP.


\newpage
\appendix
\renewcommand{\theequation}{\thesection.\arabic{equation}}
\addcontentsline{toc}{section}{Appendix}
\section*{Appendix}

\section{Energy scales from tachyon solutions}\label{app:scaling}

Here we shall demonstrate how the various dynamical energy scales of holographic QCD in the Veneziano limit arise from the background solutions for the tachyon. We will work with the V-QCD action, but as stressed in the main text, most of the results are universal, i.e., independent of the details of the action.

\subsection{Tachyon solutions}

In order to analyze the scales at finite quark mass it is essential to recall what the tachyon solution is in different regimes~\cite{jk}. For the discussion here it is enough to keep track of the power laws in $r$ so we will drop the logarithmic corrections. When the tachyon is small, it satisfies a linearized EoM. Up to irrelevant terms, we have
\be \label{taulineq}
 \t'' + 3 A' \t' - e^{2A} m_\t^2 \t \simeq 0 \ ,
\ee
where the scale factor has the AdS behavior near UV and IRFPs, $A \simeq \log(\ell/r)$, and the tachyon bulk mass is for V-QCD 
\be
 m_\t^2 = -\frac{2 a}{\kappa}
\ee
but its precise form is not important for our arguments. The bulk mass is expected to take roughly constant values near the fixed points. In the deep UV one requires that $-m_\t^2\ell^2\simeq 3$, leading to
\be \label{tauUV}
 \frac{\tau(r)}{\ell} \simeq m_q r + \sigma r^3\ , \qquad \left(r\ll \frac{1}{\LUV}\right)\ ,
\ee 
where $\sigma$ is proportional to the chiral condensate. 
Walking takes place if the coupling $\l$ flows very close to an IRFP but the tachyon is nonzero (and will eventually drive the system away from the fixed point in the deep IR). Walking can happen either right below the conformal window or in the conformal window (if the quark mass is small but finite). Let us denote 
\be
 \Delta_*(4-\Delta_*) = -\ell_*^2 m_{\t,*}^2\ ,
\ee
where $\ell_*$ and $m_{\t,*}$ are the AdS radius and tachyon mass at the IRFP, respectively.
In the vicinity of the fixed point, and if $x< x_c$ so that the (squared) tachyon mass is below the BF bound, $\ell_*^2 m_{\t,*}^2 < -4$, the solution to~\eqref{taulineq} behaves as
\be \label{tauw}
 \frac{\tau(r)}{\ell} \simeq C_w \left(r \LUV\right)^2\sin\left[\n\log(r\LUV) +\phi\right] \ ,\qquad  \left(\frac{1}{\LUV}\ll r\ll \frac{1}{\Lt}\right) \ ,
\ee
where $\n = \mathrm{Im}\Delta_*$,  If $x_c<x<x_\mathrm{BZ}$ we find instead
\bea \label{tauIRFP}
 \frac{\tau(r)}{\ell} &\simeq& C_m \left(r\LUV\right)^{\Delta_*} + C_\s \left(r\LUV\right)^{4-\Delta_*} \nn\\
  &=&  C_m \left(r\LUV\right)^{1+\gamma_*} + C_\s \left(r\LUV\right)^{3-\gamma_*} \ ,\qquad  \left(\frac{1}{\LUV}\ll r\ll \frac{1}{\Lt}\right) \ .
\eea
In the deep IR, i.e., for $r\gg 1/\Lt$, the tachyon and the dilaton have calculable IR asymptotics which depend on the details of the action~\cite{Arean:2013tja}. These asymptotics shall not be needed here.

\subsection{Scaling results}

The dependence of the various scales on the quark mass can now be found by requiring continuity (and the continuity of the derivative) of the tachyon solution. More precisely, one should require that the dominant and subdominant tachyon solutions are both continuous, but in the cases studied here this is equal to requiring that continuity of the derivatives.

\subsubsection{QCD regime}

The QCD regime is defined by $0<x<x_c$ and $x_c -x \gtrsim 1$. Let us first consider the case of small mass, $m_q/\LUV \ll 1$. In this case there is (almost) spontaneous chiral symmetry breaking, which in V-QCD means that the (vev term of the) tachyon grows large at some value of the radial coordinate (energy scale) and triggers a nontrivial IR geometry~\cite{jk}. The fields obey their IR asymptotics exactly when the tachyon is large. Therefore both the UV and IR scales are comparable to the scale of the tachyon, $\LUV\sim \LIR \sim \Lt$, and $\sigma \sim \LUV^3$.  In the dynamic AdS/QCD models~\cite{Alvares:2012kr} similar results are obtained by introducing an IR cutoff where the tachyon grows large.  From~\eqref{tauUV} one indeed sees that the mass term of the tachyon is suppressed with respect to the vev term for $r \ll 1/\LUV$ which implies that $m_q$ can be treated as small perturbation.

When the quark mass is large, $m_q/\LUV \gg 1$, the tachyon grows large at small $r\sim 1/m_q$ and therefore $\Lt \sim m_q$. The RG flow of coupling is determined by the QCD beta function for $r \ll 1/\Lt$ and by the YM beta function for $r \gg 1/\Lt$ as the growing tachyon decouples the flavor sector. Explicitly\footnote{Precise definitions of the scales in the UV asymptotics of $\l$ would require including the higher order terms, but this does not affect the results.},
\bea 
 \l &\simeq& -\frac{1}{b_0\log(r \LUV)} \ , \qquad \left(r \ll \frac{1}{\Lt}\right) \\
 \label{YMflow}
 \l &\simeq& -\frac{1}{b_0^\mathrm{YM}\log (r\LIR)} \ , \qquad \left(\frac{1}{\Lt} \ll r \ll \frac{1}{\LIR}\right) \ . 
\eea
Since $b_0/b_0^\mathrm{YM}<1$, requiring continuity leads to the counterintuitive result that the scale of the IR expansions is larger than that of the UV expansions:
\be \label{largemqscaling}
 \frac{\LUV}{\LIR} \sim \left(\frac{m_q}{\LUV}\right)^{b_0/b_0^\mathrm{YM}-1}\ , \qquad  \frac{m_q}{\LIR} \sim \left(\frac{m_q}{\LUV}\right)^{b_0/b_0^\mathrm{YM}} \ .
\ee
This is however consistent with the fact that decoupling the flavor degrees of freedom makes $\l$ to run faster than it does at small $m_q$.

\subsubsection{Walking regime}

The walking regime is defined by $x<x_c$ and $x_c -x \ll 1$. For very small quark mass it is expected that the scaling is the same as for $m_q=0$ and that $\Lt \sim \LIR$ -- the IR geometry is linked to the growth of the tachyon as in the QCD regime. Requiring continuity\footnote{There is also the continuity of the derivative involved -- this fixes the phase of the slow oscillations of the tachyon. Taking this into account one can derive the Miransky scaling of~\eqref{Mscal} as detailed in~\cite{jk} and also here in Sec.~\ref{sec:condensate}.} with generic IR boundary conditions at $r \sim 1/\LIR$ gives $C_w \sim \LIR^2/\LUV^2$ in~\eqref{tauw}. Continuity of~\eqref{tauUV} and~\eqref{tauw} at $r \sim 1/\LUV$ further leads to $\sigma \sim  \LIR^2\LUV$, and from~\eqref{tauUV} it is seen that the quark mass term is small if
\be \label{walkingmqlim}
 \frac{m_q}{\LUV} \ll \frac{\LIR^2}{\LUV^2} \sim \exp\left[-\frac{2K}{\sqrt{x_c-x}}\right] \ ,
\ee
where we used~\eqref{Mscal}.

When
\be
\exp\left[-\frac{2K}{\sqrt{x_c-x}}\right]
\ll \frac{m_q}{\LUV} \ll 1\ ,
\ee
the amount of walking is controlled by the quark mass. We still require that the IR geometry is exactly where the tachyon is large so that $\Lt\sim \LIR$ and $C_w \sim  \LIR^2/\LUV^2$, but now continuity of~\eqref{tauUV} and~\eqref{tauw} at $r\sim 1/\LUV$ yields
\be \label{walkingscaling}
 \frac{m_q}{\LUV} \sim \frac{\LIR^2}{\LUV^2} \sim \frac{\sigma}{\LUV^3}
\ee
and there is no Miransky scaling.

When $m_q/\LUV \gg 1$, the coupling never flows close to the fixed point and the walking behavior is therefore absent. One finds the same results as in the QCD regime above.

\subsubsection{Conformal window}

Conformal window is the regime with $x_c<x<\xBZ$. Let us first assume that we are not in the perturbative BZ regime, $\xBZ-x\gtrsim 1$. When the quark mass is small, $m_q/\LUV \ll 1$, one again requires that $\Lt \sim \LIR$ and that there is walking. Continuity at $r \sim 1/\LIR$ gives 
\be
 C_m \left(\frac{\LIR}{\LUV}\right)^{-\Delta_*} \sim C_\s \left(\frac{\LIR}{\LUV}\right)^{\Delta_*-4} \sim 1
\ee
for the coefficients of~\eqref{tauIRFP}. Matching the leading and subleading terms of~\eqref{tauUV} and~\eqref{tauIRFP} at $r\sim 1/\LUV$ one further obtains
\be \label{CWscaling}
 \frac{m_q}{\LUV} \sim \left(\frac{\LIR}{\LUV}\right)^{\Delta_*} \ , \qquad \frac{\sigma}{\LUV^3} \sim \left(\frac{\LIR}{\LUV}\right)^{4-\Delta_*} \sim \left(\frac{m_q}{\LUV}\right)^{\frac{4-\Delta_*}{\Delta_*}} \ .
\ee

When $m_q/\LUV \gg 1$, the flow does not become close to the fixed point, and the results are the same as in the QCD and walking regimes.

\subsubsection{BZ regime}

Let us then discuss the BZ regime ($x<\xBZ$ and $\xBZ-x \ll 1$). 
One might expect that $\Lt \sim m_q$ independently of the value of $m_q$ since $\Delta_*$ approaches one in this limit. We have, however, defined $m_q$ asymptotically in the UV, and the UV RG flow of the quark mass is singular in the BZ limit. Therefore $m_q$ and $\Lt$ are not simply related as $x \to \xBZ$ -- see the top-left plot of Fig.~\ref{fig:Escales}, where $\Lt$ increases with $x$ in the BZ region instead of approaching $m_q$. Therefore, we will use the scale $\Lt$ in the analysis below.

When the quark mass is small, the coupling flows toward the BZ fixed point when $r \ll 1/\Lt$, and like in YM when $r \gg 1/\Lt$. Since $\l_*$ is small, one can use~\eqref{YMflow} to describe the YM flow. Requiring the flow to start at $\l_*$ when $r\sim 1/\Lt$ gives the exponential scaling
\be \label{LIRmqBZ}
 \frac{\LIR}{\Lt} \sim \exp\left(-\frac{1}{b_0^\mathrm{YM}\l_*}\right) \ .
\ee

In the opposite limit, i.e., large quark mass, the reasoning leading to~\eqref{largemqscaling} applies, if one replaces $m_q$ by $\Lt$ in the formulas. There is, however, also a subtlety in the definition of $\LUV$. In~\cite{jk}, $\LUV$ was defined essentially as the scale where $b_0\l$ becomes \order{1}. However in the BZ regime the coupling reaches the fixed point well before reaching the value $1/b_0 \sim 1/(\xBZ-x)$. Consequently, $\LUV$ is exponentially suppressed with respect to the true characteristic scale of the RG flow in the UV. Let us instead denote by $\widetilde \L_\mathrm{UV}$ the scale defined as~\cite{Kiritsis:2013xj} 
(roughly corresponding to the scale where $\l/b_0$ becomes \order{1}):
\be
 \frac{b_0}{b_1 \l} + \log\left(\frac{b_0}{b_1 \l}-1\right) = -\frac{b_0^2}{b_1} \log\left(r \widetilde \L_\mathrm{UV}\right) \ . 
\ee
This formula is the RG flow given by the two-loop BZ beta function, and it is reproduced in V-QCD up to correction suppressed by $\xBZ-x$ (for the whole flow when $\t=0$ and in the UV for all backgrounds), whereas the geometry is AdS, $A \simeq -\log(r/\ell)$, up to highly suppressed \order{(\xBZ-x)^4} corrections.
Therefore~\eqref{largemqscaling} is a better estimate after the replacements $\LUV \to \widetilde \L_\mathrm{UV}$ and $m_q \to \Lt$. 

In order to take into account the extremely slow RG flow in the BZ region, the conditions for the validity of~\eqref{LIRmqBZ} and~\eqref{largemqscaling} should actually be written as
\be
 \left(\frac{m_q}{\widetilde \L_\mathrm{UV}}\right)^\frac{b_0^2}{b_1} \ll 1 \quad \mathrm{and} \quad  \left(\frac{m_q}{\widetilde \L_\mathrm{UV}}\right)^\frac{b_0^2}{b_1} \gg 1 \ ,
\ee
respectively. Here $b_1$ is the NLO coefficient of the QCD beta function and $b_0^2/b_1 \sim (\xBZ-x)^2$.
Notice that the slowness of the RG flow is already visible at not so large values of $x$ (as demonstrated by the plots in the text, see, e.g., Fig.~\ref{fig:massdep}), because we use parameters $m_q$ and $\LUV$ defined asymptotically in the UV. Indeed the equations~\eqref{massdepCW}, and~\eqref{largemqscaling} depend on the ratio $m_q/\LUV$ through combinations of the type $\left(m_q/\LUV\right)^{\morder{\xBZ-x}}$.

\section{Schr\"odinger potentials and mass scales} \label{app:masses}

Let us then analyse the behavior of mass gaps and mass splittings (separation of the lowest bound state masses) in V-QCD. For the flavor nonsinglet fluctuations this is rather straightforward, as the fluctuation equations can be transformed into the Schr\"odinger form. The singlet fluctuations are more involved, because there is in general nontrivial mixing between the meson and glueball states. Only in the probe limit $x \to 0$ and in the limit of large quark mass $m_q \to \infty$ the glueballs and mesons are decoupled. We will restrict here to the nonsinglet states and study the scalar singlet states only numerically in Sec.~\ref{sec:singlets}.

For every flavor nonsinglet sector (vectors, axial vectors, pseudoscalars, and scalars) the 
fluctuation equation can be written as
\be \label{Schrode}
 -\phi''(u) + V_S(u) \phi(u) = m^2 \phi(u) \ ,
\ee
where $V_S(u)$ is the Schr\"odinger potential, and $m^2$ is the mass of the fluctuation. This form is obtained after a coordinate transformation defined by
\be \label{udef}
 \frac{du}{dr} = \sqrt{1+ e^{-2A} \kappa \t'^2} \equiv G 
\ee
and $u(r=0)=0$. The Schr\"odinger potential is given by 
\be \label{VSdef}
 V_S(u) = \frac{\Xi''(u)}{\Xi(u)} + H(u) \ ,
\ee
where $\Xi(u)$ and $H(u)$ are different functions for each sector and can be explicitly expressed in terms of the potentials (see Appendix~A in~\cite{Arean:2013tja}). For example, in the vector sector we find that
\be
 \Xi_V  = \sqrt{V_f} w e^{A/2}\ , \qquad H_V = 0 \ ,
\ee
and for the axial vectors $\Xi_A=\Xi_V$ but 
\be \label{HAdef}
 H_A = \frac{e^{2 A}\tau^2 \kappa}{w^2} \ .
\ee

The vector/axial decay constants are given by (see Appendix~F in~\cite{Arean:2013tja})
\be \label{Fdeforig}
 f_n^2 = M^3 N_c N_f \left.\frac{\Xi^4(u)\left[\partial_u(\phi(u)/\Xi(u)) \right]^2}{m_n^2}\right|_{u=\eps} \ .
\ee
Notice that by using~\eqref{VSdef} the Schr\"odinger equation~\eqref{Schrode} may be written as
\be
 -\partial_u\left[\Xi^2(u)\partial_u\left(\frac{\phi_n(u)}{\Xi(u)}\right)\right] = \left[m_n^2 -H(u)\right] \Xi(u)\phi_n(u) \ .
\ee
Integrating this over $u$ and inserting in~\eqref{Fdeforig}, we obtain
\be \label{Fnint}
 f_n^2 = \frac{M^3 N_c N_f}{m_n^2} \left\{\int_0^\infty du \left[m_n^2 -H(u)\right] \Xi(u) \phi_n(u) \right\}^2 \ .
\ee

\subsection{Small quark mass}

Let us then analyze the dependence of the bound state masses on the quark mass in the three regimes of Fig.~\ref{fig:scalingregions}. In regime~A, the quark mass is a small perturbation and the results from~\cite{Arean:2013tja} can be used directly. The potential $V_S$ can be computed  both when $r \ll 1/\LUV$ and $r \gg 1/\LIR$ by using the asymptotic expansions of the background. The results in the UV are independent of the potentials of the V-QCD action:
\be \label{VSUV}
 V_S \simeq \frac{v_\mathrm{UV}}{u^2}\ , \qquad \left(u  \ll \frac{1}{\LUV}\right) \ ,
\ee
where $v_\mathrm{UV} = -1/4$ for the pseudoscalars and $3/4$ for other sectors, and we used the fact that $u \simeq r$ in the UV region.  

In the IR the asymptotics of the tachyon, and consequently the coordinate dependence of the Schr\"odinger potential, strongly on the choice for the potential functions in the V-QCD action. All regular choices considered in Appendices~D and~E of~\cite{Arean:2013tja} lead to a confining potential, which grows as a function of $u$ in the IR regime.

If $\LUV \sim \LIR$, as is the case at small $x$ and $m_q$, we immediately notice that the Schr\"odinger potential has its bottom for all sectors expect for pseudoscalars at $r \sim 1/\LUV \sim 1/\LIR$. The value of the Schr\"odinger potential is \order{\LIR^2} near the bottom, as can be estimated from~\eqref{VSUV} above by requiring continuity at $u \sim 1/\LIR$, and therefore the mass gap is $\sim \LIR$. Similarly it can be seen that the mass splittings and vector/axial decay constants are \order{\LIR}. This is rather evident as only one energy scale enters the definitions above, both the tachyon and $\l$ are \order{1} when $r \sim 1/\LIR$, and no cancellations are expected in the formulas. 

The pseudoscalar sector is special because of the pion modes which obey the GOR relation, as shown in Sec.~\ref{sec:gmor}. The excited pseudoscalar states appear at mass scale $\LIR$, as can be seen numerically.

In order to study regime~B and the remaining part of regime~A (near $x=x_c$ where walking is seen) we need to check the Schr\"odinger potential for $1/\LUV \ll r \ll 1/\LIR$, i.e., in the near-conformal region. The tachyon remains small also here and $r \simeq u$. The Schr\"odinger potentials have been derived in~\cite{Arean:2013tja} and read
\be \label{VSw}
 V_S \simeq \frac{v_w}{u^2}\ , \qquad \left(\frac{1}{\LUV} \ll u \ll \frac{1}{\LIR}\right) \ ,
\ee
where $v_w=3/4$ for vectors and axials, and depends on the anomalous dimension $\Delta_*$ at the (approximate) fixed point for scalars and pseudoscalars. For scalars $v_w<-1/4$ when $x<x_c$, whereas it is positive for pseudoscalars. Therefore the coefficient in the scalar potential is critical~\cite{son}, and might potentially lead to an instability or a light state (see Sec.~5.3 of~\cite{Arean:2013tja}). Numerically it is seen, however, that this is not the case and the spectrum of scalars is not qualitatively different from that of the vectors, for example.

We see from~\eqref{VSw} that $V_S$ has similar dependence when the coupling constant walks as in the UV region in~\eqref{VSUV}, for the vector and axial states. Therefore the bottom of the Schr\"odinger potential is at $u \sim 1/\LIR$. Similar arguments as above show that the mass gap, mass splitting, and decay constants are \order{\LIR}. As we pointed out above, for the scalars and pseudoscalars more careful or numerical analysis is needed. The numerical result is that the masses are similarly \order{\LIR}, with the exception of pion masses in regime~A, which obey the GOR relation. In regime~B, pions also have masses \order{\LIR}.

In conclusion, mass gaps, mass splittings, and decays constants are \order{\LIR} at small quark mass, with the sole exception of the pions in regime~A.

\subsection{Large quark mass} \label{app:largemq}

Next we shall study the dependence on the (flavor nonsinglet) meson mass spectrum on the quark mass in the regime C ($m_q/\LUV \gg 1$). Recall that in this limit some features of the bound state can be analyzed starting from field theory, because the low end of the spectrum becomes nonrelativistic. The  expected mass gap is roughly equal to $2 m_q$, and the states can be studied by using the Schr\"odinger equation. For QCD states at large mass, one expects to have two main contributions in the Schr\"odinger potential. First, at very short distances one has Coulomb potential as perturbative gluon exchange dominates. Second, there is a nonperturbative confining potential which is expected to be linear in the distance and arise from a flux tube  between the quarks. While it is difficult to derive such a linear potential from first principles, it is consistent with the observed quarkonium spectra, lattice simulations, and also found in holographic calculations.

In the limit $m_q \to \infty$, the lowest states are therefore governed by the Coulomb potential, and one finds typical Hydrogen-like spectrum with negative binding energies. For slightly higher states, the linear potential dominates. Assuming precisely linear confining potential $\sim \LIR^2 r$, where $\LIR$ is the scale of glueballs in QCD, it is straightforward to solve the Schr\"odinger equation. The masses of the states obey roughly the scaling law
\be \label{NRscaling}
 m_n -2 m_q \sim \left(\frac{\LIR^4}{m_q} \right)^{1/3} n^{2/3} \ .
\ee
For charmonium and bottomonium the observed states are in the region where both the Coulomb exchange and nonperturbative effects are important (so that~\eqref{NRscaling}, for example, is not a good approximation) and have positive binding energies.

The behavior at large quark mass in V-QCD is somewhat dependent on the choice of the potentials. As pointed out above in~\eqref{largemqscaling}, the scales $m_q$ and $\LIR$ become separated as $m_q \to \infty$. For $r \ll 1/m_q$ one obtains the usual UV asymptotic solution, and for $r \gg 1/\LIR$ the background obeys the IR asymptotics, leading to the confining Schr\"odinger potential. In the middle, however, the behavior of the Schr\"odinger potential is nontrivial. To compute it, we first need to solve the tachyon from its EoM. 

\subsubsection{The tachyon and the mass gap}

The tachyon background EoM may be written as~\cite{jk,Arean:2013tja}
\begin{align} \label{taueom}
 \tau'' + \left[3 A' + \l' \frac{\partial}{\partial \l} \log (V_f \kappa)\right] \tau' &+ e^{-2A} \kappa \left[4 A' + \l' \frac{\partial}{\partial \l} \log (V_f \sqrt{\kappa})\right] (\tau')^3  \nonumber \\
 - \frac{1+e^{-2A} \kappa (\tau')^2}{e^{-2A} \kappa} & \frac{\partial}{\partial \tau} \log V_f = 0 \ .
\end{align}

When $1/m_q \ll r \ll 1/\LIR$, the tachyon has already grown large, and decoupled the quarks from the glue. The dilaton is still small, and its evolution is governed by the YM RG flow, as the quarks are decoupled. More precisely, in this regime
\be \label{YMRG}
 A = - \log r + \log \ell_0 +\morder{\frac{1}{\log(r\LIR)}} \ , \quad \l = -\frac{1}{b_0^\mathrm{YM} \log(r\LIR) } +\morder{\frac{1}{\log(r\LIR)^2}} \ ,
\ee
where $\ell_0 = \ell(x=0)$ is the UV AdS radius for YM, and $b_0^\mathrm{YM}$ is the leading coefficient of the YM $\beta$-function.
The appearance of the IR scale $\LIR$ in these expressions, which are the UV expansions for YM theory, may be surprising. In fact, it would perhaps be more appropriate to denote the scale of the expansions by a new quantity $\LUV^\mathrm{YM}$ (which is not the same as $\LUV$, the scale of the UV expansions for $r \ll 1/m_q$). But YM has only one energy scale, which is therefore the only scale in the model smaller than $m_q$ thanks to the decoupling of the quarks. That is, $\LUV^\mathrm{YM} \sim \LIR$, and for simplicity we have already used this in~\eqref{YMRG}. 

Let us then insert the expansions in~\eqref{taueom} in order to solve for the tachyon. We keep the leading behavior of all coefficients in $1/\log(r\LIR)$. In particular, since $\l'/A' = \morder{\log(r\LIR)^{-2}}$, the terms involving $\l'$ will be dropped. As we shall see below, this works at least for $r$ close to the lower end of the scaling region. We also expect that the tachyon grows large, and therefore $1 \ll e^{-2A} \kappa (\tau')^2$. The last term on the first line of~\eqref{taueom} and the term on the second line dominate. Consequently the tachyon satisfies
\be \label{taulargemq}
  4 \frac{r}{\ell_0^2} \kappa_0  \tau' + \frac{\partial}{\partial \tau} \log V_f \simeq 0 \ ,
\ee
where we used the expansions~\eqref{YMRG}, and $\kappa_0 = \kappa(\l=0)$. 

Let us study the asymptotics
\be
 \log V_f \simeq - a_0 \tau^{v_p} \ , \qquad (\tau \to \infty)
\ee
where we require $v_p>1$ in order to ensure that the tachyon grows fast enough\footnote{The analysis can be extended to $0<v_p \le 1$ where some assumptions made above fail and a more careful analysis is needed. The tachyon still grows only logarithmically and the conclusions are similar to the case $1<v_p<2$.}.
Then the ``asymptotic'' solution of~\eqref{taulargemq} reads\footnote{Taking account the logarithmic flow of~\eqref{YMRG} would result in terms in~\eqref{taupowerlargemq} and~\eqref{tauloglargemq} which have subleading (logarithmic) $r$ dependence but are leading with respect to the constant $\t_0$. For simplicity, we have omitted such corrections, as they do not affect our conclusions. The same applies to the formulas~\eqref{tauexppowerlargemq} and~\eqref{tauexploglargemq} below. }
\begin{align} 
\label{taupowerlargemq}
 \tau &\simeq \tau_0\ (rm_q)^\xi \ ,& \qquad &(v_p=2) \\
\label{tauloglargemq}
 \tau &\simeq \left[\xi(2-v_p)\log (rm_q) +\tau_0\right]^\frac{1}{2-v_p} \ ,& \qquad &(1<v_p<2) 
\end{align}
where
\be 
 \xi = \frac{v_p a_0 \ell_0^2}{4 \kappa_0} \ .
\ee
For $v_p>2$ there are no regular solutions. 

Let us then compute the Schr\"odinger potentials. First, the Schr\"odinger coordinate is given by
\be\label{dudrlargemq}
 \frac{du}{dr} \simeq e^{-A}\sqrt{\kappa}\tau' \simeq \frac{r}{\ell_0} \sqrt{\kappa_0} \tau' \ .
\ee  
The behavior of $\Xi$ is dominated by its dependence on the tachyon in all sectors (flavor nonsinglet vectors, axials, scalars, and pseudoscalars):
\be
 \log \Xi \simeq \mp \frac 12 a_0 \tau^{v_p} \ .
\ee
Since $\log \Xi$ increases fast enough with $\t$, the first term in~\eqref{VSdef} can be approximated by
\be
 \frac{\Xi''(u)}{\Xi(u)} =\left[\frac{d}{du}\log \Xi(u)\right]^2 +\frac{d^2}{du^2}\log \Xi(u) \simeq \left[\frac{d}{du}\log \Xi(u)\right]^2 \ .
\ee
Combining these we obtain
\be \label{Xireslargemq}
 \frac{\Xi''(u)}{\Xi(u)} \simeq \left(\frac{dr}{du}\right)^2 \left(\tau'\right)^2 \left[\frac{d}{d\tau} \log \Xi\right]^2 \simeq \frac{\ell_0^2 a_0^2v_p^2}{4 \kappa_0} \frac{\tau^{2v_p-2}}{r^2} \ .
\ee
This result can be expressed in terms of the Schr\"odinger coordinate by using~\eqref{taupowerlargemq}, \eqref{tauloglargemq} and~\eqref{dudrlargemq}, but the above form turns out to be more useful.
The second term of~\eqref{VSdef} is negligible for the vectors and scalars, but important for the axials (and pseudoscalars), for which we find
\be \label{Hreslargemq}
 H_A \simeq \frac{\ell_0^2 \kappa_0}{w_0^2}\frac{\tau^2}{r^2} \ ,
\ee
where $w_0 =w(\l=0)$.

Let us first assume that the results~\eqref{taupowerlargemq} and~\eqref{tauloglargemq} hold in the whole regime $1/m_q \ll r \ll 1/\LIR$. We will later discuss when this is not the case. For $1<v_p<2$ we then obtain that up to logarithmic corrections, $V_S \sim 1/r^2$. The potential decreases with $r$, and reaches its bottom at $r\sim 1/\LIR$, where $V_S \sim \LIR^2$. Therefore the meson mass gaps are characterized by $\LIR$, and only logarithmically enhanced with increasing $m_q$.

For $v_p=2$, the estimates~\eqref{Xireslargemq} and~\eqref{Hreslargemq} match up to the multiplicative coefficient. The Schr\"odinger potentials behave as 
\be
 V_S \sim \frac{(r m_q)^{2\xi}}{r^2}\ , \qquad \left(\frac{1}{m_q} \ll r \ll \frac{1}{\LIR}\right) \ .
\ee
If $\xi<1$, the result decreases with $r$. The bottom of the potential is reached at $r\sim 1/\LIR$, which leads to the mass gap
\be
 m_\mathrm{gap} \sim \LIR \left(\frac{m_q}{\LIR}\right)^{\xi}
\ee
If $\xi>1$ the result increases with $r$. The bottom of the potential is therefore at $r \sim 1/m_q$, and
\be
 m_\mathrm{gap} \sim m_q \ .
\ee
In the marginal case $\xi=1$ the Schr\"odinger potential is flat and \order{m_q^2} in the whole regime. The mass gap is therefore also
\be
 m_\mathrm{gap} \sim m_q \ .
\ee

In conclusion, the mas gap has the power law expected for nonrelativistic states when $\xi \ge 1$. However, as we shall see below, for $\xi>1$ the mass splitting is \order{m_q}, i.e., larger than that of nonrelativistic bound states. Therefore only the marginal case $\xi=1$ may potentially reproduce both realistic mass gap and splitting.

Let us then discuss logarithmic corrections to the critical case $v_p=2$. That is, we assume 
\be
 \log V_f \simeq -a_0 \tau^ 2 (\log \tau)^{v_\ell} 
\ee
as $\tau \to \infty$. In this case, the tachyon solution is given by
\begin{align} 
\label{tauexppowerlargemq}
 \tau &\simeq \exp\left[\tau_0\ (r m_q)^\xi\right] \ ,& \qquad &(v_\ell=1) \\
\label{tauexploglargemq}
 \tau &\simeq \exp\left\{\left[\xi(1-v_\ell)\log (rm_q) +\tau_0\right]^\frac{1}{1-v_\ell}\right\} \ ,& \qquad &(v_\ell<1) 
\end{align}
and no regular solution is found when $v_\ell>1$. The leading term of the Schr\"odinger potential is now
\be
 \frac{\Xi''(u)}{\Xi(u)} \simeq \frac{\ell_0^2 a_0^2v_p^2}{4 \kappa_0} \frac{\tau^{2} (\log \tau)^{2 v_\ell}}{r^2} \ .
\ee
By similar analysis as above, the bottom of the Schr\"odinger potentials lies at $r \sim 1/\LIR$ and the mass gap is \order{\LIR} when $v_\ell<0$, up to corrections which grow slower than any power of $m_q$ as $m_q \to \infty$. When $0<v_\ell \le 1$, the bottom of the the potential is at $r \sim 1/m_q$, and the mass gap is \order{m_q}.

Recall also that, as we pointed out above, in some cases the above analysis is not valid in the whole regime $1/m_q \ll r \ll 1/\LIR$. This can happen if the terms involving $\l'(r)$ in~\eqref{taueom} start to dominate as $r$ approaches the value $1/\LIR$. The growth of these terms requires that the factors in the square brackets depend on the tachyon, because $\l$ is known to obey the RG flow of YM in this regime, and remains small -- actually $r\sim 1/\LIR$ is exactly the region, where $\l$ finally reaches values \order{1}. Such a dependence on the tachyon may arise from the logarithmic derivative $\frac{\partial}{\partial \l} \log V_f$. The most natural Ansatz which leads to this is an exponential 
\be
 V_f \propto \exp(-a(\l) \tau^{v_p})
\ee
where the crucial point is that the factor $a(\l)$ depends on $\l$. For the sake of generality, we however assume that
\be \label{dlogVfas}
 \frac{\partial}{\partial \l} \log V_f \sim  - a_1 \tau^{d_p} 
\ee
as $\tau \to \infty$, where $d_p$ may differ from $v_p$.

From~\eqref{taueom} we see that the terms which were neglected above become important when
\be \label{changecondorig}
 A' \sim \l' \frac{\partial}{\partial \l} \log V_f \ .
\ee
Inserting here~\eqref{YMRG} and~\eqref{dlogVfas}, we obtain the condition
\be \label{changecond}
 \left[\log(r\LIR)\right]^2 \sim \tau^{d_p} \ .
\ee
If $d_p>0$ and for any growing tachyon solution this condition is saturated within the range $1/m_q \ll r \ll 1/\LIR$, because at $r \sim 1/\LIR$ the left hand side is \order{1} and the tachyon is already $\gg 1$ (and for $r\sim 1/m_q$, the left hand side is sizeable whereas the tachyon is, by definition of $m_q$, \order{1}).

For the sake of concreteness, let us consider the case $v_p=2$ so that the tachyon has the $r$-dependence of~\eqref{taupowerlargemq} when the additional terms $\propto \l'$ are still small. Then~\eqref{changecond} is saturated for
\be
 r \sim \frac{1}{m_q} \left[\log \frac{m_q}{\LIR}\right]^\frac{2}{\xi d_p} \equiv r_c \ .
\ee
That is, the tachyon behaves as $r^\xi$ only up to $r \sim r_c$, which is only enhanced with respect to $r \sim 1/m_q$ by a logarithmic term. For $r_c \ll r \ll 1/\LIR$ (i.e., for almost the whole interval $1/m_q \ll r \ll 1/\LIR$) we need to solve the tachyon EoM with different dominant terms. Assuming that the right hand side in~\eqref{changecondorig} dominates over the left hand side, we obtain
\be
 e^{-2A} \kappa \l' \t'\ \frac{\partial}{\partial \l} \log V_f  \simeq \frac{\partial}{\partial\t} \log V_f 
\ee   
which becomes for the current case
\be
 \t' \t^{d_p-1} \simeq \frac{4 \xi b_0^\mathrm{YM}}{r a_1} \left[\log(r\LIR)\right]^2 \ .
\ee
This equation is solved by
\be
 \t \simeq \left[\frac{4 d_p\xi b_0^\mathrm{YM}}{3 a_1}\left[\log(r\LIR)\right]^3 +\t_0 \right]^\frac{1}{d_p}
\ee
so that the tachyon only increases logarithmically for $r_c \ll r \ll 1/\LIR$. Inserting this expression in~\eqref{Xireslargemq}, and recalling that $r_c$ is close to $1/m_q$, we see that the bottom of the Schr\"odinger potential is found at $r \sim 1/\LIR$, and the mass gap is again \order{\LIR} up to logarithmic corrections. 

In conclusion, the phenomenologically interesting large mass gap (with power law dependence on $m_q$) is obtained only when $v_p=2$, $0 \le v_\ell \le 1$, and when $\frac{\partial}{\partial \l} \log V_f$ is suppressed (e.g., $d_p \le 0$). The last requirement means that for the critical exponential asymptotics $\log V_f \sim - a \tau^2$, the factor $a$ cannot\footnote{More precisely, only dependence which is highly suppressed in the UV is allowed, for example terms $\sim \exp(-\#/\l)$.} depend on $\l$.

Interestingly we notice that potentials I, the construction of which was motivated by using completely independent arguments in~\cite{jk,Arean:2013tja}, do produce a large mass gap. In this case we have 
\be
 V_f(\l,\t) = V_{f0}(\l) \exp(-a_0 \t^2)
\ee
where $a_0$ is indeed independent of $\l$. The value of $a_0$ was fixed in the UV (where the tachyon is small) to reproduce the UV dimension of the chiral condensate and quark mass so that
\be
 \xi = \frac{v_p a_0 \ell_0^2}{4 \kappa_0} = \frac{3 \ell_0^2}{4 \ell^2} \ .
\ee
The value of $\xi$ is therefore slightly below the critical value $\xi = 1$ (because also $\ell_0^2/\ell^2 < 1$). This is not surprising since holographic models are not expected to work perfectly in the UV region, where the coupling $\l$ is small, so fixing the parameters by UV arguments may lead to some tension in the model. Recall, for example, that the UV asymptotics of the pressure and the UV asymptotics of the correlators for the energy momentum tensor give slightly different numbers for the normalization of the glue action~\cite{Kajantie:2013gab}. 

For potentials II, we used instead $a(\l)$ with nontrivial dependence on $\l$, and consequently the meson mass gap will be small.

Finally, let us point out that in all cases classified above, $H_A$ of~\eqref{HAdef} contributes at leading order to the Schr\"odinger potentials near their bottoms. This is seen from the estimates~\eqref{Xireslargemq} and~\eqref{Hreslargemq} when the bottom is at $r \sim 1/\LIR$, and by inserting the UV expansions in~\eqref{HAdef} if the bottom is at $r \sim 1/m_q$. Consequently, axial vectors and pseudoscalar mesons have larger mass gaps than vectors and scalars: the ratio of, say, the axial and the vector mass gap approaches a finite number which is larger than one in the limit of large quark mass\footnote{When $1<v_p<2$, the term $H_A$ is actually leading by a logarithmic factor, so the mass gaps of the axials and pseudoscalars are logarithmically enhanced with respect to those of the vectors and the scalars.}.

\subsubsection{Mass splittings}

Let us then discuss the mass splittings between the lowest meson states for the various cases described above. When the Schr\"odinger potential $V_S(u)$ has a clear and regular minimum, the splittings can be estimated by computing the second derivative $V_S''(u_0)$ at the minimum $u=u_0$. Most of the potentials discussed above fall into two categories, where the Schr\"odinger potential has a clear minimum either at $r \sim 1/m_q$ or at $r \sim 1/\LIR$.

When the minimum is at $r \sim 1/m_q$, we found that the mass gap was \order{m_q}. From the definition~\eqref{udef} of the Schr\"odinger coordinate we see that $u \sim r$ and the scale of the derivatives of all relevant functions is given by $d/dr \sim m_q$. Therefore $m_q$ is the only scale that enters the analysis, and the splitting is also \order{m_q}, i.e., much larger than expected for nonrelativistic bound states. Notice that this also includes some of the asymptotics with critical $v_p=2$, i.e., those with $0<v_\ell \le 1$, and those with $v_\ell=0$ and $\xi>1$.

When the minimum is at $r \sim 1/\LIR$, the computation is more involved. We shall only discuss the case $v_p=2$ and $v_\ell=0$, for which $\tau \sim (r m_q)^\xi$ and the mass gap is large,
\be
V_S(u_0) \sim \left.\frac{\t^2}{r^2}\right|_{r\sim 1/\LIR} \sim \LIR^2 \left(\frac{m_q}{\LIR}\right)^{2 \xi} \ .
\ee 
Recall that the minimum is at $r \sim 1/\LIR$ when $0<\xi <1$. At the minimum, all fields ($\t$, $\l$, and $A$) have logarithmic or power-law dependence on the coordinate $r$ so the scale of the $r$-derivatives is $\LIR$. The relevant quantities are, however, the $u$-derivatives of the  potential, which can be estimated by using the chain rule and $du/dr \sim \t \sim (m_q/\LIR)^\xi$. We find that
\be \label{VSders}
 V_S''(u_0) \sim \LIR^4\ , \qquad V_S^{(n)}(u_0) \sim \LIR^{2+n}\left(\frac{m_q}{\LIR}\right)^{(2-n)\xi}
\ee
where $n>2$. The extent of the Schr\"odinger wave functions around the minimum is determined for the lowest fluctuation modes by $V_S''(u_0)$, and roughly given by $\Delta u = u - u_0 \sim 1/\LIR$. Outside this region, the wave functions vanish very fast. The higher order derivatives in~\eqref{VSders} vanish as $m_q \to \infty$, and therefore higher order terms in the Taylor expansion of the Schr\"odinger potential are suppressed (for $u - u_0 \sim 1/\LIR$) and the potential takes the Harmonic oscillator form. In particular, the masses are given by
\be
 m_n^2 = V_S(u_0) + \sqrt{2 V_S''(u_0)}\left(n+\frac 12 \right) + \morder{\frac{\LIR^{2+\xi}}{m_q^\xi}} \ ,
\ee
where $n=0,1,2,\ldots$. From here we see that the mass splitting is suppressed as $\sim \LIR \left(\LIR/m_q\right)^{\xi}$ at large $m_q$.

In the remaining case ($v_p=2$, $v_\ell=0$, and $\xi=1$) the bottom of the Schr\"odinger potential, as obtained from the leading tachyon solution, is flat for $1/m_q \ll r \ll 1/\LIR$, suggesting a relatively small level splitting. There are, however, subleading logarithmic corrections to the solution due to the YM RG flow, which we have not computed. They will be important, and are expected to cause a minimum either at $r \sim 1/m_q$ or at $r \sim 1/\LIR$. Then the mass splittings of the very lowest states will be as in one of the cases discussed above (with $\xi=1$ if the minimum is at $r \sim 1/\LIR$), up to logarithmic corrections, unless the subleading corrections cancel miraculously. 

\subsubsection{Decay constants}

Finally let us discuss the mass dependence of the (vector/axial) decay constants, given in~\eqref{Fnint}.
The Schr\"odinger wave function vanishes very rapidly whenever $m_n^2<V_S(u)$, which limits the integral to the ``classically allowed'' region. 

First we notice that when the Schr\"odinger has its bottom at $r\sim 1/\LIR$, the decay constants of the lowest states are very small, because $\Xi$ in the above integral formula contains the factor $\sqrt{V_f}$ which is exponentially suppressed because the tachyon is also large near $r \sim 1/\LIR$. For the interesting case of $v_p=2$, $v_\ell=0$, and $0<\xi<1$ we find that
\be \label{fnpotI}
 \frac{f_n^2}{N_fN_c} \sim \exp\left[-\#\left(\frac{m_q}{\LIR}\right)^{2\xi} \right] \ .
\ee
One can also show that the pion decay constant has similar dependence on $m_q$.
Recall that $H_V=0$ for the vectors and $H_A$ is given in~\eqref{HAdef}.
The low-lying states are therefore asymptotically decoupled. Sizeable decay constants are only found for highly excited states. More precisely, the suppression factor disappears when the classically allowed region extends to $r \sim 1/m_q$ where the tachyon is no longer small. Since $V_S \sim 1/u^2$ in the UV, this requires $m_n \sim m_q$ (whereas the lowest states had $m_n \sim \LIR (m_q/\LIR)^\xi$). Therefore the coupled states appear at the scale where mesons are expected in QCD. By using~\eqref{Fnint}, it is possible to show that for such states the decay constants are $f_n^2/(N_cN_f) \sim \LIR^2$.

When the Schr\"odinger potential has its bottom at $r \sim 1/m_q$, it is easy to see from the above expressions that
\be
 \frac{f_n^2}{N_cN_f} \sim m_q^2 
\ee
for the lowest states, because $m_q$ is the only scale which enters the formulas.

\section{Analysis of the free energy} \label{app:freeencoll}

\subsection{Chiral condensate as the derivative of free energy} \label{app:qbarqnorm}

The chiral condensate (including the sum over flavor) can be defined as\footnote{Recall that the quark mass is defined only up to a proportionality constant in the holographic model, which we have set to unity for simplicity. The inverse of this constant would appear in the definition of the chiral condensate.}
\be \label{condasderdef}
 \langle\bar q q\rangle = \frac{\partial \mathcal{E}}{\partial m_q} = \frac{1}{V_4}\frac{\partial S_{\mathrm{on-shell},\mathrm{E}}}{\partial m_q} \ ,
\ee
where $\mathcal{E}$ is the (zero temperature) energy density of QCD, and the subscript $\mathrm{E}$ denotes that Euclidean signature was used (so that the sign of the action is opposite with respect to the Minkowski signature, which was used in the main text).

By computing the renormalized on-shell action using the identity~\eqref{condasderdef} one can find the relation between the chiral condensate and the coefficient $\sigma$ in the vev term of the tachyon. This is, however, slightly complicated in V-QCD. First, since there is full backreaction: changing the value of the quark mass will affect the geometry, possibly leading to nontrivial contributions in the $m_q$ derivative of~\eqref{condasderdef}. Second, the counterterms needed to regularize the on-shell action depend on the quark mass, and may also contribute in the derivative. 
As it turns out, these issues can be fully solved in the limit of zero quark mass. 

Below we will first demonstrate how the free energy, and the backreaction in particular, can be analyzed by using the EoMs for the background. This will be compared to the direct computation of the regularized on-shell action done in Appendix~\ref{app:freeen}.
In the calculations below the fields shall be decomposed as
\be \label{UVdecomp}
 A = A_0+A_\t +A_1\,\qquad \l=\l_0+\l_\t+\l_1\, \qquad \t = \t_0+\t_1+\t_q
\ee
asymptotically in the UV (where $r \to 0$). Here the terms with subscript zero (one) are the source (vev) terms. The terms $A_\t$ and $\l_\t$ and are sourced by the leading quadratic terms in the tachyon in the UV ($\propto \t_0^2$) and are therefore $\propto m_q^2$. The term $\t_q$ is the leading nonlinear term of the tachyon. It is $\propto m_q^3$ and  computed in Appendix~\ref{app:nltachyon}. It is shown to be subleading with respect to the vev term $\t_1$ and therefore irrelevant in the calculations below. It is also argued that the same is true for similar terms due to quartic and higher order tachyon perturbation in the expansions of $A$ and $\l$. That is, the terms sourced by $\t_0 \t_q$ or $\t_0^4$ are subleading with respect to the vev terms $A_1$ and $\l_1$. The UV expansions of the various terms will be given in Appendix~\ref{app:freeen}.

\subsubsection{Chiral condensate at vanishing quark mass}

Let us start by computing the chiral condensate at zero quark mass. This is the simplest case because the square of the source term of the tachyon does not contribute in the on-shell action, and as we shall see, there is no issue with extracting value of $\sigma$ (the proportionality coefficient of the vev term) from the UV asymptotics.

Consider a perturbation of a generic background (around $m_q=0$) which keeps $\LUV$ (but not the quark mass) fixed. 
The variation of the background solution in the UV is found by using the UV expansions from Appendix~\ref{app:freeen}: 
\bea 
\label{Ainf}
 \d A(r) &\simeq& \d A_1(r) = \d \mG \frac{r^4}{\ell^3}\left[1+\morder{\frac{1}{\log r\Lambda}}\right] \\
\label{dlamq0}
  b_0\d  \l(r) &\simeq & b_0\d  \l_1(r) = \left[-\frac{45}{8} \d \mG -\frac{9}{32}B_\s  \sigma\, \d m_q  \right] \frac{r^4}{\ell^3}\left[1+\morder{\frac{1}{\log r\Lambda}}\right] \\
  \frac{\d \t(r)}{\ell} &=&  \d m_q\,  r  (-\log r\Lambda)^{-\rho}\left[1+\morder{\frac{1}{\log r\Lambda}}\right] \nonumber\\
 && +\  \d \sigma\,  r^3  (-\log r\Lambda)^{\rho}\left[1+\morder{\frac{1}{\log r\Lambda}}\right]\ ,
\eea
where
\be
 B_\s = x W_0 \kappa_0 \ell^5 \ .
\ee

The chiral condensate may then be computed by studying the variation of the on-shell action as seen from~\eqref{condasderdef}. The computation is analogous to the holographic derivation of the first law of thermodynamics at finite temperature and chemical potential: the differential of the free energy equals the variation of the action, and expressing this in terms of the various UV and IR boundary terms gives the terms in the desired expression. 
In order to be as precise as possible, we will formulate the computation in terms of a conserved infinitesimal current, which can also be easily used in the computations later. It can be found as follows.

The on-shell Lagrangian can be expressed as a total derivative~\cite{jk}:
\be \label{Lonshell}
  L_\mathrm{on-shell} \propto \frac{d}{dr}\left[-2  A' e^{3 A}\right]\ .
\ee
But any leading variation of a generic Lagrangian around its on-shell value is a total derivative as well, given formally by\footnote{In the case of the gravitational action there is a complication because $R$ contains second derivatives. It is, however, well known that the second derivatives can be isolated in another total derivative term, related to the Gibbons-Hawking boundary term.}
\be
 \frac{d}{dr}\left[\sum_i\frac{\pa L}{\pa \varphi_i'}\delta \varphi_i\right] \ ,
\ee
where the sum is over all fields in the Lagrangian. Requiring equality of the generic expression and the variation of~\eqref{Lonshell}, one identifies the following infinitesimal conserved ``current'':
\be \label{Jdef}
 J =  6 e^{3 A} A' \d A - 6 e^{3 A} \d A' -\frac{8 e^{3 A} \l' \d \l}{3 \l^2} - \frac{x e^{3 A} V_f \kappa \t'\d \t }{\sqrt{1+e^{-2 A} \kappa (\t')^2 }} \ .
\ee
Indeed it is straightforward to check that $J'=0$ by using the EoMs (and their variations).

We will then require that the variation of the background is regular in the IR. By using the IR expansions of the background~\cite{jk} in the expression~\eqref{Jdef} one sees that $J$ vanishes in the IR  ($r \to \infty$) so it must vanish everywhere for regular variations:
\be
 J = 0 \ .
\ee
In the UV  the above expansions are inserted, leading to
\be \label{mqvari}
0 = \lim_{r\to 0} J =  
- 15\,  \d \mG  -\frac{9}{4}B_\s \sigma \,  \d m_q \ .
\ee
The expression for the free energy density in terms of $\mathcal{G}$, $\sigma$, and $m_q$ can be extracted from the finite temperature computation in Appendix~\ref{app:freeen}
(see the expression for the free energy~\eqref{pmqne0}). 
At small quark mass and at $T=0$ (so that also $C=0$) one obtains the expression\footnote{We omitted contributions from the reference background used to regulate this expression, but as we shall argue below, this does not affect the result at $m_q=0$.}
\be \label{mErenorm}
 \mathcal{E} =M^3N_c^2 \left(15\mG + \frac{1}{4}B_\s m_q \sigma\right) \ .
\ee
Inserting here~\eqref{mqvari},
\be
  \d \mathcal{E} = -2 M^3N_c^2 B_\s \sigma \,  \d m_q \ .
\ee
Therefore the chiral condensate is given by 
\be
 \langle \bar q q\rangle = -2 M^3N_c^2 B_\s \sigma  \ .
\ee
The coefficient in this equation can be fixed by using the asymptotics of the scalar-scalar correlator (see Appendix~C in~\cite{Arean:2013tja}), which leads to
\be
 \langle \bar q q\rangle = -\frac{N_fN_c}{2\pi^2}\, \sigma\ .
\ee

\subsubsection{Chiral condensate at finite quark mass}

Let us then try to generalize the above computation to finite $m_q$. Recall that there is an issue in the definition of the vevs $\mG$ and $\sigma$. In principle, they are well defined as the coefficients of the vev terms $A_1$ and $\t_1$, respectively. However such definitions are useless in practice, because the source terms $A_0$ and $\tau_0$ cannot be solved perturbatively to high enough orders in order to separate them from the vev terms in the UV. 
In practice only differences of the vevs can be computed. Therefore we will define the values of $\mG$ and $\s$ with respect to some reference solutions.
Let us first discuss how we define a reference solution for all values of $m_q$.

First a (IR regular) solution with exactly zero tachyon (and therefore zero quark mass) is picked. This solution is chosen to have, by definition, $\mG=0$ (and trivially $\sigma=0$). Therefore $A_\t$ and $A_1$ in~\eqref{UVdecomp} are zero, and the solution for the scale factor $A$ defines the source term $A_0$. Let us call this solution~1. Then solution~2 is chosen which has finite $m_q$,  defines the nonnormalizable term $\t_0$ of the tachyon in the UV, and has $\sigma=0$ by definition. Further one requires that $\mG=0$ also for solution~2. Since $A_0$ was already defined, this choice also defines $A_\t$.  

For $0<x<x_c$, the solution~2 can be identified with the solution that was used to define the vanishing of $\sigma$ in the plots of Fig.~\ref{fig:spiral}. In principle any solution can be picked, but as we argued in Sec.~\ref{sec:condensate}, a choice which avoids fine tuning is to pick the solution~2 is the crossover between regimes~A and B when $0<x<x_c$ and a solution with tiny quark mass in the conformal window.

We have argued that ``nonlinear'' terms of the tachyon are suppressed with respect to the vev terms. Similarly, the terms sourced by quartic tachyons in the UV expansions for $A$ and $\l$ are suppressed with respect to the vev terms. Therefore fixing the values of $\mG$ and $\sigma$ for solutions~1 and 2 is enough to define the vevs for all backgrounds. It is also possible construct a reference background which has vanishing $\mG$ and $\sigma$ for arbitrary $m_q$, 
as a (generally not IR regular)  combination of the two IR regular solutions 1 and 2 constructed above by appropriately scaling the constructed $A_\t$ and $\t_0$ to have the desired quark mass. 

The above construction implies that the vevs can be written as
\be
 \mG = \hat \mG - \mG_1 - m_q^2 k_A \ , \qquad \sigma = \hat \sigma - m_q k_\t
\ee
where the hatted quantities are the exact coefficients of the vev terms, and the quantities without hat are given by the above subtraction procedure. The coefficients $k_i$ and $\mG_1$ are independent of $m_q$.

Let us then proceed with the calculation. Consider again a perturbation of a background with $\d m_q \ne 0$, 
but now at generic value of $m_q$. The UV expansion of the perturbation has additional terms, in particular 
\bea
 \d A &=&  -\frac{1}{18} B_\s m_q\, \d m_q \frac{r^2}{\ell^3} \left(-\log r\Lambda\right)^{-2\rho}\left[1+\morder{\frac{1}{\log r\Lambda}}\right] \nn\\
&&+\d \hat \mG \frac{r^4}{\ell^3}\left[1+\morder{\frac{1}{\log r\Lambda}}\right] \ ,
\eea
where we chose to use the hatted vevs.
The variation of $b_0 \l$ includes a term $\propto m_q\, dm_q$ which turns out to only contribute at subleading orders in the UV, and the term (compare to~\eqref{dlamq0})
\be
 b_0\delta \l_1 = \left[-\frac{45}{8} \d \hat \mG -\frac{9}{32}B_\s  \left(\hat \sigma\, \d m_q +m_q\,\d \hat \sigma\right) \right] \frac{r^4}{\ell^3}\left[1+\morder{\frac{1}{\log r\Lambda}}\right] \ .
\ee

It is again useful to study the current~\eqref{Jdef}. As above, $J=0$ because it vanishes in the IR limit. In the UV one obtains
\be \label{fin1stlaw}
0 = \lim_{r\to 0} J =  
- 15\,  \d \hat \mG  -\frac{1}{4}B_\s m_q \,  \d \hat \sigma -\frac{9}{4}B_\s \hat \sigma \,  \d m_q \ .
\ee
In particular, we expect that no terms $\propto m_q \d m_q$ arise in the UV because $\hat \sigma$ and $\hat \mG$ are exactly the coefficients of the vev terms. 
This identity may be rewritten as
\be
 \d\left[ 15 \hat \mG + \frac{1}{4}B_\s m_q \, \hat \sigma\right] = -2 B_\s \hat \sigma \,  \d m_q \ ,
\ee
where the expression in the square brackets has similar structure to the regularized vacuum energy given above in~\eqref{mErenorm}. The natural expectation is that indeed
\be \label{mErenorm2}
 \mathcal{E} =M^3N_c^2 \left(15\hat \mG + \frac{1}{4}B_\s m_q \hat \sigma\right) \ ,
\ee
but in principle the expression could also contain additional terms  $\propto m_q^2$ (or constants) which would cancel in the regularization. 

We see that
\be
 \d\mathcal{E} = -2 M^3 N_c^2 B_\s \hat \sigma \,  \d m_q = - 2  M^3 N_c^2 B_\s \left( \sigma + k_\t m_q\right) \, \d m_q \  ,
\ee
where extra terms $\propto m_q^2$ in the definition of $\mE$ would effectively change the value of $k_\t$ which remains unknown in any case\footnote{A possible handle to control the value of this coefficient is to fix somehow the subleading linear corrections to the GOR relation discussed in Appendix~\ref{app:gor}.}. Therefore the chiral condensate is given by
\be \label{qqbarmqfin}
 \langle \bar q q\rangle = -2 M^3N_c^2B_\s \hat \sigma=-2 M^3N_c^2B_\s \left(\sigma + k_\t m_q \right) \ .
\ee

\subsubsection{Free energy on the Efimov spiral} 

Let us then discuss how the free energy is computed for the configurations of Fig.~\ref{fig:spiral}, which are found when $0<x<x_c$. Let us assume for simplicity 
that we are able to define the vevs such that the constant $k_\t$ of~\eqref{qqbarmqfin} vanishes.
In this case the chiral condensate is simply given by the derivative of the free energy density with respect to the quark mass:
\be \label{qqbarderef}
  \langle \bar q q\rangle = \frac{\partial \mE}{\partial m_q} = -2 M^3N_c^2 B_\s \sigma \equiv - N_f N_c c_\sigma \sigma \ .
\ee
Inserting the asymptotic result of~\eqref{spiraleqs} in~\eqref{qqbarderef} and integrating, one can readily find the free energy for asymptotically small quark mass:
\bea
\label{freeenspiral}
 \frac{1}{N_fN_c\LUV^4}(\mE-\mE_0) 
&=& -\frac{c_\sigma K_\mathrm{IR}^2}{2 K_m K_\sigma \sin(\phi_m\!-\!\phi_\sigma)^2}\ e^{-4 \vs}\\\nn
&&\times \left[ -\sin\left(\phi_\mathrm{IR}\!-\!\phi_\sigma \!-\! k \vs\right) \sin\left(\phi_\mathrm{IR}\!-\!\phi_m \!-\! k \vs\right)  +\frac{\nu}{4}\sin(\phi_m\!-\!\phi_\sigma)\right] \\
\label{freeenspiral2}
&=& -\frac{c_\sigma}{2}\frac{m_q}{\LUV}\frac{\sigma}{\LUV^3} -\frac{\nu c_\sigma K_\mathrm{IR}^2}{8 K_m K_\sigma \sin(\phi_m\!-\!\phi_\sigma)}\ e^{-4 \vs}  \ ,
\eea
where $\mE_0$ is the free energy of the solution having $m_q=0=\sigma$, and $\vs = \log \LUV/\LIR$. In general, free energy differences are given by the area between the spiral and the horizontal axis (see, for example, Fig.~\ref{fig:doubletr}).

Interestingly, the result for both zero $m_q$ and zero $\sigma$ simplifies to 
\be
  \frac{1}{N_fN_c\LUV^4}(\mE-\mE_0) = -\frac{\n c_\sigma K_\mathrm{IR}^2}{8 K_m K_\sigma \sin(\phi_m\!-\!\phi_\sigma)}\ e^{-4 \vs} < 0 \ ,
\ee
where we used the handedness of the spiral in~\eqref{handedness}: $\sin(\phi_m-\phi_\sigma)>0$. Therefore these solutions are dominant over the solution with zero tachyon.

\subsubsection{Free energy with multi-trace deformations}

Let us then study how the free energy is computed in the presence of multi-trace deformations. Recall that the UV boundary conditions are
\bea
 \a_m &=& m_q + \sum_{n=2}^{n_\mathrm{max}} g_n c_\s^{n-1} \b_m^{n-1} \\
 \b_m &=& \s
\eea
and that the identity~\eqref{fin1stlaw} is interpreted as
\be
 0 = - 15\,  \d \mG  -\frac{1}{4}B_\s \a_m \,  \d \b_m -\frac{9}{4}B_\s \b_m \,  \d \a_m \ .
\ee
By using the UV boundary conditions, this identity can be rearranged to read
\be
 \d \left[\frac{15 M^3}{x} \mG + \frac{1}{8} c_\s \a_m \s +\sum_{n=2}^{n_\mathrm{max}}\left(\frac{n-1}{n} 
\right) g_n c_\s^n \s^n\right]
 = - c_\s \s \d m_q - \sum_{n=2}^{n_\mathrm{max}} \frac{1}{n} c_\s^n \s^n \d g_n \ .
\ee
Therefore the quantity in the square brackets is identified as the new free energy (over $N_f N_c$). It matches with the free energy obtained by renormalizing the action~\eqref{mErenorm} up to the last term involving the couplings $g_n$. This extra term is proportional to 
\be
 W - \b_m \frac{\delta W}{\delta \b_m(x)} \ ,
\ee
where $W$ is as in~\eqref{Wbulk}. Therefore the term agrees with that found in~\cite{Mueck:2002gm}.

The condensates are given by
\bea \label{condasder}
 \langle \mO \rangle &=& \frac{1}{N_fN_c} \frac{\pa \mE}{\pa m_q} = - c_\s \s \\
 \langle \mO^n \rangle &=& \frac{(-1)^n n}{N_fN_c} \frac{\pa \mE}{\pa m_q} = \left(- c_\s \s\right)^n \ ,
\eea
where the normalization factors were read from~\eqref{Wdef}. Therefore we find agreement with the large $N$ factorization of the expectation values.

The free energy on the Efimov spiral asymptotically close to the origin can be computed as above. That is, by inserting the spiral equations~\eqref{spiraleqsab} to~\eqref{condasder} and integrating, we obtain
\bea \label{spiralEmulti}
  \frac{1}{N_fN_c\LUV^4}(\mE-\mE_0) &=& -\frac{c_\sigma}{2}\frac{m_q}{\LUV}\frac{\sigma}{\LUV^3} + \sum_{n=2}^{n_\mathrm{max}} \frac{n-2}{2 n} g_n c_\s^n \s^n \nn\\
&&-\frac{\n c_\sigma K_\mathrm{IR}^2}{8 K_m K_\sigma \sin(\phi_m\!-\!\phi_\sigma)}\ e^{-4 \vs} \ .
\eea
Notice that the term involving $g_n$ in~\eqref{spiralEmulti} vanishes for $n=2$, which is consistent with nonzero $g_2$ amounting to a redefinition of the parameters of the Efimov spiral. The higher order terms are suppressed as $u$ grows because $\s \sim \exp(-2u)$. Therefore we conclude as above that the solutions with $m_q=0$ but $\sigma \ne 0$ dominate over the solution with $\s=0$, and that the solution with the smallest $\vs$ (for $\vs\gg 1$ so that the parametrization of the spiral is accurate) has the lowest free energy.

\subsection{Contributions $\propto m_q^3$ in the on-shell action} \label{app:nltachyon}

The on-shell action may involve terms $\propto m_q^3$ which arise due to the nonlinear nature of the tachyon EoM. We will now check how these terms behave. We will need the UV asymptotics of the source and the vev terms of the tachyon which are given in equations~\eqref{tau0exp} and in~\eqref{tau1exp} in Appendix~\ref{app:freeen}.
Using results from~\cite{jk}, the tachyon EoM can be written in the UV as
\be
 \t'' - \frac{3}{r} \t' + \frac{1}{r^2}\left(3-\frac{2\rho}{\log(r\L)}\right) \t -\frac{4 \kappa_0 r}{\ell^2} (\t')^3 + \frac{3 \kappa_0}{\ell^2}\t (\t')^2 \simeq 0\ ,
\ee
where $\kappa_0 =\kappa(0)$. All contributions suppressed by $1/\log(r\L)^2$ or more are hidden in the coefficients of the terms linear in the tachyon, as well as all contributions suppressed by $1/\log(r\L)$ in the nonlinear terms (including the complete terms $\propto \t^2\t'$ and $\propto \t^2(\t')^3$). 

The ``nonlinear'' tachyon solution is then found by writing
\be
 \t(r) = \t_0(r)+ \t_q(r) \ ,
\ee
where the UV expansion of  $\t_0$ is given in~\eqref{tau0exp} and $\t_q$ is the term which needs to be solved. Using the fact the $\tau_0$ solves the linear equation, one finds that
\bea
 \t_q'' - \frac{3}{r} \t_q' + \frac{1}{r^2}\left(3-\frac{2\rho}{\log(r\L)}\right) \t_q &\simeq&  -\frac{4 \kappa_0 r}{\ell^2} (\t_0')^3 + \frac{3 \kappa_0}{\ell^2}\t_0 (\t_0')^2 \nn \\
&\simeq& \kappa_0 m_q^3 \ell \left(-\log(r\L)\right)^{-3\rho} \ .
\eea
One finds that (dropping the terms corresponding to the source and vev terms, which arise as solutions to the homogeneous equation)
\be
 \frac{\t_q}{\ell} = \frac{\kappa_0 m_q^3}{2(4\rho-1)} r^3 \left(-\log(r\L)\right)^{-3\rho+1}\left[1 +\morder{\frac{1}{\log(r\L)}}\right] \ ,
\ee
which agrees with the result of~\cite{ikp} when $\rho=0$. 

Notice that the nonlinear term is subleading to the vev term in~\eqref{tau1exp} provided that
\be
 \rho > \frac{1}{4} \ ,
\ee
which is satisfied for QCD in the Veneziano limit as
\be
 \rho = \frac{\gamma_0}{b_0} = \frac{9}{22-4 x} \ge \frac{9}{22} > \frac{1}{4} \ ,
\ee
where $b_0$ and $\gamma_0$ are the leading coefficients of the beta function and the anomalous dimension of the quark mass, respectively. Therefore one can conclude that only linear terms of the tachyon EoM are relevant in the computation of the on-shell action\footnote{Notice that even though the nonlinearities in the tachyon asymptotics do not appear directly in the computation of the on-shell action, nonlinear terms of the EoM are important in general because they affect the values of the vevs such as the chiral condensate through IR boundary conditions.}. The conclusion is different from that of~\cite{ikp} where the quark mass did not run. Notice, however, that these terms are only logarithmically suppressed and would be important for large values of $m_q$ if a finite value of the UV cutoff was used.

\subsection{Regularized on-shell action} \label{app:freeen}

The free energy is given by the on-shell value of the action. It is a UV divergent quantity and needs to be renormalized\footnote{For detailed analysis of the holographic renormalization of dilaton gravity, see~\cite{Papadimitriou:2011qb,Kiritsis:2014kua}}. We work here at finite temperature, and subtract the UV divergences by considering the difference with respect to a reference (zero temperature) background with the same values for the sources. We could also consider the difference between two different zero temperature backgrounds. 

It is not difficult to show that the Lagrangian is a total derivative even in the presence of the tachyon (as already shown at zero temperature in~\cite{jk}). After a straightforward computation, and taking into account the Gibbons-Hawking term, one finds that
\be \label{Sonshell}
 S_\mathrm{on-shell} = M^3 N_c^2\beta V_3 \, \left.(6 A'f+f')e^{3A}\right|_{r=\eps} + \mathrm{counterterms} \ .
\ee

When the quark mass is finite, the tachyon contributes at \order{r^2} (terms $\propto m_q^2$) and at \order{r^4} (terms  $\propto m_q \sigma$) in the UV expansions. These contributions need to be taken into account in the holographic renormalization procedure. The calculation of~\cite{ihqcd2} at $N_f=0$ used a reference (thermal gas) solution to subtract the divergences. We will use the same technique here. Denoting the fields and other variables of the reference solution by tildes, the following conditions need to be fulfilled at the UV cutoff:
\begin{align}
\label{betaVconds}
 \tilde \beta e^{\tilde A(\tilde \eps)} &= \beta e^{ A(\eps)}\sqrt{f(\eps)}\ ,&  \qquad \tilde V_3 e^{3 \tilde A( \tilde \eps)} &= V_3 e^{3 A(\eps)}\ , &\\
\label{latauconds}
 \tilde \l(\tilde \eps) &= \l(\eps)\ , & \tilde \tau (\tilde \eps) &= \tau(\eps) \ ,
\end{align}
where the possibility that the cutoffs of the two solutions are different was included, $\tilde \eps \ne \eps$, as in~\cite{ihqcd2}. In the absence of the tachyon this was convenient since the UV scales $\Lambda=\LUV$ of the two solutions could be chosen to be the same, $\tilde \Lambda = \Lambda$. The fact that the present system has two scalars suggest that it is better to choose here $\tilde \eps =\eps$ and satisfy the last two conditions by varying the sources $\Lambda$ and $m_q$. However, it turns out to be convenient to still require that $\tilde \Lambda = \Lambda$ and vary the cutoff (and $m_q$) instead. This is so because the tachyon contributions are independent on whether one chooses to vary $\Lambda$ or $\eps$: The variation if suppressed by \order{\eps^4}, i.e., $\tilde \eps/\eps = 1 + \morder{\eps^4}$, so that the effect on the tachyon contributions will be down by \order{\eps^6} and therefore negligible. Keeping $\Lambda$ fixed one can maintain very close contact to the computation of~\cite{
ihqcd2}.

Turning on the quark mass will modify the UV expansions of $A$ and $\l$. at this point it is useful to write down the complete UV expansion including all relevant terms. One can decompose, as $r \to 0$,
\be
 A=A_0+A_\t+A_1\ , \qquad \l = \l_0 +\l_\t + \l_1\ , \qquad f = 1 + f_1 \ , \qquad \t = \t_0 + \t_1 \ .
\ee
The source terms have the expansions~\cite{jk}
\bea
\label{A0exp}
 A_0 &=& -\log r + \log \ell +\frac{4}{9\log r\Lambda}+ \morder{\frac{1}{(\log r\Lambda)^2}} \\
 b_0 \l_0 &=& -\frac{1}{\log r\Lambda} + \morder{\frac{1}{(\log r\Lambda)^2}}\\
\label{tau0exp}
 \frac{\t_0}{\ell} &=& m_q\,  r (-\log r\Lambda)^{-\rho}\left[1+\morder{\frac{1}{\log r\Lambda}}\right]  \ .
\eea
and $\rho=\gamma_0/b_0$ with $\gamma_0$ being the leading coefficient of the anomalous dimension of the quark mass. The \order{r^2} terms $A_\t$ and $\l_\t$ were also included here which are sourced by the \order{r^2} tachyon perturbation\footnote{There are also \order{r^4} terms which arise due to the perturbation from the quartic terms in the tachyon. As we argued above, these contributions are subleading with respect to those arising from the vev terms.}. As it turns out, $\l_\t$ is suppressed by logarithms of $\log r \Lambda$ with respect to $A_\t$ so that it will not enter the calculation and its expansion will not be needed. $A_\t$ is given by
\be
 A_\t = -\frac{1}{36} B_\s m_q^2 \frac{r^2}{\ell^3} \left(-\log r\Lambda\right)^{-2\rho}\left[1+\morder{\frac{1}{\log r\Lambda}}\right] \ ,
\ee  
where $W_0=V_{f0}(0)$, $\kappa_0 =\kappa(0)$, and as above
\be
 B_\s = x W_0 \kappa_0 \ell^5 \ .
\ee
The vev terms have the expansions (compare to~\cite{ihqcd2})
\bea 
\label{A1exp}
 A_1(r) &=& \mG \frac{r^4}{\ell^3}\left[1+\morder{\frac{1}{\log r\Lambda}}\right] \\
 b_0 \l_1(r) &=& \left(-\frac{45}{8} \mG -\frac{9}{32} B_\s m_q \sigma \right) \frac{r^4}{\ell^3}\left[1+\morder{\frac{1}{\log r\Lambda}}\right] \\
 f_1(r) &=& -\frac{C}{4}\frac{r^4}{\ell^3}\left[1+\morder{\frac{1}{\log r\Lambda}}\right] \\
\label{tau1exp}
 \frac{\t_1(r)}{\ell} &=& \sigma\,  r^3  (-\log r\Lambda)^{\rho}\left[1+\morder{\frac{1}{\log r\Lambda}}\right]\ .
\eea

Let us then go on with the renormalization procedure. In the expressions below only the differences of the vevs $\mG$ and $\sigma$ with respect to the values of the reference solutions will appear. Therefore, without loss of generality, one can set the vevs of the reference solution to zero.
As motivated above, one can choose $\tilde \Lambda = \Lambda$. Then the relation between $\tilde \eps$ and $\eps$ is fixed\footnote{It can be checked that the variation of $m_q$, which would enter through the term $\l_\t$ is subleading.} by the first condition in~\eqref{latauconds}:
\be \label{tepsdef}
 \frac{\tilde \eps}{\eps} = 1 + \left(-\frac{45}{8} \mG -\frac{9}{32} B_\s m_q \sigma\right) \frac{\eps^4}{\ell^3}(-\log\eps\Lambda)^2\left[1+\morder{\frac{1}{\log \eps\Lambda}}\right]\ ,
\ee
whereas the second condition sets
\be
 \tilde m_q = m_q +\sigma \eps^2 (-\log\eps\Lambda)^{2\rho}\left[1+\morder{\frac{1}{\log \eps\Lambda}}\right]\ .
\ee

It is then straightforward to calculate the renormalized pressure. Inserting the expression of the (unrenormalized) on-shell action from Eq.~\eqref{Sonshell} one finds that
\bea
 -\beta V_3 p &=&\lim_{\eps \to 0}\left[S(\eps)-\tilde S(\tilde \eps)\right] \\
&=&M^3N_c^2\lim_{\eps \to 0} \left[ \beta V_3 (6 A'(\eps)f(\eps)+f'(\eps))e^{3A(\eps)}- 6 \tilde \beta \tilde V_3  \tilde A'( \tilde\eps)e^{3 \tilde A(\tilde \eps)}\right] \ .
\eea
After eliminating $\tilde \beta$ and $\tilde V_3$ by using the conditions~\eqref{betaVconds}, the expression for the pressure reads
\be
\label{prenorm}
 p =M^3N_c^2\lim_{\eps \to 0}e^{3A(\eps)}\left[6 \tilde A'(\tilde \eps) e^{A(\eps)- \tilde A(\tilde \eps)}\sqrt{f(\eps)}-6 A'(\eps)f(\eps)-f'(\eps)\right]\ .
\ee
The relation~\eqref{tepsdef} implies that
\be
 \tilde A'(\tilde \eps) e^{- \tilde A(\tilde \eps)} = \tilde A'(\eps) e^{- \tilde A(\eps)} + \left(\frac{5}{2} \mG +\frac{1}{8} B_\s m_q \sigma \right)\frac{\eps^4}{\ell^4}\left[1+\morder{\frac{1}{\log \eps\Lambda}}\right]
\ee
where the expansion~\eqref{A0exp} was used. Further, notice that the variation of $m_q$ enters through
\be
 A_\t(\eps) =  \tilde A_\t(\eps) +\frac{1}{18} B_\s m_q \sigma \frac{\eps^4}{\ell^3}\left[1+\morder{\frac{1}{\log \eps\Lambda}}\right]
\ee
so that
\be
  A(\eps) = \tilde A(\eps) +\left[\mG+\frac{1}{18} B_\s m_q \sigma\right] \frac{\eps^4}{\ell^3}\left[1+\morder{\frac{1}{\log \eps\Lambda}}\right] \ .
\ee
A similar result can be found for the derivative of $A$. Inserting these relations and the expansion of $f$ in~\eqref{prenorm} one obtains the final result\footnote{Recall that the vevs here should be interpreted as their differences with respect to the values of the reference solutions, because we set the vevs of the reference solution to zero.}
\be
 \label{pmqne0}
 p =M^3N_c^2 \left(\frac{1}{4}C-15\mG-\frac{1}{4}B_\s m_q \sigma\right) \ .
\ee

\section{Gell-Mann-Oakes-Renner relation} \label{app:gor}

In this Appendix we check the GOR relation explicitly. The starting point is the normalization integral~\eqref{PSintnorm} for the pion wave function.  
When $m_q$ is small, it is dominated\footnote{As can be verified numerically, this holds for the pion state but not for higher pseudoscalar states.} at small $u\simeq r$. Also the wave function $\hat \psi_P$ is constant in the UV up to corrections \order{r^2} (see Appendix~G in~\cite{Arean:2013tja}).
Therefore integral one needs to compute is (the UV contribution to)
\be \label{Idef}
 I = \int_0 dr \frac{1}{V_{f}\kappa e^{3A} \t^2} \ .
\ee

Let us first try to compute the integral by using the UV expansions of the various fields, given in Appendix~\ref{app:freeen}. We obtain
\bea
 I &=& \frac{1}{W_0\kappa_0 \ell^3}\int_0^\infty dr \frac{r^3}{\t(r)^2}\left[1+\morder{\frac{1}{\log r\L}}\right]\\\nn
 &=& \frac{1}{W_0\kappa_0 \ell^5}
\int_0^\infty \!\!\!\! \frac{dr\  r^3}{\left\{m_q r (-\!\log r\L)^{-\rho}\left[1\!+\!\morder{\frac{1}{\log r\L}}\right]+\s r^3 (-\!\log r\L)^{\rho}\left[1\!+\!\morder{\frac{1}{\log r\L}}\right]\!\right\}^2} 
\eea
where $\L=\LUV$ and $\kappa_0 =\kappa(0)$.
As $m_q \to 0$ this integral is dominated by the regime with $r \sim \sqrt{m_q/\s}$. Substituting here $v \simeq r(-\log r\L)^\rho$ one obtains 
\be \label{Ires}
 I =\frac{1}{W_0\kappa_0 \ell^5} \int_0^\infty dv  \frac{v^3\left[1+\morder{\frac{1}{\log v\L}}\right]}{\left(m_q v +\s v^3 \right)^2} = \frac{1}{2W_0\kappa_0 \ell^5 m_q \s}\left[1+\morder{\frac{1}{\log (m_q\L^2/\s)}}\right] \ .
\ee
Notice that in the walking regime corrections are suppressed only when $m_q$ is a small perturbation, i.e., \eqref{walkingmqlim} holds. Also, the result~\eqref{Ires} is not valid in the conformal window for any $m_q$.

From~\eqref{Ires} we see that the UV RG flow leads to correction terms which are suppressed only by logarithms of $m_q$. It is however possible to show that such corrections vanish to all orders. First we notice that the tachyon EoM~\eqref{taueom} can be written as
\be
 \tau'' + \frac{d}{dr}\log\left(e^{3A}V_{f0}\kappa\right) \t' + \frac{2 e^{2A}a}{\kappa} \t = 0 
\ee
up to corrections suppressed by $\t^2$ in the UV. Let us denote by $\t_\s$ the solution having $m_q=0$ and $\sigma=1$, and let $\t$ be a generic solution with $m_q \ne 0$. The Wronskian
\be
 W \equiv \t \t_\s' - \t_\s \t'
\ee
satisfies
\be
 W' + \frac{d}{dr}\log\left(e^{3A}V_{f0}\kappa\right) W = 0
\ee
with the solution
\be \label{Wid}
 W = \t \t_\s' - \t_\s \t'= \frac{C_W}{e^{3A}V_{f0}\kappa} \ .
\ee
By inserting the UV expansions for all fields we see that the constant is given by
\be
 C_W = 2 W_0 \kappa_0 \ell^5 m_q \ .
\ee
Notice that~\eqref{Wid} may be written as
\be
 \frac{d}{dr} \frac{\t_\s}{\t} = \frac{C_W}{e^{3A}V_{f0}\kappa\t^2} \ .
\ee
By integrating this identity, the UV contribution to the integral~\eqref{Idef} becomes
\be
  \int_0^{r_\mathrm{cut}}\frac{1}{e^{3A}V_{f0}\kappa\t^2} = \frac{1}{C_W} \left.\frac{\t_\s}{\t}\right|_0^{r_\mathrm{cut}}  = \frac{1}{C_W} \frac{\t_\s(r_\mathrm{cut})}{\t(r_\mathrm{cut})}\ .
\ee
Choosing the cutoff such that $r_\mathrm{cut} \gg \sqrt{m_q/\s}$ (but so that it is not too large so that our approximation are still valid) we obtain
\be \label{IresUV}
 I = \int_0^{r_\mathrm{cut}}\frac{1}{e^{3A}V_{f0}\kappa\t^2} =\frac{1}{C_W \s}\left[1+\morder{\frac{m_q}{\s r_\mathrm{cut}^2}}\right] \ . 
\ee
Our approximations will break down at $r_\mathrm{cut} \sim 1/\LUV$, where the $r$-dependence of the tachyon changes qualitatively. In order to complete the calculation, one should estimate the contributions to the normalization integral~\eqref{PSintnorm} for $r \gtrsim 1/\LUV$. But in this regime the dependence on the tachyon is regular and one can just do a Taylor expansion at $m_q=0$. The cutoff dependence should cancel against the UV contribution given in~\eqref{IresUV}. As the dust clears, we can effectively set $r_\mathrm{cut} \sim 1/\LUV$, in~\eqref{IresUV}, obtaining
\be
 I = \frac{1}{C_W \s}\left[1+\morder{\frac{m_q \LUV^2}{\s}}\right] \ .
\ee

The normalization condition~\eqref{PSintnorm} becomes for small quark mass
\be
 1 = \hat {\psi}_{P,\pi}^2(0) I \left[1+\morder{\frac{m_q \LUV^2}{\s}}\right] = \frac{\hat {\psi}_{P,\pi}^2(0) }{2 \ell^5 W_0 \kappa_0 m_q \s}\left[1+\morder{\frac{m_q \LUV^2}{\s}}\right] \ ,
\ee
where we restricted to the lowest pseudoscalar mode, the pion. 
Solving for $\hat {\psi}_{P,\pi}(0)$ and inserting in~\eqref{fpimqne0} one obtains the GOR relation
\be
 f_\pi^2m_\pi^2 \simeq 2 M^3 N_f N_c \ell^5 W_0 \kappa_0 m_q \s = 2 M^3 N_c^2 B_\s m_q\s \simeq - m_q \langle\bar q q\rangle 
\ee
with corrections suppressed by $m_q \LUV^2/\s$.
It can be checked that the proportionality factor is correct for our definitions of $f_\pi$ and $\langle\bar q q\rangle$.

\section{Fluctuation equations, $f_\pi$, and the S-parameter}\label{app:fpi}

The radial wave functions for the flavor nonsinglet fluctuations satisfy the following equations~\cite{Arean:2013tja}:
\bea \label{flucteqs}
  \partial_u( C_V\partial_u  \psi_V) &=& C_V q^2 \psi_V \\
  \partial_u( C_V\partial_u  \psi_A) &=& C_V (H_A+q^2) \psi_A \\
  \partial_u( C_V \partial_u \psi_L) &=& C_V H_A(\psi_L-\psi_P) \\
  H_A \partial_u \psi_P &=& -q^2 \partial_u \psi_L
\eea
where
\begin{align} \label{GHCdefs}
 \frac{du}{dr} &= \sqrt{1+e^{-2A}\kappa (\t')^2} \equiv G\ ,&H_A &=\frac{4 \t^2 \kappa e^{2A} }{w^2}\ ,   \\
 C_V &= V_f w^2 e^A\ . & &  
\end{align}
In addition it is convenient to define the pseudoscalar wave function
\be \label{psiPdef}
 \hat \psi_P = -C_V \partial_u \psi_L
\ee
which satisfies the single equation
\be
  \partial_u( C_P \partial_u \hat \psi_P) = C_P(H_A+q^2) \hat \psi_P
\ee
where
\be
 C_P = \frac{4}{C_V H_A}  = \frac{1}{V_f \t^2 \kappa e^{3 A} }\ .
\ee

At nonzero quark mass the pion decay constant may be defined in terms of the pole of $\Pi_L$ in~\eqref{Pidef} at $q^2 = -m_\pi^2$. As the computation is similar also for higher modes, a generic pseudoscalar fluctuation can be considered. To compute the decay constants one needs to study the fluctuations for small values of 
\be
 \delta q^2 = q^2 +m_n^2
\ee
where $m_n^2$ is the mass of the fluctuation mode. The wave functions are written as 
\bea \label{psiexpdef}
 \psi_L &=& k_L\left[\psi_{L,n} +\delta q^2 \tilde \psi_{L}   + \morder{(\delta q^2)^2}\right] \\
 \psi_P &=& k_L\left[\psi_{P,n} + \delta q^2 \tilde \psi_{P}  + \morder{(\delta q^2)^2}\right]  \ ,
\eea
where $k_L$ is a normalization constant which will be fixed below.

When $\delta q^2 = 0$ the wave functions are normalizable. One can therefore choose the fields $\psi_{I,n}$ to satisfy the usual normalization condition
\be \label{PSintnorm}
 1 = \int_0^\infty du C_P \hat {\psi}_{P,n}^2 = \int_0^\infty du \frac{4 C_V}{H_A} (\partial_u  \psi_{L,n})^2 \ ,
\ee
with $\hat {\psi}_{P,n}$ defined as in~\eqref{psiPdef} for the normalizable mode.
At finite but small $\delta q^2$ we impose the standard normalization condition in the UV: 
\be \label{PSnormcond}
 1 = \psi_L(0) \simeq  k_L \delta q^2 \tilde \psi_{L}(0)\ ; \qquad  0 = \psi_P(0) \propto \tilde \psi_{P}(0) \ ,
\ee
where we used the fact that the normalizable wave functions vanish in the UV.

Expanding the fluctuation equations at small $\delta q^2$ gives
\bea
   \partial_u( C_V \partial_u \tilde \psi_L) &=& C_V H_A(\tilde \psi_L-\tilde \psi_P) \\
  H_A \partial_u \tilde \psi_P &=& m_n^2 \partial_u \tilde \psi_L- \partial_u \psi_{L,n} \ .
\eea
By using these equations and the fluctuation equations for $\psi_{I,n}$ one finds the identity
\bea
 &&\partial_u\left[m_n^2 C_V \left(\tilde \psi_L\partial_u \psi_{L,n}- \psi_{L,n}\partial_u \tilde \psi_L\right)+ H_A C_V  \left(\tilde \psi_P\partial_u \psi_{P,n}-\psi_{P,n}\partial_u \tilde \psi_P\right)\right] \nn\\
&=& H_A C_V \psi_{P,n}(\psi_{P,n}-\psi_{L,n})
\eea
Integrating this over $u$, using the boundary conditions from above, and further using the fluctuation equations to simplify the result one obtains
\be
 C_V \tilde \psi_L \partial_u \psi_{L,n}\big|_{u=0} = - \frac{1}{m_n^2} \int_0^\infty du\  \psi_{P,n} \partial_u \hat {\psi}_{P,n} \ .
\ee
Inserting~\eqref{PSnormcond}, integrating partially, and using again fluctuation equations this relation simplifies to
\be
   C_V \partial_u \psi_{L,n}\big|_{u=0} =  - \frac{k_L \delta q^2}{4  } \int_0^\infty du C_P  \hat { \psi}_{P,n}^2 = -\frac{k_L\delta q^2}{4 } \ .
\ee
By using this result and the definitions from~\eqref{psiexpdef}, $\Pi_L$ from~\eqref{Pidef} can be written as
\bea \label{PiLresfin}
 \Pi_L(q^2) &=& \frac{M^3 N_f N_c}{4 q^2} C_V \partial_u \psi_L\big|_{u=0} \simeq  \frac{M^3 N_f N_c}{4 q^2} C_V k_L \partial_u \psi_{L,n}\big|_{u=0} \nn\\
 &=& - \frac{M^3 N_f N_c}{q^2} \frac{1}{\delta q^2}  (C_V \partial_u \psi_{L,n})^2\big|_{u=0} 
\eea
for small $\delta q^2$. The decay constants are given in terms of the residue at $\delta q^2 = 0$: 
\be \label{fpimqne0}
 f_n^2 m_n^2 = M^3 N_f N_c (C_V \partial_u \psi_{L,n})^2\big|_{u=0} = M^3 N_f N_c ( \hat {\psi}_{P,n})^2\big|_{u=0} \ .
\ee

From~\eqref{PiLresfin} we see that for $q^2 \sim m_\pi^2$, as $m_q \to 0$,
\be
  \Pi_L(q^2) \simeq \frac{f_\pi^2m_\pi^2}{q^2(q^2+m_\pi^2)} = \frac{f_\pi^2}{q^2} -\frac{f_\pi^2}{q^2+m_\pi^2}  
\ee
so that $S_0$ (defined as the zero momentum value of $q^2\Pi_L(q^2)$) approaches $f_\pi^2$. The same factor for the axial-axial correlator does not have a pion node, so that simply
\be
\Pi_A(q^2) \simeq \frac{f_\pi^2}{q^2}
\ee
in the same scaling limit. Consequently, the axial-axial correlator has the correct structure at small momentum:
\be
 \left( q^2\eta^{\m\n}-q^\m q^\n\right) \Pi_A(q^2) + q^\m q^\n \Pi_L(q^2) \simeq f_\pi^2 \left(\eta^{\m\n} - \frac{q^\m q^\n}{q^2+m_\pi^2}\right) \ .
\ee
That is, the longitudinal term arises from the coupling of the pion to the axial current.

Let us then derive a few integral representations which are useful when computing observables numerically. First we analyze
\be
 D(q^2) \equiv q^2 \Pi_A(q^2)-q^2 \Pi_A(q^2) = \frac{1}{4} M^3 N_f N_c C_V \left[\pa_u\psi_V(u,q^2)-\pa_u\psi_V(u,q^2)\right]_{u=\eps} \ ,
\ee
where we inserted \eqref{Pidef}, and used the fact that $r\simeq u$ near the boundary. 
Recalling the normalization $\psi_V(u=\eps)=1=\psi_V(u=\eps)$ we obtain
\begin{align}
C_V\left[\pa_u\psi_V-\pa_r\psi_V\right]_{u=\eps} &= \int_\eps^\infty du\ \pa_u\left[\psi_V C_V \pa_u\psi_A-\psi_A C_V \pa_u\psi_V \right] \nn\\
 &= \int_\eps^\infty du\  \psi_V H_A C_V \psi_A \ ,
\end{align}
where fluctuation equations~\eqref{flucteqs} were used at the last step. Taking $\eps \to 0$, it follows that
\be
 D(q^2) = \frac{1}{4} M^3 N_f N_c \int_0^\infty du\  \psi_V H_A C_V \psi_A \ .
\ee

A rather similar formula can be derived for the S-parameter. Let us denote 
\be
 \left.\frac{\pa}{\pa q^2} \psi_{V/A}\right|_{q^2=0} = \dot \psi_{V/A} \ .
\ee
Then the S-parameter~\eqref{Sdiffform} can be written as
\be
 S = \pi M^3 N_f N_c C_V \pa_u\left(\dot \psi_A -\dot \psi_V\right)_{u=\eps} \ .
\ee
Derivating the fluctuation equations with respect to $q^2$ results in
\begin{align} \label{fluctdots}
 \pa_u(C_V\pa_u \dot \psi_V) & = C_V \psi_V \nn \\
 \pa_u(C_V\pa_u \dot \psi_A) & = C_V \psi_A+ C_V H_A \dot \psi_A 
\end{align}
where it is understood that $\psi_{V,A}$ are evaluated at $q^2=0$. The combination appearing in the S-parameter can then be written as
\begin{align}
C_V \pa_u\left[\dot \psi_A -\dot \psi_V\right]_{u=\eps} &\simeq \int_\eps^\infty du\ \pa_u\Big[\psi_V C_V \pa_u \dot \psi_V - \dot \psi_V C_V \pa_u \psi_V \nn\\
&\ \ \ \  \ -\psi_A C_V \pa_u \dot \psi_A + \dot \psi_A C_V \pa_u \psi_A\Big] \\
 &= \int_\eps^\infty du \ C_V \left(\psi_A^2-\psi_V^2 \right) \ ,
\end{align}
where we dropped higher order terms in $\eps$ in the first step (noticing that $\dot \psi_{V,A}$ vanish fast in the UV) and used the fluctuation equations~\eqref{flucteqs} as well as equations~\eqref{fluctdots} in the second step. Taking $\eps \to 0$, we obtain the final result
\be 
 S= \pi M^3 N_f N_c \int_0^\infty du \ C_V \left[\psi_A^2-\psi_V^2\right]_{q^2=0} \ .
\ee
Here one could also insert that $\psi_V = 1$ when $q^2=0$.

\section{Free field (one-loop) computation of the S-parameter} \label{app:SFT}

The free field result for the vector-vector and axial-axial correlators is given by
\bea
 &&\int d^4x\ e^{-iqx}\ \langle 0| \ T\!\left\{ J_{\mu}^{a\,(V/A)}(x) J_{\n}^{b\,(V/A)}(0)\right\} |0\rangle  \nn\\
 &&= \ \frac{N_c\delta^{a b}}{2}\int \frac{d^4 k}{(2\pi)^4} \frac{\mathrm{Tr}\left[\gamma^\m(\gamma_5)(\aslash{k}+m)\gamma^\n(\gamma_5)(\aslash{k}-\aslash{q} + m)\right]}{(k^2+m^2-i\eps)\left[(k-q)^2+m^2-i\eps\right]} \ ,
\eea
where the $\gamma_5$'s are only present in the axial-axial correlator. The loop is divergent but the contribution to the S-parameter will be finite.

Doing the integral with dimensional regularization one obtains
\bea
 \Pi_V(q^2) &=& \frac{2N_fN_c}{(4\pi)^2} \int_0^1 dx \, x(1-x)\nn\\
&&\times \left[\frac{2}{\eps}-\log(m^2+x(1-x)q^2)-\gamma + \log 4\pi +\morder{\eps}\right]\\
 \Pi_A(q^2) &=& \frac{2N_fN_c}{(4\pi)^2} \int_0^1 dx \left[x(1-x)+\frac{m^2}{q^2}\right]\nn\\
&&\times  \left[\frac{2}{\eps}-\log(m^2+x(1-x)q^2)-\gamma + \log 4\pi +\morder{\eps}\right]\\
 \Pi_L(q^2) &=& \frac{2N_fN_c}{(4\pi)^2} \frac{m^2}{q^2}  \nn\\
&&\times  \int_0^1 dx \left[\frac{2}{\eps}-\log(m^2+x(1-x)q^2)-\gamma + \log 4\pi +\morder{\eps}\right] \ .
\eea
Consequently
\be
 D(q^2)-D(0) = -\frac{2N_fN_c\,m^2}{(4\pi)^2}  \int_0^1 dx\, \log\left[1+x(1-x)\frac{q^2}{m^2}\right]\ ,
\ee
where the value at $q^2=0$ was subtracted in order to remove the logarithmic divergence.

For the S-parameter one obtains the well-known result
\be
 S = -4\pi D'(0) = \frac{N_fN_c}{2\pi} \int_0^1 dx\, x(1-x) = \frac{N_fN_c}{12\pi}\ ,
\ee
whereas the corresponding finite difference reads
\bea
-4\pi\frac{D(q^2)-D(0)}{q^2} &=&  \frac{N_fN_c}{2\pi} \frac{m^2}{q^2}  \int_0^1 dx\, \log\left[1+x(1-x)\frac{q^2}{m^2}\right]\\
&=&  \frac{N_fN_c}{\pi} \frac{m^2}{q^2}\left[\sqrt{1+4\frac{m^2}{q^2}}\,\mathrm{arctanh}\frac{1}{\sqrt{1+4\frac{m^2}{q^2}}}-1 \right]
\eea
in agreement with~\cite{sannino}. The series as $q^2 \to 0$ and as $m^2\to 0$ are given by
\bea
 -4\pi\frac{D(q^2)-D(0)}{q^2} &=& \frac{N_fN_c}{12\pi}\left[1-\frac{1}{10}\frac{q^2}{m^2} + \frac{1}{70}\frac{q^4}{m^4} +\cdots \right] \\\nn
&=& \frac{N_fN_c}{\pi}\left[-\frac{m^2}{q^2}-\frac{m^2}{2q^2}\log\frac{m^2}{q^2} +\frac{m^4}{q^4} - \frac{m^4}{q^4}\log\frac{m^2}{q^2} + \cdots\right]
\eea

\section{Details on numerics}\label{app:numerics}

The numerical results in this paper were computed for sets of potentials termed ``potentials I'' and ``potentials II''. They are exactly the sets given in~\cite{Arean:2013tja}, but we repeat their definitions here for completeness:
\begin{itemize}
 \item \textbf{Both Potentials I \& II.}
  \begin{align} \label{potIandIIcommon}
    V_{g}(\l)  & = V_0\left[1+V_1 \l + V_2 \l^2 \frac{\sqrt{1+\log(1+\frac{\lambda}{\l_0})}}{\left(1+\frac{\lambda}{\l_0}\right)^{2/3}}\right]\ , \nonumber\\
    V_{f0}(\l) & = W_0\left[1+W_1 \l + W_2 \l^2\right]\ .
  \end{align}
 \item \textbf{Potentials I.}
  \be \label{potIdefs}
    a(\l)   = a_0\ ,\qquad   \h(\l)  = \frac{1}{\left(1+\frac{3a_1}{4}\l\right)^{4/3}}\ .
  \ee
 \item \textbf{Potentials II.}
  \be \label{potIIdefs}
    a(\l)   = a_0\,\frac{1+a_1 \l + \frac{\l^2}{\l_0^2}}{\left(1+\frac{\l}{\l_0}\right)^{4/3}}\ , \qquad    \h(\l)  = \frac{1}{\left(1+\frac{\l}{\l_0}\right)^{4/3}}\ .
  \ee
\end{itemize}

Here the most of the coefficients are fixed by matching to perturbative QCD. Exceptions include the normalizations factors $V_0$, which fixes the UV AdS radius, and $W_0$, which remains as a free parameter. We also set $\ell(x=0)=1$, and choose the parameter $\l_0$, which only affects the higher order coefficients of the UV expansions, such that the higher order coefficients have approximately the same relative size as with standard scheme choices in perturbative QCD. Explicitly, the coefficients satisfy
\begin{align}
 V_0 &= 12\ , \qquad V_1 = \frac{11}{27 \pi^2}\ ,\qquad  V_2= \frac{4619}{46656 \pi ^4}\ ; \nonumber \\
 W_1 &= \frac{24+(11-2 x) W_0}{27 \pi ^2 W_0}\ ,\qquad W_2 = \frac{24 (857-46 x)+\left(4619-1714 x+92 x^2\right) W_0}{46656 \pi ^4 W_0}\ ; \nonumber\\
 a_0 &= \frac{12-x W_0}{8} \ , \qquad a_1 =  \frac{115-16 x}{216 \pi ^2} \ , \qquad \l_0 = {8 \pi^2} \ .
\end{align}

Two qualitatively different choices for $W_0$ are possible: either constant $W_0$, which satisfies
\be \label{W0range}
0<W_0<24/11\ ,
\ee
or $W_0$ fixed such that the pressure agrees with the Stefan-Boltzmann (SB) result at high temperatures \cite{alho} (without the need to include $x$ dependence in the normalization of the action). The latter option is given explicitly (when $\ell(x=0)=1$) by
\be
 W_0 = \frac{12}{x}\left[1-\frac{1}{(1+\frac{7}{4}x)^{2/3}}\right]\qquad \textrm{(Stefan-Boltzmann)}\ ,
\ee
so that the AdS radius is
\be
 \ell(x) = \sqrt[3]{1+\frac{7}{4}x}\ .
\ee
In this article we have always chosen $W_0=3/11$ for potentials~I, and the SB normalized $W_0$ for potentials~II. For the coupling of the gauge fields $w(\l)$ which is required for the computation of the vector correlators, we used $w(\l)=\kappa(\l)$ for potentials~I and $w(\l)=1$ for potentials~II.

The numerical result were mostly computed as detailed in~\cite{jk,Arean:2013tja,alho}: coupled ordinary differential equations were solved by shooting from the IR, either starting from near the IR singularity (at zero temperature) or near the horizon (at finite temperature). In order to obtain accurate and reliable results, some tricks had to be used in various cases,  in particular near the BZ point $x=\xBZ$ and at large quark masses. 

In general, the difficulties of the numerics in V-QCD arise from two sources: the IR singularity and the logarithmic corrections to the UV asymptotics. The latter easily lead to sizeable numerical errors when one tries to extract the values of the sources or vevs near the boundary. To improve the accuracy of the numerical analysis, we did the following:
\begin{itemize}
 \item The scale factor $A$ was used as the coordinate instead of $r$. This makes it easier to analyze the UV asymptotics, because the mapping from $r$ to $A$ is logarithmic and expands the details near the boundary.
 \item Near the IR singularity (at zero temperature), where possible, EoMs were implemented such that they do not contain the extremely small factors $\propto \exp(-a\t^2)$. In many cases, this could be done by writing all expression in terms of $V_\mathrm{log}(\l,\t) = \log V_f(\l,\t) = -a\t^2 + \log V_{f0}(\l)$ rather than in terms of $V_f$.
 \item Near the BZ region and at very large $m_q$, the zero temperature backgrounds were constructed by shooting from the IR toward the UV in four steps. Very close to the singularity, where the tachyon is large and completely decouples the flavors, YM solution was used for the geometry and the dilaton, whereas the tachyon was solved by using a simplified EoM with only the dominant terms on top of the YM background. In the next step, the complete tachyon EoM together with the decoupled YM flow for the geometry was used. In the third step, all fields were solved from the full coupled EoMs. In the final step, the tachyon was again solved separately since it is decoupled from the other fields near the boundary. Actually $\hat \t = e^{A} \t$ was used as the field, because it decreases much slower than $\t$, and the flow can then be tracked closer to the boundary. 

The first two steps were necessary because the tachyon EoM becomes stiff in the IR, in particular for large $m_q$, so that the tachyon takes larger values in the IR than 
otherwise. The last step was useful in particular at large $x$, where the RG flow of the fields is slower, and one needs to solve them closer to the boundary in order to control the leading logarithmic behavior of the various terms. 
 \item For the finite temperature backgrounds in the BZ region and at large $m_q$, the background was solved in two steps, which were analogous to the two last steps of the zero temperature construction. That is, the full system was used in the IR, and the tachyon was treated as a decoupled field near the UV. 
 \item The S-parameter was computed by using the integral formula~\eqref{Sparamint} rather than UV asymptotics of the fluctuation wave functions.
 \item The nonsinglet meson wave functions were computed by rewriting the fluctuation equations into a system of two coupled first order equations (rather than a second order equation). Again the equations were written in a form which does not contain explicitly the tiny factors $\exp(-a\t^2)$.
\end{itemize}

\end{document}